\documentclass{emulateapj}
\usepackage{amsfonts}
\usepackage{lineno}
\usepackage{multirow}
\usepackage{color}
\usepackage{hyperref}
\usepackage{breakurl}

\shorttitle{Multi-wavelength Study of Quiescent States of Mrk\,421 in 2013}
\shortauthors{M. Balokovi\'{c} {\em et~al.}}

\def \etal {{et~al.\,}}
\def \note #1 {{\tt NOTE: #1}}
\def \tbd #1 {{\textcolor{red}{! #1 !} }}

\def \swift {{\em Swift\ }}
\def \swiftxrt {{\em Swift}\,-XRT\ }

\def \swiftuvot {{\em Swift}\,-UVOT\ }

\def \rxte {{\em RXTE\ }}
\def \rxtepca {{\em RXTE}\,-PCA\ }

\def \veritas {{VERITAS\ }}
\def \magic {{MAGIC\ }}

\def \xspec {{\tt Xspec\,}}
\def \elgamma {{\gamma}}

\slugcomment{  }

\begin{document}

\title{{\bf Multiwavelength Study of Quiescent States of Mrk\,421\ \\ with Unprecedented Hard X-ray Coverage Provided by {\em NuSTAR}\ in 2013}}

\author{
M.\,Balokovi\'{c}\altaffilmark{1,a},
D.\,Paneque\altaffilmark{2,b},
G.\,Madejski\altaffilmark{3,c},
A.\,Furniss\altaffilmark{4,d},
J.\,Chiang\altaffilmark{3,e},\\ 
M.\,Ajello\altaffilmark{5},
D.\,M.\,Alexander\altaffilmark{6},
D.\,Barret\altaffilmark{7,8},
R.\,D.\,Blandford\altaffilmark{3}, 
S.\,E.\,Boggs\altaffilmark{9},
F.\,E.\,Christensen\altaffilmark{10},
W.\,W.\,Craig\altaffilmark{9,11},
K.\,Forster\altaffilmark{1},
P.\,Giommi\altaffilmark{12},
B.\,Grefenstette\altaffilmark{1},
C.\,Hailey\altaffilmark{13},
F.\,A.\,Harrison\altaffilmark{1}, 
A.\,Hornstrup\altaffilmark{10},
T.\,Kitaguchi\altaffilmark{14,15},
J.\,E.\,Koglin\altaffilmark{3},
K.\,K.\,Madsen\altaffilmark{1},
P.\,H.\,Mao\altaffilmark{1},
H.\,Miyasaka\altaffilmark{1},
K.\,Mori\altaffilmark{16},
M.\,Perri\altaffilmark{12,17},
M.\,J.\,Pivovaroff\altaffilmark{9},
S.\,Puccetti\altaffilmark{12,17},
V.\,Rana\altaffilmark{1},
D.\,Stern\altaffilmark{18}, 
G.\,Tagliaferri\altaffilmark{19},
C.\,M.\,Urry\altaffilmark{20}, 
N.\,J.\,Westergaard\altaffilmark{10},
W.\,W.\,Zhang\altaffilmark{21},
A.\,Zoglauer\altaffilmark{9} \\ 
(The {\em NuSTAR} Team), \\
S.\,Archambault\altaffilmark{22},
A.\,Archer\altaffilmark{23},
A.\,Barnacka\altaffilmark{24},
W.\,Benbow\altaffilmark{25},
R.\,Bird\altaffilmark{26},
J.\,H.\,Buckley\altaffilmark{23},
V.\,Bugaev\altaffilmark{23},
M.\,Cerruti\altaffilmark{25},
X.\,Chen\altaffilmark{27,28},
L.\,Ciupik\altaffilmark{29},
M.\,P.\,Connolly\altaffilmark{30},
W.\,Cui\altaffilmark{31},
H.\,J.\,Dickinson\altaffilmark{32},
J.\,Dumm\altaffilmark{33},
J.\,D.\,Eisch\altaffilmark{32},
A.\,Falcone\altaffilmark{34},
Q.\,Feng\altaffilmark{31},
J.\,P.\,Finley\altaffilmark{31},
H.\,Fleischhack\altaffilmark{28},
L.\,Fortson\altaffilmark{33},
S.\,Griffin\altaffilmark{22},
S.\,T.\,Griffiths\altaffilmark{35},
J.\,Grube\altaffilmark{29},
G.\,Gyuk\altaffilmark{29},
M.\,Huetten\altaffilmark{28},
N.\,H{\aa}kansson\altaffilmark{27},
J.\,Holder\altaffilmark{36},
T.\,B.\,Humensky\altaffilmark{37},
C.\,A.\,Johnson\altaffilmark{37},
P.\,Kaaret\altaffilmark{35},
M.\,Kertzman\altaffilmark{38},
Y.\,Khassen\altaffilmark{26},
D.\,Kieda\altaffilmark{39},
M.\,Krause\altaffilmark{28},
F.\,Krennrich\altaffilmark{32},
M.\,J.\,Lang\altaffilmark{30},
G.\,Maier\altaffilmark{28},
S.\,McArthur\altaffilmark{40},
K.\,Meagher\altaffilmark{41},
P.\,Moriarty\altaffilmark{30},
T.\,Nelson\altaffilmark{33},
D.\,Nieto\altaffilmark{13},
R.\,A.\,Ong\altaffilmark{42},
N.\,Park\altaffilmark{40},
M.\,Pohl\altaffilmark{27,28},
A.\,Popkow\altaffilmark{42},
E.\,Pueschel\altaffilmark{26},
P.\,T.\,Reynolds\altaffilmark{43},
G.\,T.\,Richards\altaffilmark{41},
E.\,Roache\altaffilmark{25},
M.\,Santander\altaffilmark{44},
G.\,H.\,Sembroski\altaffilmark{31},
K.\,Shahinyan\altaffilmark{33},
A.\,W.\,Smith\altaffilmark{21,45},
D.\,Staszak\altaffilmark{22},
I.\,Telezhinsky\altaffilmark{27,28},
N.\,W.\,Todd\altaffilmark{23},
J.\,V.\,Tucci\altaffilmark{31},
J.\,Tyler\altaffilmark{22},
S.\,Vincent\altaffilmark{28},
A.\,Weinstein\altaffilmark{32},
A.\,Wilhelm\altaffilmark{27,28},
D.\,A.\,Williams\altaffilmark{37},
B.\,Zitzer\altaffilmark{46} \\ 
(The VERITAS Collaboration), \\
M.\,L.\,Ahnen\altaffilmark{47},
S.\,Ansoldi\altaffilmark{48},
L.\,A.\,Antonelli\altaffilmark{49},
P.\,Antoranz\altaffilmark{50},
A.\,Babic\altaffilmark{51},
B.\,Banerjee\altaffilmark{52},
P.\,Bangale\altaffilmark{2},
U.\,Barres\,de\,Almeida\altaffilmark{2,70},
J.\,A.\,Barrio\altaffilmark{53},
J.\,Becerra\,Gonz\'alez\altaffilmark{21,45,54}
% ...
\\ 
(The MAGIC Collaboration), \\
J.\,Perkins\altaffilmark{21},
F.\,Verrecchia\altaffilmark{12,17},
C.\,Leto\altaffilmark{12,17},
M.\,B{\" o}ttcher\altaffilmark{77},\\ 
M.\,Villata\altaffilmark{78},
C.\,M.\,Raiteri\altaffilmark{78},
J.\,A.\,Acosta-Pulido\altaffilmark{79,80},
R.\,Bachev\altaffilmark{81},
A.\,Berdyugin\altaffilmark{82},
D.\,A.\,Blinov\altaffilmark{83,84,85},
M.\,I.\,Carnerero\altaffilmark{78,79,80},
W.\,P.\,Chen\altaffilmark{86},
P.\,Chinchilla\altaffilmark{79,80},
G.\,Damljanovic\altaffilmark{87},
C.\,Eswaraiah\altaffilmark{86},
T.\,S.\,Grishina\altaffilmark{83},
S.\,Ibryamov\altaffilmark{81},
B.\,Jordan\altaffilmark{88},
S.\,G.\,Jorstad\altaffilmark{83,89},
M.\,Joshi\altaffilmark{83},
E.\,N.\,Kopatskaya\altaffilmark{83},
O.\,M.\,Kurtanidze\altaffilmark{90,91,92},
S.\,O.\,Kurtanidze\altaffilmark{90},
E.\,G.\,Larionova\altaffilmark{83},
L.\,V.\,Larionova\altaffilmark{83},
V.\,M.\,Larionov\altaffilmark{83,93},
G.\,Latev\altaffilmark{94},
H.\,C.\,Lin\altaffilmark{86},
A.\,P.\,Marscher\altaffilmark{89},
A.\,A.\,Mokrushina\altaffilmark{83,93},
D.\,A.\,Morozova\altaffilmark{83},
M.\,G.\,Nikolashvili\altaffilmark{90},
E.\,Semkov\altaffilmark{81},
P.S.\, Smith\altaffilmark{105},
A.\,Strigachev\altaffilmark{81},
Yu.\,V.\,Troitskaya\altaffilmark{83},
I.\,S.\,Troitsky\altaffilmark{83},
O.\,Vince\altaffilmark{87},
J.\,Barnes\altaffilmark{95},
T.\,G\"uver\altaffilmark{96},
J.\,W.\,Moody\altaffilmark{97},
A.\,C.\,Sadun\altaffilmark{98},
S.\,Sun\altaffilmark{99}, \\ 
T.\,Hovatta\altaffilmark{1,100},
J.\,L.\,Richards\altaffilmark{31},
W.\,Max-Moerbeck\altaffilmark{101},
A.\,C.\,R.\,Readhead\altaffilmark{1},
A.\,L\"ahteenm\"aki\altaffilmark{100,102},
M.\,Tornikoski\altaffilmark{100},
J.\,Tammi\altaffilmark{100},
V.\,Ramakrishnan\altaffilmark{100},
R.\,Reinthal\altaffilmark{82},
E.\,Angelakis\altaffilmark{103},
L.\,Fuhrmann\altaffilmark{103}
I.\,Myserlis\altaffilmark{103}
V.\,Karamanavis\altaffilmark{103},
A.\,Sievers\altaffilmark{104},
H.\,Ungerechts\altaffilmark{104},
J.\,A.\,Zensus\altaffilmark{103}
}

\altaffiltext{1}{Cahill Center for Astronomy and Astrophysics, California Institute of Technology, Pasadena, CA 91125, USA}
\altaffiltext{A}{{\tt mislavb@astro.caltech.edu}}
\altaffiltext{2}{Max-Planck-Institut f\"ur Physik, D-80805 M\"unchen, Germany}
\altaffiltext{B}{{\tt dpaneque@mppmu.mpg.de}}
\altaffiltext{3}{W.\,W. Hansen Experimental Physics Laboratory, Kavli Institute for Particle Astrophysics and Cosmology, Department of Physics and SLAC National Accelerator Laboratory, Stanford University, Stanford, CA 94305, USA}
\altaffiltext{C}{{\tt madejski@stanford.edu}}
\altaffiltext{4}{Department of Physics, Stanford University, Stanford, CA 94305, USA}
\altaffiltext{D}{{\tt amy.furniss@gmail.com}}
\altaffiltext{E}{{\tt jchiang@slac.stanford.edu}}
\altaffiltext{5}{Department of Physics and Astronomy, Clemson University, Kinard Lab of Physics, Clemson, SC 29634-0978, USA}
\altaffiltext{6}{Department of Physics, Durham University, Durham DH1 3LE, UK}
\altaffiltext{7}{Universit\'{e} de Toulouse, UPS - OMP, IRAP, Toulouse, France}
\altaffiltext{8}{CNRS, Institut de Recherche en Astrophysique et Plan\'{e}tologie, 9 Av.\ colonel Roche, BP 44346, F-31028 Toulouse Cedex 4, France}
\altaffiltext{9}{Space Science Laboratory, University of California, Berkeley, CA 94720, USA}
\altaffiltext{10}{DTU Space, National Space Institute, Technical University of Denmark, Elektrovej 327, DK - 2800 Lyngby, Denmark}
\altaffiltext{11}{Lawrence Livermore National Laboratory, Livermore, CA 94550, USA}
\altaffiltext{12}{ASI Science Data Center, Via del Politecnico snc I-00133, Roma, Italy}
\altaffiltext{13}{Physics Department, Columbia University, New York, NY 10027, USA}
\altaffiltext{14}{Department of Physical Science, Hiroshima University, Higashi-Hiroshima, Hiroshima 739-8526, Japan}
\altaffiltext{15}{Core of Research for the Energetic Universe, Hiroshima University, Higashi-Hiroshima, Hiroshima 739-8526, Japan}
\altaffiltext{16}{Columbia Astrophysics Laboratory, Columbia University, New York, NY 10027, USA}
\altaffiltext{17}{INAF - Osservatorio Astronomico di Roma, via di Frascati 33, I-00040 Monteporzio, Italy}
\altaffiltext{18}{Jet Propulsion Laboratory, California Institute of Technology, Pasadena, CA 91109, USA}
\altaffiltext{19}{INAF - Osservatorio Astronomico di Brera, Via E, Bianchi 46, I-23807 Merate, Italy}
\altaffiltext{20}{Yale Center for Astronomy and Astrophysics, Physics Department, Yale University, PO Box 208120, New Haven, CT 06520-8120, USA}
\altaffiltext{21}{NASA Goddard Space Flight Center, Greenbelt, MD 20771, USA}
\altaffiltext{22}{Physics Department, McGill University, Montreal, QC H3A 2T8, Canada}
\altaffiltext{23}{Department of Physics, Washington University, St. Louis, MO 63130, USA}
\altaffiltext{24}{Harvard-Smithsonian Center for Astrophysics, 60 Garden Street, Cambridge, MA 02138, USA}
\altaffiltext{25}{Fred Lawrence Whipple Observatory, Harvard-Smithsonian Center for Astrophysics, Amado, AZ 85645, USA}
\altaffiltext{26}{School of Physics, University College Dublin, Belfield, Dublin 4, Ireland}
\altaffiltext{27}{Institute of Physics and Astronomy, University of Potsdam, 14476 Potsdam-Golm, Germany}
\altaffiltext{28}{Deutsches Elektronen-Synchrotron (DESY), Platanenallee 6, 15738 Zeuthen, Germany}
\altaffiltext{29}{Astronomy Department, Adler Planetarium and Astronomy Museum, Chicago, IL 60605, USA}
\altaffiltext{30}{School of Physics, National University of Ireland Galway, University Road, Galway, Ireland}
\altaffiltext{31}{Department of Physics and Astronomy, Purdue University, West Lafayette, IN 47907, USA} 
\altaffiltext{32}{Department of Physics and Astronomy, Iowa State University, Ames, IA 50011, USA}
\altaffiltext{33}{School of Physics and Astronomy, University of Minnesota, Minneapolis, MN 55455, USA}
\altaffiltext{34}{Department of Astronomy and Astrophysics, 525 Davey Lab, Pennsylvania State University, University Park, PA 16802, USA}
\altaffiltext{35}{Department of Physics and Astronomy, University of Iowa, Van Allen Hall, Iowa City, IA 52242, USA}
\altaffiltext{36}{Department of Physics and Astronomy and the Bartol Research Institute, University of Delaware, Newark, DE 19716, USA}
\altaffiltext{37}{Santa Cruz Institute for Particle Physics and Department of Physics, University of California, Santa Cruz, CA 95064, USA}
\altaffiltext{38}{Department of Physics and Astronomy, DePauw University, Greencastle, IN 46135-0037, USA}
\altaffiltext{39}{Department of Physics and Astronomy, University of Utah, Salt Lake City, UT 84112, USA}
\altaffiltext{40}{Enrico Fermi Institute, University of Chicago, Chicago, IL 60637, USA}
\altaffiltext{41}{School of Physics and Center for Relativistic Astrophysics, Georgia Institute of Technology, 837 State Street NW, Atlanta, GA 30332-0430}
\altaffiltext{42}{Department of Physics and Astronomy, University of California, Los Angeles, CA 90095, USA}
\altaffiltext{43}{Department of Applied Science, Cork Institute of Technology, Bishopstown, Cork, Ireland}
\altaffiltext{44}{Department of Physics and Astronomy, Barnard College, Columbia University, NY 10027, USA}
\altaffiltext{45}{University of Maryland, College Park, MD 20742, USA}
\altaffiltext{46}{Argonne National Laboratory, 9700 S. Cass Avenue, Argonne, IL 60439, USA}
\altaffiltext{47}{ETH Zurich, CH-8093 Zurich, Switzerland}
\altaffiltext{48}{Universit\`a di Udine, and INFN Trieste, I-33100 Udine, Italy}
\altaffiltext{49}{INAF National Institute for Astrophysics, I-00136 Rome, Italy}
\altaffiltext{50}{Universit\`a  di Siena, and INFN Pisa, I-53100 Siena, Italy}
\altaffiltext{51}{Croatian MAGIC Consortium, Rudjer Boskovic Institute, University of Rijeka and University of Split, HR-10000 Zagreb, Croatia}
\altaffiltext{52}{Saha Institute of Nuclear Physics, 1\textbackslash{}AF Bidhannagar, Salt Lake, Sector-1, Kolkata 700064, India}
\altaffiltext{53}{Universidad Complutense, E-28040 Madrid, Spain}
\altaffiltext{54}{Inst. de Astrof\'isica de Canarias, E-38200 La Laguna, Tenerife, Spain; Universidad de La Laguna, Dpto. Astrof\'isica, E-38206 La Laguna, Tenerife, Spain}
\altaffiltext{55}{University of \L\'od\'z, PL-90236 Lodz, Poland}
\altaffiltext{56}{IFAE, Campus UAB, E-08193 Bellaterra, Spain}
\altaffiltext{57}{Universit\"at W\"urzburg, D-97074 W\"urzburg, Germany}
\altaffiltext{58}{Centro de Investigaciones Energ\'eticas, Medioambientales y Tecnol\'ogicas, E-28040 Madrid, Spain}
\altaffiltext{59}{Universit\`a di Padova and INFN, I-35131 Padova, Italy}
\altaffiltext{60}{Institute for Space Sciences (CSIC\textbackslash{}IEEC), E-08193 Barcelona, Spain}
\altaffiltext{61}{Technische Universit\"at Dortmund, D-44221 Dortmund, Germany}
\altaffiltext{62}{Unitat de F\'isica de les Radiacions, Departament de F\'isica, and CERES-IEEC, Universitat Aut\`onoma de Barcelona, E-08193 Bellaterra, Spain}
\altaffiltext{63}{Universitat de Barcelona, ICC, IEEC-UB, E-08028 Barcelona, Spain}
\altaffiltext{64}{Japanese MAGIC Consortium, ICRR, The University of Tokyo, Department of Physics and Hakubi Center, Kyoto University, Tokai University, The University of Tokushima, KEK, Japan}
\altaffiltext{65}{Finnish MAGIC Consortium, Tuorla Observatory, University of Turku and Department of Physics, University of Oulu, Finland}
\altaffiltext{66}{Inst. for Nucl. Research and Nucl. Energy, BG-1784 Sofia, Bulgaria}
\altaffiltext{67}{Universit\`a di Pisa, and INFN Pisa, I-56126 Pisa, Italy}
\altaffiltext{68}{ICREA and Institute for Space Sciences (CSIC\textbackslash{}IEEC), E-08193 Barcelona, Spain}
\altaffiltext{69}{Universit\`a dell'Insubria and INFN Milano Bicocca, Como, I-22100 Como, Italy}
\altaffiltext{70}{Centro Brasileiro de Pesquisas F\'isicas (CBPF\textbackslash{}MCTI), R. Dr. Xavier Sigaud, 150 - Urca, Rio de Janeiro - RJ, 22290-180, Brazil}
\altaffiltext{71}{Humboldt University of Berlin, Istitut f\"ur Physik  Newtonstr. 15, 12489 Berlin Germany}
\altaffiltext{72}{Ecole polytechnique f\'ed\'erale de Lausanne (EPFL), Lausanne, Switzerland}
\altaffiltext{73}{Laboratoire AIM, Service d'Astrophysique, DSM\textbackslash{}IRFU, CEA\textbackslash{}Saclay FR-91191 Gif-sur-Yvette Cedex, France}
\altaffiltext{74}{Finnish Centre for Astronomy with ESO (FINCA), Turku, Finland}
\altaffiltext{75}{INAF - Osservatorio Astronomico di Trieste, Trieste, Italy}
\altaffiltext{76}{ISDC - Science Data Center for Astrophysics, 1290, Versoix (Geneva), Switzerland}
\altaffiltext{77}{Centre for Space Research, Private Bag X6001, North-West University, Potchefstroom Campus, Potchefstroom, 2520, South Africa}
\altaffiltext{78}{INAF - Osservatorio Astrofisico di Torino, 10025 Pino Torinese (TO), Italy}
\altaffiltext{79}{Instituto de Astrofisica de Canarias (IAC), La Laguna, Tenerife, Spain}
\altaffiltext{80}{Departamento de Astrofisica, Universidad de La Laguna, La Laguna, Tenerife, Spain}
\altaffiltext{81}{Institute of Astronomy, Bulgarian Academy of Sciences, 72 Tsarigradsko shosse Blvd., 1784 Sofia, Bulgaria}
\altaffiltext{82}{Tuorla Observatory, Department of Physics and Astronomy, V\" ais\" al\" antie 20, FIN-21500 Piikki\"o, Finland}
\altaffiltext{83}{Astronomical Institute, St.\,Petersburg State University, Universitetskij Pr. 28, Petrodvorets, 198504 St.\,Petersburg, Russia}
\altaffiltext{84}{Department of Physics and Institute for Plasma Physics, University of Crete, 71003, Heraklion, Greece}
\altaffiltext{85}{Foundation for Research and Technology - Hellas, IESL, Voutes, 71110 Heraklion, Greece}
\altaffiltext{86}{Graduate Institute of Astronomy, National Central University, 300 Zhongda Road, Zhongli 32001, Taiwan}
\altaffiltext{87}{Astronomical Observatory, Volgina 7, 11060 Belgrade, Serbia}
\altaffiltext{88}{School of Cosmic Physics, Dublin Institute For Advanced Studies, Ireland}
\altaffiltext{89}{Institute for Astrophysical Research, Boston University, 725 Commonwealth Avenue, Boston, MA 02215}
\altaffiltext{90}{Abastumani Observatory, Mt. Kanobili, 0301 Abastumani, Georgia}
\altaffiltext{91}{Engelhardt Astronomical Observatory, Kazan Federal University, Tatarstan, Russia}
\altaffiltext{92}{Center for Astrophysics, Guangzhou University, Guangzhou 510006, China}
\altaffiltext{93}{Pulkovo Observatory, St.-Petersburg, Russia}
\altaffiltext{94}{Institute of Astronomy and NAO, Sofia, Bulgaria}
\altaffiltext{95}{Department of Physics, Salt Lake Community College, Salt Lake City, Utah 84070 USA}
\altaffiltext{96}{Istanbul  University, Science Faculty, Department  of Astronomy and Space Sciences, Beyaz\i t, 34119, Istanbul, Turkey}
\altaffiltext{97}{Department of Physics and Astronomy, Brigham Young University, Provo Utah 84602 USA}
\altaffiltext{98}{Department of Physics, University of Colorado Denver, Denver, Colorado, CO 80217-3364, USA}
\altaffiltext{99}{Center for Field Theory and Particle Physics and Department of Physics, Fudan University, 200433 Shanghai, China}
\altaffiltext{100}{Aalto University, Mets\"ahovi Radio Observatory, Mets\"ahovintie 114, 02540 Kylm\"al\"a, Finland}
\altaffiltext{101}{National Radio Astronomy Observatory, P.O. Box 0, Socorro, NM 87801, USA}
\altaffiltext{102}{Aalto University Department of Radio Science and Engineering, P.O. BOX 13000, FI-00076 Aalto, Finland}
\altaffiltext{103}{Max-Planck-Institut f{\" u}r Radioastronomie, Auf dem H{\" u}gel 69, 53121 Bonn, Germany}
\altaffiltext{104}{Instituto de Radio Astronomía Milim{\' e}trica, Avenida Divina Pastora 7, Local 20, 18012 Granada, Spain}
\altaffiltext{105}{Steward Observatory, University of Arizona, 933 N. Cherry Ave., Tucson, AZ 85721, USA}

\begin{abstract} 

We present coordinated multiwavelength observations of the bright, nearby BL\,Lac object Mrk\,421 taken in 2013 January-March, involving GASP-WEBT, {\em Swift}, {\em NuSTAR}, {\em Fermi}\,-LAT, MAGIC, VERITAS, and other collaborations and instruments, providing data from radio to very-high-energy~(VHE) $\gamma$-ray bands. {\em NuSTAR} yielded previously unattainable sensitivity in the 3--79~keV range, revealing that the spectrum softens when the source is dimmer until the X-ray spectral shape saturates into a steep $\Gamma\approx3$ power law, with no evidence for an exponential cutoff or additional hard components up to $\sim$80~keV. For the first time, we observed both the synchrotron and the inverse-Compton peaks of the spectral energy distribution (SED) simultaneously shifted to frequencies below the typical quiescent state by an order of magnitude. The fractional variability as a function of photon energy shows a double-bump structure which relates to the two bumps of the broadband SED. In each bump, the variability increases with energy which, in the framework of the synchrotron self-Compton model, implies that the electrons with higher energies are more variable. The measured multi-band variability, the significant X-ray-to-VHE correlation down to some of the lowest fluxes ever observed in both bands, the lack of correlation between optical/UV and X-ray flux, the low degree of polarization and its significant (random) variations, the short estimated electron cooling time, and the significantly longer variability timescale observed in the {\em NuSTAR} light curves point toward in-situ electron acceleration, and suggest that there are multiple compact regions contributing to the broadband emission of Mrk\,421 during low-activity states.

\end{abstract} 



\section{Introduction} 

\label{sec:intro}

Markarian 421 (Mrk\,421\ hereafter) is a nearby active galaxy with a featureless optical spectrum devoid of prominent emission or absorption lines, with strongly polarized variable optical and radio flux, and compact (milli-arcsecond-scale) radio emission. As such, it is classified as a BL\,Lacertae type. Its spectral energy distribution (SED) is well described by a characteristic two-peak shape (for a review, see, e.g., \citealt{urry+padovani-1995,ulrich+1997}). In the more general context of blazars, Mrk\,421\ belongs to a subclass of the so-called high-energy-peaked BL~Lacertae (HBL) objects, relatively low-luminosity sources with both peaks located at relatively high energies (respectively at $\sim$1~keV and $\sim$100~GeV). Mrk\,421\ is among the closest and most comprehensively studied objects of the HBL class and is also the first extragalactic source detected in the very high energy $\gamma$-ray band ($E>$100~GeV, VHE hereafter; \citealt{punch+1992}).

The observed properties of Mrk\,421, as well as other similar blazars, are best explained as arising from a relativistic jet seen at a small angle to our line of sight \citep{urry+padovani-1995}. The nonthermal and polarized continuum observed from the radio band to the soft X-ray band suggests that this part of the SED is due to a distribution of relativistic electrons radiating via the synchrotron process. The radiation in the $\gamma$-ray band is likely due to inverse-Compton scattering by energetic electrons responsible for the synchrotron radiation, as confirmed by simultaneous, correlated variations in the low- and high-energy SED components (e.g., \citealt{giebels+2007,fossati+2008,aleksic+2015b}). The seed photons are most likely the synchrotron photons internal to the jet. Such ``synchrotron self-Compton'' (SSC) models, developed by many authors (for early examples see, e.g.,~\citealt{jones+1974,ghisellini+1985,marscher+gear-1985}) have been successfully invoked to describe the full SED of HBL objects (e.g., \citealt{ulrich+1997,fossati+2008,tavecchio+2010}).

The range of measured flux of Mrk\,421\ spans up to two orders of magnitude in some spectral bands, with flares occurring on very short timescales (a day or less; e.g., \citealt{gaidos+1996,tanihata+2003,fossati+2008}). Possibly the best bands to study such variability are the X-ray and VHE $\gamma$-ray bands: in the context of the SSC model, they represent radiation from the most energetic electrons, which have the shortest timescales for radiative losses. The cross-correlation of time series measured in various bands provides additional clues to the radiative processes, the acceleration and energy distribution of radiating particles, and the structure and intrinsic power of the relativistic jet. The relative temporal variability in different spectral bands, from radio through VHE $\gamma$-rays, provides an important handle on the location of the energy dissipation with respect to the central black hole \citep{sikora+2009,janiak+2012}. In the context of a specific model for the radiation, the underlying particle distributions may be determined more or less directly from the observed multiwavelength SEDs. Particle-acceleration mechanisms can then be constrained by the shape of those particle distributions. Diffusive shock acceleration, an example of a first-order Fermi (Fermi~I) process, is generally associated with single power-law distributions (e.g., \citealt{blandford+eichler-1987,jones+ellison-1991}). In contrast, log-parabolic distributions are produced in models of stochastic acceleration (e.g.,~\citealt{massaro+2004,tramacere+2011}), which can be considered equivalent to a second-order Fermi (Fermi~II) process.

Mrk\,421\ and other HBL-type blazars have been extensively studied in the soft X-ray band (e.g.,~\citealt{makino+1987,takahashi+1996,ravasio+2004,tramacere+2007b,tramacere+2009}), revealing a range of spectral slopes in various quiescent and flaring states. Less is known about the hard X-ray ($\gtrsim10$~keV) properties of blazar jet emission: the data are far fewer and available mostly for flaring episodes, or averaged over relatively long timescales (e.g.,~\citealt{guainazzi+1999,giebels+2007,fossati+2008,ushio+2009,abdo+2011}). This energy band probes the most energetic and fastest varying tail of the distribution of synchrotron-radiating particles, and therefore represents an important diagnostic of the content of the jet and the processes responsible for the acceleration of particles to the highest energies. The inverse-Compton component increases with energy, and could potentially contribute significantly to the hard X-ray band. If so, it would also provide a strong constraint on the low-energy part of the electron distribution, which is a significant, if not dominant, part of the total kinetic power of the jet.

Mrk\,421\ observations were part of the Nuclear Spectroscopic Telescope Array ({\em NuSTAR}; \citealt{harrison+2013}) blazar program, aimed at advancing our understanding of astrophysical jets. The multiwavelength campaign focused on Mrk\,421\ was carried out between December 2012 and May 2013, with three to four pointings per month, designed to maximize strictly simultaneous overlap with observations by the VHE $\gamma$-ray facilities \veritas and \magic$\!\!$. We also secured nearly simultaneous soft X-ray, optical and UV observations from the \swift satellite. The $\gamma$-ray data from {\em Fermi}\,-LAT\, which observes Mrk\,421\ every 3 hours, was also used together with all the coordinated multiwavelength data.  Mrk\,421\ varied in flux throughout the campaign, with a relatively low flux in the X-ray and VHE bands at the beginning, increasing to a major flare toward the end of the campaign. In this paper, we present part of the data collected during the first three months of the campaign with particular emphasis on the detailed shape of the X-ray spectrum, its variability and the correlated variability observed in VHE $\gamma$-rays. We also report briefly on the observations of Mrk\,421\ prior to the start of the campaign, in July 2012, when the object was used for calibration purposes during the in-orbit verification phase of {\em NuSTAR}. During this period, Mrk\,421\ emission was broadly consistent with previously observed quiescent states, which we define here to be characterized by relatively low flux at all frequencies and by the absence of significant flaring (see, e.g., \citealt{abdo+2011}). The flaring period of the 2013 campaign will be covered in a separate publication.

The outline of the paper is as follows. In \S\,\ref{sec:obs+data} we describe the multiwavelength observations and data used in this paper. We dedicate \S\,\ref{sec:nustar} to a detailed characterization of the hard X-ray spectrum of Mrk\,421\ with {\em NuSTAR}. The results of the multiwavelength campaign in 2013 January--March are presented in \S\,\ref{sec:results}. Discussion of the empirical results and modeling of the broadband properties are given in \S\,\ref{sec:discussion}, and in \S\,\ref{sec:summary} we summarize the main results. We adopt a distance of 141~Mpc to Mrk\,421, calculated from its measured redshift $z=0.0308$ (based on absorption lines in the spectrum of the host galaxy; \citealt{ulrich+1975}) and the cosmological parameters recently refined by the Planck Collaboration \citep{planck-cosmology-2014}: $h_0=0.67$, $\Omega_{\Lambda}=0.685$.

\section{Observations and Data Analysis} 

\label{sec:obs+data}

\subsection{Radio} 

\label{sec:data-radio}

The Owens Valley Radio Observatory (OVRO) 40-meter telescope was used for observation at 15~GHz, as a part of a long-term blazar monitoring program. Additional observations were scheduled at times of coordinated observations with X-ray and VHE $\gamma$-ray observatories. The data were reduced using standard processing and calibration techniques described in detail in \citet{richards+2011}. Radio observations of Mrk\,421\ between 2.64 and 142~GHz have been obtained within the framework of the F-GAMMA program \citep{fuhrmann+2007,angelakis+2010,fuhrmann+2014}, a $\gamma$-ray blazar monitoring program related to the {\em Fermi} Gamma-ray Space Telescope. Observations with the Effelsberg 100-meter and Pico Veleta 30-meter telescopes are performed roughly once per month. The Effelsberg measurements are conducted with heterodyne receivers at 2.64, 4.85, 8.35, 10.45, 14.60, 23.05, and 32.0~GHz, while the Pico Veleta telescope is used with the EMIR receiver to provide the high-frequency (86.2 and 142.3~GHz) flux measurements. Standard data processing and calibration were performed as described in \citet{angelakis+2008} and \citet{angelakis+2015}. The Mets{\"a}hovi Radio Observatory 14-meter telescope also participated in this multi-instrument campaign, providing observations of Mrk\,421\ at 37~GHz every few days. Details of the observing strategy and data reduction for this monitoring program can be found in \citet{terasranta+1998}.

\subsection{Optical} 

\label{sec:data-optical}

The coverage at optical frequencies was provided by various telescopes
around the world within the GASP-WEBT program (e.g.,
\citealt{Villata2008}, \citealt{Villata2009}). In particular, the
following observatories contributed to this campaign: Teide (IAC80),
Crimean, Lowell (Perkins telescope), Roque de los Muchachos (KVA and
Liverpool telescopes), Abastumani, Pulkovo, St.\,Petersburg,
Belogradchik, Rozhen (50/70~cm, 60~cm and 200~cm telescopes),
Vidojevica and Lulin. Additionally, many observations were performed
with iTelescopes, Bradford Robotic Telescope, ROVOR, and the TUBITAK
National Observatory. In this paper, we use only R-band photometry. The
calibration stars reported in \citet{Villata1998} were used for
calibration, and the Galactic extinction was corrected with the
reddening corrections given in \citet{schlafly+2011}. The flux from
the host galaxy was estimated using the R-band flux from
\citet{nilsson+2007} for the apertures of 5\arcsec\ and 7.5\arcsec\
used by various instruments. We applied an offset of $-5$~mJy to the
fluxes from ROVOR in order to achieve better agreement with the light
curves from the other instruments. This difference may be related to
the specific spectral response of the filters used, or the different
analysis procedures that were employed. Additionally, a point-wise
fluctuation of 2\% on the measured flux was added in quadrature to the
statistical uncertainties in order to account for potential day-to-day
differences in observations with any of the instruments.

Polarization measurements are utilized from four observatories:  Lowell
(Perkins telescope), St.\,Petersburg, Crimean, and Steward
(Bok telescope). The polarization measurements from Lowell and St.\,Petersburg
observatories are derived from R-band imaging polarimetry.  The
measurements from Steward Observatory are derived from 4000--7550\,\AA\ band
spectropolarimetry with a resolution of $\sim$15\,\AA. 
The reported values are constructed from the median Q/I and U/I in the
5000--7000\,\AA\ band.  The effective wavelength of this bandpass is not too
different from the Kron-Cousins R-band and the wavelength dependence in
the polarization of Mrk\,421\ seen in the spectropolarimetry during this
period is not strong enough to significantly affect the variability analysis
of the measurements from various telescopes. The observing and data-processing
procedures for the polarization measurements are described in \citet{2008A&A...492..389L,smith+2009,2010ApJ...715..362J}.

\begin{deluxetable*}{cccccccccc} 
\tabletypesize{\scriptsize}
\tablewidth{0pt}

\tablecaption{ Summary of the \swift observations of Mrk\,421\ (January--March 2013) \label{tab:observations-swift} }

\tablehead{
  \colhead{\multirow{2}{*}{\shortstack[c]{\, \\ \bf Sequence \\ \bf ID}}} &
  \colhead{\bf Start Date} &
  \colhead{\bf Start Time} &
  \colhead{\multirow{2}{*}{\shortstack[c]{\, \\ \bf Number \\ \bf of \\ \bf Orbits}}} &
  \multicolumn{2}{c}{{\bf Exposure}\tablenotemark{a} ( ks )} &
  \multicolumn{3}{c}{{\bf UV Flux Density}\tablenotemark{b} ( mJy )} &
  \colhead{\multirow{2}{*}{\shortstack[c]{\, \\ \bf Count \\ {\bf Rate}\tablenotemark{c} \\ \rm ( s$^{-1}$ )}}} \\
  \cline{5-6} \cline{7-9} \\
  \colhead{} &
  \colhead{( UTC )} &
  \colhead{( MJD )} &
  \colhead{} &
  \colhead{UVOT} &
  \colhead{XRT} &
  \colhead{W1} &
  \colhead{M1} &
  \colhead{W2} &
  \colhead{} \\
}

\startdata

00080050001 & 2013-Jan-02 & 56294.7961 & 2 & 1.6 & 1.8 & $26.8\pm0.9$ & $27.0\pm0.9$ & \nodata      & $12.6\pm0.1$ \\
00035014024 & 2013-Jan-04 & 56296.9370 & 1 & 1.0 & 1.0 & $23.6\pm0.8$ & $23.7\pm0.8$ & $20.4\pm0.7$ & $18.8\pm0.2$ \\
00035014025 & 2013-Jan-08 & 56300.1523 & 1 & 0.8 & 0.8 & $21.1\pm0.7$ & $20.3\pm0.7$ & $17.6\pm0.6$ & $7.8\pm0.1$ \\
00035014026 & 2013-Jan-10 & 56302.1557 & 2 & 1.1 & 1.7 & $22.5\pm0.7$ & $23.1\pm0.8$ & $19.7\pm0.7$ & $9.1\pm0.1$ \\
00035014028 & 2013-Jan-10 & 56302.3418 & 2 & 0.8 & 1.3 & $22.7\pm0.8$ & $22.3\pm0.7$ & $19.2\pm0.6$ & $9.0\pm0.1$ \\
00035014027 & 2013-Jan-10 & 56302.4752 & 1 & 1.3 & 1.3 & $22.1\pm0.7$ & $22.1\pm0.7$ & $18.8\pm0.6$ & $8.33\pm0.09$ \\
00035014029 & 2013-Jan-10 & 56302.6764 & 3 & 2.9 & 3.8 & $21.5\pm0.7$ & $21.3\pm0.7$ & $18.3\pm0.6$ & $9.52\pm0.06$ \\
00035014031 & 2013-Jan-10 & 56302.9601 & 1 & 0.7 & 0.7 & $21.3\pm0.7$ & $20.9\pm0.7$ & $18.1\pm0.6$ & $11.0\pm0.2$ \\
00035014032 & 2013-Jan-12 & 56304.4790 & 1 & 1.1 & 1.1 & $18.7\pm0.6$ & $18.5\pm0.6$ & $16.1\pm0.5$ & $14.1\pm0.1$ \\
00035014034 & 2013-Jan-15 & 56307.0928 & 3 & 3.8 & 4.0 & $17.2\pm0.6$ & $17.4\pm0.6$ & $15.0\pm0.5$ & $22.4\pm0.1$ \\
00035014033 & 2013-Jan-15 & 56307.3519 & 5 & 5.0 & 6.3 & $17.7\pm0.6$ & $17.8\pm0.6$ & $15.4\pm0.5$ & $22.59\pm0.08$ \\
00035014035 & 2013-Jan-18 & 56310.1675 & 1 & 1.0 & 1.0 & $18.9\pm0.6$ & $18.7\pm0.6$ & $16.1\pm0.5$ & $8.9\pm0.1$ \\
00080050002 & 2013-Jan-20 & 56312.2389 & 7 & 3.9 & 8.8 & $19.6\pm0.6$ & $19.4\pm0.6$ & $16.4\pm0.6$ & $9.17\pm0.04$ \\
00035014036 & 2013-Jan-22 & 56314.5070 & 1 & 1.1 & 1.1 & $21.7\pm0.7$ & $21.7\pm0.7$ & $19.2\pm0.6$ & $10.5\pm0.1$ \\
00035014038 & 2013-Jan-25 & 56317.3009 & 3 & 0.6 & 7.8 & $15.7\pm0.5$ & $15.4\pm0.6$ & $13.4\pm0.4$ & $11.4\pm0.2$ \\
00035014039 & 2013-Jan-28 & 56320.3057 & 1 & 1.0 & 1.0 & $13.7\pm0.5$ & $13.7\pm0.5$ & $12.0\pm0.4$ & $17.6\pm0.2$ \\
00035014040 & 2013-Feb-01 & 56324.6601 & 1 & 1.1 & 1.1 & $13.3\pm0.4$ & $13.3\pm0.4$ & $11.7\pm0.4$ & $28.5\pm0.2$ \\
00035014041 & 2013-Feb-04 & 56327.1409 & 2 & 0.3 & 0.8 & $15.3\pm0.6$ & $15.0\pm0.6$ & $13.4\pm0.5$ & $28.7\pm0.3$ \\
00080050003 & 2013-Feb-06 & 56329.0586 & 6 & 2.1 & 9.5 & $14.5\pm0.5$ & $14.4\pm0.5$ & $12.8\pm0.4$ & $21.54\pm0.05$ \\
00035014043 & 2013-Feb-10 & 56333.1279 & 1 & 1.0 & 1.0 & $13.2\pm0.4$ & $13.0\pm0.4$ & $11.3\pm0.4$ & $21.1\pm0.2$ \\
00080050005 & 2013-Feb-12 & 56335.0700 & 6 & 2.6 & 6.3 & $17.4\pm0.6$ & $17.7\pm0.6$ & $15.7\pm0.5$ & $22.82\pm0.08$ \\
00035014044 & 2013-Feb-15 & 56338.0045 & 1 & 1.0 & 0.8 & $18.1\pm0.6$ & $18.2\pm0.6$ & $15.8\pm0.5$ & $8.9\pm0.6$ \\
00080050006 & 2013-Feb-17 & 56340.0047 & 7 & 2.9 & 9.2 & $18.1\pm0.6$ & $18.7\pm0.6$ & $16.1\pm0.5$ & $13.18\pm0.05$ \\
00035014045 & 2013-Feb-19 & 56342.1393 & 2 & 0.6 & 1.1 & $16.0\pm0.5$ & $15.7\pm0.6$ & $13.4\pm0.4$ & $12.2\pm0.2$ \\
00035014046 & 2013-Feb-23 & 56346.3481 & 1 & 1.0 & 1.0 & $19.6\pm0.7$ & $19.9\pm0.7$ & $17.5\pm0.6$ & $15.0\pm0.2$ \\
00035014047 & 2013-Feb-27 & 56350.3573 & 1 & 1.1 & 1.1 & $19.6\pm0.7$ & $19.2\pm0.6$ & $17.0\pm0.6$ & $12.3\pm0.1$ \\
00035014048 & 2013-Mar-01 & 56352.3675 & 1 & 1.1 & 1.0 & $19.4\pm0.6$ & $19.2\pm0.6$ & $16.5\pm0.5$ & $16.9\pm0.1$ \\
00080050007 & 2013-Mar-04 & 56355.9845 & 1 & 1.0 & 1.0 & $23.8\pm0.9$ & $24.9\pm0.8$ & $21.2\pm0.7$ & $33.8\pm0.3$ \\
00080050009 & 2013-Mar-05 & 56356.0538 & 5 & 2.5 & 3.9 & $24.0\pm0.8$ & $24.2\pm0.8$ & $21.0\pm0.7$ & $30.4\pm0.1$ \\
00035014049 & 2013-Mar-07 & 56358.3190 & 1 & 0.9 & 0.6 & $27.0\pm0.9$ & $27.9\pm0.9$ & $25.3\pm0.9$ & $25.5\pm0.3$ \\
00080050011 & 2013-Mar-12 & 56363.0045 & 7 & 6.1 & 8.3 & $25.6\pm0.9$ & $25.8\pm0.9$ & $22.0\pm0.7$ & $17.2\pm0.8$ \\
00035014051 & 2013-Mar-15 & 56366.2540 & 1 & 0.8 & 0.8 & $20.5\pm0.7$ & $20.7\pm0.7$ & $18.0\pm0.6$ & $23.43\pm0.06$ \\
00080050013 & 2013-Mar-17 & 56368.0609 & 6 & 7.7 & 8.9 & $22.5\pm0.7$ & $22.5\pm0.7$ & $19.5\pm0.6$ & $19.6\pm0.3$ \\
00035014052 & 2013-Mar-18 & 56369.0665 & 1 & 1.0 & 1.0 & $21.1\pm0.7$ & $21.3\pm0.7$ & $18.6\pm0.6$ & $21.95\pm0.07$ \\
00035014053 & 2013-Mar-19 & 56370.0675 & 1 & 1.0 & 1.0 & $20.7\pm0.7$ & $21.1\pm0.7$ & $19.2\pm0.6$ & $30.3\pm0.2$ \\
00035014054 & 2013-Mar-23 & 56374.2797 & 1 & 0.9 & 1.0 & $20.7\pm0.7$ & $20.7\pm0.7$ & $18.5\pm0.6$ & $58.0\pm0.3$ \\

\enddata

\tablenotetext{a}{For \swiftxrt$\!\!$, sum of all good time intervals after standard filtering; for \swiftuvot$\!\!$ the total integration time, summed over all bands.}
\tablenotetext{b}{Extinction-corrected flux in \swiftuvot filters (see text for details).}
\tablenotetext{c}{\swiftxrt source count rate in the 0.3--10~keV band averaged over the exposure time. Background has been subtracted, and PSF and pile-up corrections have been applied. The uncertainty is quoted at 68\% significance (1\,$\sigma$).}

\end{deluxetable*} 

\begin{deluxetable*}{c cc ccc ccc} 
\tabletypesize{\scriptsize}
\tablewidth{0pt}

\tablecaption{ Models fitted to the \swiftxrt spectra of each observation \label{tab:swiftxrt_models} }

\tablehead{
  \colhead{\multirow{2}{*}{\shortstack[c]{\, \\ \bf Start Time \\ \rm [ MJD ]\,}}} &
  \multicolumn{2}{c}{\bf Power Law} &
  \multicolumn{3}{c}{{\bf Log-parabola} ($E_{*}=1$~keV)} &
  \multicolumn{3}{c}{{\bf Time-averaged Flux}\tablenotemark{a}} \\
  \cline{2-3} \cline{4-6} \cline{7-9} \\
  \colhead{} &
  \colhead{$\Gamma$} &
  \colhead{$\chi^2$/d.o.f.} &
  \colhead{$\alpha$} &
  \colhead{$\beta$} &
  \colhead{$\chi^2$/d.o.f.} &
  \colhead{$0.3-3$~keV} &
  \colhead{$3-7$~keV} &
  \colhead{$2-10$~keV} \\
}

\startdata

$56294.7906$ & $2.86\pm0.03$ & $254/199$  & $2.85\pm0.03$ & $0.23\pm0.09$  & $233/198$ & $20.0\pm0.5$ & $2.7\pm0.2$  & $5.6\pm0.4$ \\
$56296.9370$ & $2.68\pm0.02$ & $281/227$  & $2.64\pm0.02$ & $0.28\pm0.07$  & $235/226$ & $27.0\pm0.3$ & $4.8\pm0.2$  & $9.9\pm0.6$ \\
$56300.1523$ & $2.75\pm0.04$ & $171/144$  & $2.74\pm0.04$ & $0.1\pm0.1$    & $168/143$ & $10.0\pm0.4$ & $1.8\pm0.1$  & $3.6\pm0.3$ \\
$56302.1557$ & $2.80\pm0.03$ & $219/198$  & $2.79\pm0.03$ & $0.16\pm0.09$  & $208/197$ & $13.0\pm0.2$ & $2.0\pm0.2$  & $4.1\pm0.3$ \\
$56302.3418$ & $2.82\pm0.03$ & $199/183$  & $2.79\pm0.03$ & $0.2\pm0.1$    & $187/182$ & $14.0\pm0.2$ & $2.1\pm0.2$  & $4.3\pm0.3$ \\
$56302.4751$ & $2.89\pm0.03$ & $175/172$  & $2.87\pm0.03$ & $0.3\pm0.1$    & $159/171$ & $13.0\pm0.4$ & $1.5\pm0.1$  & $3.3\pm0.2$ \\
$56302.6764$ & $2.74\pm0.02$ & $335/275$  & $2.71\pm0.02$ & $0.19\pm0.06$  & $304/274$ & $14.0\pm0.2$ & $2.4\pm0.1$  & $5.0\pm0.2$ \\
$56302.9601$ & $2.75\pm0.04$ & $139/155$  & $2.73\pm0.04$ & $0.2\pm0.1$    & $132/154$ & $15.0\pm0.2$ & $2.5\pm0.3$  & $5.1\pm0.2$ \\
$56304.4790$ & $2.60\pm0.02$ & $252/218$  & $2.56\pm0.03$ & $0.26\pm0.08$  & $219/217$ & $21.0\pm0.6$ & $4.2\pm0.2$  & $8.4\pm0.3$ \\
$56307.0928$ & $2.49\pm0.01$ & $500/394$  & $2.45\pm0.01$ & $0.22\pm0.03$  & $385/393$ & $33.0\pm0.2$ & $8.3\pm0.4$  & $16.0\pm0.4$ \\
$56307.3519$ & $2.60\pm0.01$ & $728/431$  & $2.56\pm0.01$ & $0.26\pm0.03$  & $477/430$ & $33.0\pm0.2$ & $6.6\pm0.2$  & $13.0\pm0.3$ \\
$56310.1675$ & $2.85\pm0.03$ & $198/161$  & $2.83\pm0.04$ & $0.3\pm0.1$    & $180/160$ & $12.0\pm0.3$ & $1.6\pm0.1$  & $3.5\pm0.3$ \\
$56312.2389$ & $2.72\pm0.01$ & $403/351$  & $2.69\pm0.01$ & $0.17\pm0.04$  & $350/350$ & $14.0\pm0.1$ & $2.5\pm0.1$  & $5.1\pm0.1$ \\
$56314.5070$ & $2.69\pm0.03$ & $232/197$  & $2.68\pm0.03$ & $0.07\pm0.10$  & $229/196$ & $16.0\pm0.3$ & $3.2\pm0.2$  & $6.4\pm0.3$ \\
$56317.3009$ & $2.71\pm0.03$ & $165/171$  & $2.69\pm0.04$ & $0.2\pm0.1$    & $158/170$ & $17.0\pm0.3$ & $3.0\pm0.3$  & $6.2\pm0.4$ \\
$56320.3057$ & $2.58\pm0.02$ & $278/216$  & $2.53\pm0.03$ & $0.31\pm0.08$  & $238/215$ & $25.0\pm0.6$ & $5.1\pm0.3$  & $10.0\pm0.5$ \\
$56324.6601$ & $2.35\pm0.02$ & $302/261$  & $2.31\pm0.03$ & $0.21\pm0.06$  & $270/260$ & $49.0\pm0.7$ & $15.0\pm0.5$ & $30\pm1$ \\
$56327.1409$ & $2.51\pm0.02$ & $307/250$  & $2.45\pm0.02$ & $0.33\pm0.07$  & $231/249$ & $40.0\pm0.5$ & $9.1\pm0.3$  & $18.0\pm0.9$ \\
$56329.0586$ & $2.39\pm0.01$ & $1123/530$ & $2.33\pm0.01$ & $0.24\pm0.02$  & $687/529$ & $31.0\pm0.1$ & $9.0\pm0.1$  & $18.0\pm0.3$ \\
$56333.1279$ & $2.41\pm0.02$ & $307/256$  & $2.37\pm0.02$ & $0.21\pm0.06$  & $277/255$ & $31.0\pm0.6$ & $8.8\pm0.4$  & $17.0\pm0.6$ \\
$56335.0700$ & $2.51\pm0.01$ & $528/444$  & $2.49\pm0.01$ & $0.10\pm0.03$  & $496/443$ & $34.0\pm0.3$ & $8.9\pm0.1$  & $18.0\pm0.2$ \\
$56338.0045$ & $2.76\pm0.03$ & $172/169$  & $2.74\pm0.04$ & $0.2\pm0.1$    & $166/168$ & $17.0\pm0.3$ & $2.8\pm0.3$  & $5.8\pm0.6$ \\
$56340.0047$ & $2.61\pm0.01$ & $586/419$  & $2.58\pm0.01$ & $0.17\pm0.03$  & $493/418$ & $20.0\pm0.1$ & $4.2\pm0.1$  & $8.4\pm0.2$ \\
$56342.1393$ & $2.53\pm0.03$ & $243/187$  & $2.51\pm0.04$ & $0.1\pm0.1$    & $236/186$ & $19.0\pm0.3$ & $4.5\pm0.5$  & $9.0\pm0.3$ \\
$56346.3481$ & $2.75\pm0.03$ & $223/196$  & $2.72\pm0.03$ & $0.27\pm0.09$  & $198/195$ & $21.0\pm0.6$ & $3.3\pm0.2$  & $6.9\pm0.4$ \\
$56350.3573$ & $2.60\pm0.03$ & $221/196$  & $2.57\pm0.03$ & $0.20\pm0.09$  & $206/195$ & $18.0\pm0.4$ & $3.8\pm0.3$  & $7.6\pm0.5$ \\
$56352.3675$ & $2.46\pm0.02$ & $269/243$  & $2.44\pm0.03$ & $0.14\pm0.07$  & $256/242$ & $24.0\pm0.5$ & $6.5\pm0.4$  & $13.0\pm0.8$ \\
$56355.9845$ & $2.52\pm0.02$ & $279/256$  & $2.48\pm0.02$ & $0.21\pm0.07$  & $249/255$ & $55.0\pm0.8$ & $13.0\pm0.6$ & $26.0\pm0.6$ \\
$56356.0538$ & $2.55\pm0.01$ & $692/444$  & $2.52\pm0.01$ & $0.21\pm0.03$  & $518/443$ & $45.0\pm0.4$ & $10.0\pm0.2$ & $20.0\pm0.4$ \\
$56358.3190$ & $2.39\pm0.03$ & $230/213$  & $2.41\pm0.03$ & $-0.07\pm0.08$ & $228/212$ & $48\pm1$     & $17\pm1$     & $33\pm2$ \\
$56363.0045$ & $2.57\pm0.01$ & $837/476$  & $2.53\pm0.01$ & $0.24\pm0.02$  & $517/475$ & $34.0\pm0.2$ & $7.3\pm0.1$  & $15.0\pm0.2$ \\
$56366.2540$ & $2.40\pm0.02$ & $230/245$  & $2.38\pm0.03$ & $0.12\pm0.07$  & $222/244$ & $27.0\pm0.3$ & $8.4\pm0.4$  & $16.0\pm0.9$ \\
$56368.0609$ & $2.40\pm0.01$ & $758/516$  & $2.37\pm0.01$ & $0.14\pm0.02$  & $643/515$ & $34.0\pm0.2$ & $10.0\pm0.2$ & $20.0\pm0.2$ \\
$56369.0665$ & $2.37\pm0.02$ & $352/301$  & $2.33\pm0.02$ & $0.19\pm0.05$  & $315/300$ & $44.0\pm0.5$ & $14.0\pm0.6$ & $27.0\pm0.8$ \\
$56370.0675$ & $2.36\pm0.02$ & $323/308$  & $2.34\pm0.02$ & $0.09\pm0.05$  & $313/307$ & $43.0\pm0.6$ & $14.0\pm0.4$ & $28.0\pm0.8$ \\
$56374.2797$ & $2.14\pm0.01$ & $539/425$  & $2.08\pm0.02$ & $0.18\pm0.03$  & $457/424$ & $78.0\pm0.7$ & $35.0\pm0.6$ & $68\pm2$ \\

\enddata

\tablenotetext{a}{Flux calculated from the best-fit model, in units of $10^{-11}$~erg~s$^{-1}$~cm$^{-2}$.}

\end{deluxetable*} 

\begin{deluxetable*}{cccccccc} 
\tabletypesize{\scriptsize}
\tablewidth{0pt}

\tablecaption{ Summary of the {\em NuSTAR}\ observations of Mrk\,421\ (January--March 2013) \label{tab:observations-nustar} }

\tablehead{
  \colhead{\multirow{2}{*}{\shortstack[c]{\, \\ \bf Sequence \\ \bf ID}}} &
  \colhead{\bf Start Date} &
  \colhead{\bf Start Time} &
  \colhead{\multirow{2}{*}{\shortstack[c]{\, \\ \bf Number \\ \bf of \\ \bf Orbits}}} &
  \colhead{\multirow{2}{*}{\shortstack[c]{\, \\ \bf Duration \\ \, \\ \rm ( ks ) }}} &
  \colhead{\multirow{2}{*}{\shortstack[c]{\, \\ \bf Exposure\tablenotemark{a} \\ \, \\ \rm ( ks ) }}} &
  \multicolumn{2}{c}{{\bf Count Rate}\tablenotemark{b} ( s$^{-1}$ )} \\
  \cline{7-8} \\
  \colhead{} &
  \colhead{( UTC )} &
  \colhead{( MJD )} &
  \colhead{} &
  \colhead{} &
  \colhead{} &
  \colhead{FPMA} &
  \colhead{FPMB} \\
}

\startdata
10002015001 & 2012-Jul-07 & 56115.1353 & 14 & 81.0 & 42.0 & $3.71\pm0.01$ & $3.84\pm0.01$ \\
10002016001 & 2012-Jul-08 & 56116.0732 & 8 & 46.2 & 25.4 & $4.18\pm0.01$ & $4.45\pm0.01$ \\
60002023002 & 2013-Jan-02 & 56294.7778 & 3 & 15.6 & 9.2 & $1.162\pm0.009$ & $1.155\pm0.008$ \\
60002023004 & 2013-Jan-10 & 56302.0533 & 8 & 44.6 & 22.6 & $0.785\pm0.007$ & $0.751\pm0.006$ \\
60002023006 & 2013-Jan-15 & 56307.0386 & 8 & 45.9 & 22.4 & $2.79\pm0.01$ & $2.74\pm0.01$ \\
60002023008 & 2013-Jan-20 & 56312.0980 & 8 & 45.2 & 24.9 & $0.923\pm0.006$ & $0.899\pm0.006$ \\
60002023010 & 2013-Feb-06 & 56329.0116 & 8 & 42.2 & 19.3 & $3.52\pm0.01$ & $3.55\pm0.01$ \\
60002023012 & 2013-Feb-12 & 56335.0106 & 6 & 35.4 & 14.8 & $4.39\pm0.02$ & $4.43\pm0.02$ \\
60002023014 & 2013-Feb-17 & 56339.9828 & 7 & 41.7 & 17.4 & $1.50\pm0.01$ & $1.54\pm0.01$ \\
60002023016 & 2013-Mar-04 & 56355.9631 & 6 & 35.0 & 17.3 & $4.11\pm0.02$ & $4.13\pm0.02$ \\
60002023018 & 2013-Mar-11 & 56362.9690 & 6 & 31.9 & 17.5 & $3.04\pm0.01$ & $3.02\pm0.01$ \\
60002023020 & 2013-Mar-17 & 56368.0210 & 6 & 35.1 & 16.6 & $4.33\pm0.02$ & $4.38\pm0.02$ \\
\enddata

\tablenotetext{a}{Livetime-corrected sum of all good time intervals comprising the observation.}
\tablenotetext{b}{PSF-corrected source count rate and its uncertainty in the 3--30 keV band averaged over the exposure time.}

\end{deluxetable*} 

\subsection{Swift {\em UVOT} and {\em XRT}} 

\label{sec:data-swift}

\swift observations with the UV/Optical Telescope (UVOT; \citealt{roming+2005}) were performed only with the UV filters (namely W1, M2, and W2). Observations with the optical filters were not needed because we had organized extensive coverage with ground-based optical telescopes, which have better sensitivity and angular resolution than \swiftuvot$\!\!$. We performed aperture photometry for all filters in all the observations using the standard UVOT software distributed within the HEAsoft package (version~6.10) and the calibration included in the latest release of the CALDB. Counts were extracted from an aperture of 5\arcsec\ radius for all filters and converted to fluxes using the standard zero points from \citet{breeveld+2011}. The fluxes were then dereddened using the value of $E(B-V)=0.014$ \citep{schlegel+1998,schlafly+2011} with $A_{\lambda}/E(B-V)$ ratios calculated using the mean Galactic interstellar extinction curve from \citet{fitzpatrick-1999}. No variability was detected within single exposures in any filter. The results of the processing were carefully verified, checking for possible contaminations from nearby objects within source apertures and from objects falling within background apertures. In almost all observations, Mrk\,421\ is on the ``ghost wings'' \citep{li+2006} from the nearby star 51\,UMa, so we estimated the background from two circular apertures of 16\arcsec\ radius off the source but on the wings, excluding stray light and support structure shadows.

The complete list of \swift X-ray Telescope (XRT; \citealt{burrows+2005}) and UVOT observations used here is given in Table~\ref{tab:observations-swift}. The observations were organized to be taken simultaneously with (or as close as possible to) the \magic$\!\!$/\veritas and {\em NuSTAR}\ observations, following the fruitful monitoring campaign practice since 2009. \swift observed the source 33 times in 2013 up to the end of March. All \swiftxrt observations were carried out using the Windowed Timing (WT) readout mode. The data set was processed with the XRTDAS software package (version~2.9.0) developed at ASDC and distributed with the HEASoft package (version~6.13). Event files were calibrated and cleaned with standard filtering criteria with the {\tt xrtpipeline} task using the latest calibration files available in the \swift CALDB. The average spectrum was extracted from the summed cleaned event file. Events for the spectral analysis were selected within a circle of 20-pixel ($\simeq$46\arcsec) radius, which encloses about 80\% of the PSF, centered on the source position. The background was extracted from a nearby circular region of 40-pixel radius. The ancillary response files (ARFs) were generated with the {\tt xrtmkarf} task applying corrections for PSF losses and CCD defects using the cumulative exposure map. The latest response matrices (version~14) available in the \swift CALDB were used.

Before the spectral fitting, the 0.3--10~keV source spectra were binned using the {\tt grppha} task to ensure a minimum of 20 counts per bin. Spectra were modeled in \xspec (version 12.8.0) using power-law and log-parabolic models, identical to the modeling presented in detail in \S\,\ref{sec:nustar-fullobs}. The models include photoelectric absorption by a fixed column density estimated to be $N_{\rm H}=1.92\times10^{20}$~cm$^{-2}$ \citep{kalberla+2005}. The log-parabolic model was found to fit the data better in each observation (though statistical improvement is marginal in some cases), and was therefore used to compute fluxes in various subbands. Spectral parameters are provided for each observation in Table~\ref{tab:swiftxrt_models}.

\subsection{NuSTAR} 

\label{sec:data-nustar}

{\em NuSTAR}\ (Nuclear Spectroscopic Telescope Array; \citealt{harrison+2013}) is a focusing hard X-ray telescope operating in the band from 3~to 79~keV. It is the first X-ray observatory to extend the sensitivity beyond the $\simeq$10~keV cutoff shared by virtually every current focusing X-ray satellite. The inherently low background associated with concentrating target X-rays enables {\em NuSTAR}\ to achieve approximately a 100-fold improvement in sensitivity over the collimated and coded-mask instruments that operate, or have operated, in the same bandpass. All observations are conducted in parallel with two coaligned, independent telescopes called FPMA and FPMB (for Focal Plane Module A and B).

The {\em NuSTAR}\ primary mission includes monitoring of several types of blazars; Mrk\,421\ has been selected for this program as a representative of the high-peaked BL~Lac (HBL) class. In order to maximize the strictly simultaneous overlap of observations by {\em NuSTAR}\ and ground-based VHE $\gamma$-ray observatories during the {5-month} campaign, three observations per month were scheduled according to visibility of Mrk\,421\ at the \magic and \veritas sites. A typical {\em NuSTAR}\ observation spanned 10~hours, resulting in 15--20~ks of source exposure after accounting for orbital modulation of visibility and filtering out South Atlantic Anomaly crossings where the background radiation is high. In addition to those observations, Mrk\,421\ was observed as a bright calibration target in July 2012 and early January 2013. The total exposure time over 88 orbits of {\em NuSTAR}\ observations in this period is $\simeq$250~ks. A list of all {\em NuSTAR}\ observations considered in this paper is given in Table~\ref{tab:observations-nustar}. Analysis of the remainder of the campaign data will be presented elsewhere.

The raw data have been reduced using the NuSTARDAS software version~1.3.1, as a part of the HEAsoft package version 6.12. The spectra of Mrk\,421\ were extracted from a circular region of 100\arcsec\ radius centered on the peak of the distribution of cleaned events. Background spectra were extracted from a region encompassing the same detector on which the source was focused, excluding the circular region from which the source counts were extracted. As the background generally differs between different detectors and may be variable on few-orbit timescales, extraction from a region of maximal area on the same detector where the source is present provides the best background estimate over the {\em NuSTAR}\ band. Nevertheless, other background extractions have been attempted and no significant differences have been observed in the results. 

The response files were generated using the standard {\tt nupipeline} and {\tt nuproducts} scripts, and the calibration files from CALDB version 20131223. All flux values reported in this paper have been corrected for the finite extraction aperture by the processing software. The dominant background component above 25~keV is the internal detector background. With good background characterization, the data may be used for spectral modeling up to the high-energy end of the {\em NuSTAR}\ band at 79~keV. The spectra of all {\em NuSTAR}\ observations of Mrk\,421\ are above the background level at least up to 25~keV and up to $\approx$40~keV in observations at high flux. For this reason, we quote count rates only up to 30~keV in the remainder of the paper. Three faint serendipitous sources have been found in the {\em NuSTAR}\ field of view (detected only in the deep co-added image using all observations presented in G.\,B.\,Lansbury \etal$\!$, {\em in preparation}); however, they do not represent a contamination problem due to the overwhelming brightness of Mrk\,421 in all epochs.

\subsection{Fermi-{\em LAT}} 

\label{sec:data-fermi}

The Large Area Telescope (LAT) on board the {\em Fermi} satellite is a
pair-conversion telescope with energy coverage from 20\,MeV to
$>300$~GeV. The LAT has a $\sim2.4$~sr field of view and provides
all-sky monitoring coverage on a $\sim3$ hour time
scale~\citep{atwood+2009}. For the analyses presented in this paper,
we have selected {\tt Source} class events with energies in the range
0.1--300\,GeV and within $15^\circ$ of the position of Mrk\,421. In
order to greatly reduce contamination from Earth limb photons, we have excluded
events at zenith angles $>100^\circ$ and any events collected when the
spacecraft rocking angle was $>52^\circ$. The data were analyzed using
the {\tt P7REP\_SOURCE\_V15} instrument-response functions and the
standard unbinned-likelihood software provided with version 09-33-00
of the {\em Fermi} Science Tools\footnote{\url{http://fermi.gsfc.nasa.gov/ssc/data/analysis/}}.

The analyses considered data in day-long and week-long intervals
contemporaneous with the {\em NuSTAR}\ observation windows. The likelihood
model used for all intervals included the sources from the second
{\em Fermi}\,-LAT\ catalog~\citep{nolan+2012} located within a 15$^\circ$
region-of-interest centered on Mrk\,421, as well as the standard
Galactic diffuse, isotropic and residual instrumental background
emission models provided by the {\em Fermi} Science Support
Center\footnote{\raggedright \url{http://fermi.gsfc.nasa.gov/ssc/data/access/lat/Background_Models.html}}. 
For all epochs, the spectrum of Mrk\,421\ was fitted with a
power-law model, with both the flux normalization and photon index
being left as free parameters in the likelihood fit. 
We summarize the spectral parameters for four selected epochs (discussed in detail in \S\,\ref{sec:discussion}) in Table~\ref{tab:observations-fermi}. The systematic uncertainty on the flux is estimated as approximately 5\% at 560~MeV and under 10\% at 10~GeV and above \citep{ackermann+2012}. As variability in the {\em Fermi}\,-LAT\ band was not significant, these epochs may be considered representative of the entire 2013 January--March period.

\begin{deluxetable*}{ccccc} 
\tabletypesize{\scriptsize}
\tablewidth{0pt}

\tablecaption{ Spectral parameters of the {\em Fermi}\,-LAT\ observations of Mrk\,421\ for four selected epochs in January--March 2013 \label{tab:observations-fermi} }

\tablehead{
  \colhead{\bf Start Time} &
  \colhead{\bf Stop Time} &
  \colhead{\bf Photon Flux\tablenotemark{a}} &
  \colhead{\multirow{2}{*}{\bf Photon Index}} &
  \colhead{\bf Energy Flux\tablenotemark{a}} \\
  \colhead{( MJD )} &
  \colhead{( MJD )} &
  \colhead{( $10^{-7}$~s$^{-1}$\,cm$^{-2}$ )} &
  \colhead{} &
  \colhead{( $10^{2}$~eV\,s$^{-1}$\,cm$^{-2}$ )}
}

\startdata
56112.1500 & 56119.6000 & $2.3\pm0.4$ & $1.74\pm0.09$ & $3.4\pm0.8$ \\
56291.7900 & 56298.5200 & $2.1\pm0.5$ & $1.8\pm0.1$   & $2.9\pm0.8$ \\
56298.5200 & 56304.8200 & $1.5\pm0.4$ & $1.6\pm0.1$   & $3.1\pm0.9$ \\
56304.8200 & 56309.8400 & $2.6\pm0.6$ & $1.9\pm0.1$   & $3\pm1$     \\
\enddata

\tablenotetext{a}{Fluxes in the 0.1--100\,GeV band.}

\end{deluxetable*} 


\begin{deluxetable*}{ccccccccc} 
\tabletypesize{\scriptsize}
\tablewidth{0pt}

\tablecaption{ Summary of the \magic observations of Mrk\,421\ (January--March 2013) \label{tab:observations-magic} }

\tablehead{
  \colhead{\multirow{2}{*}{\shortstack[c]{\, \\ \bf Start Time \\ \rm ( MJD ) }}} &
  \colhead{\multirow{2}{*}{\shortstack[c]{\, \\ \bf Exp. \\ \rm ( min. ) }}} &
  \colhead{\multirow{2}{*}{\shortstack[c]{\, \\ \bf Zenith Angle \\ \bf Range \rm ( $^{\circ}$ ) }}} &
  \colhead{\multirow{2}{*}{\shortstack[c]{\, \\ \bf $\sigma$\,\tablenotemark{a} }}} &
  \multicolumn{4}{c}{{\bf Model Fit\,\tablenotemark{b,c} }} &
  \colhead{\multirow{2}{*}{\shortstack[c]{\, \\ \bf Flux $>$200~GeV\,\tablenotemark{c} \\ \rm ( $10^{-11}$~s$^{-1}$\,cm$^{-2}$ ) }}} \\
  \cline{5-8} \\
  \colhead{} & \colhead{} & \colhead{} & \colhead{} &
  \colhead{$F_0$} &
  \colhead{$\Gamma$ or $\alpha$} &
  \colhead{$\beta$} &
  \colhead{$\chi^2/$d.o.f.} &
  \colhead{}
}

\startdata
\multirow{2}{*}{56302.1365} & \multirow{2}{*}{122} & \multirow{2}{*}{9--23} & \multirow{2}{*}{16.5} & $1.7\pm0.1$ & $2.87\pm0.07$ & \nodata & 19.7 / 16 & \nodata \\
 & & & & $1.9\pm0.1$ & $3.2\pm0.2$ & $0.8\pm0.3$ & 9.7 / 15 & $5.7\pm0.4$ \\
56307.2556 & 39 & 21--29 & 14.9 & $3.0\pm0.2$ & $2.48\pm0.09$ & \nodata & 18.8 / 15 & $10.4\pm0.9$ \\
56310.2441 & 54 & 20--31 & 8.3 & $1.3\pm0.2$ & $2.8\pm0.1$ & \nodata & 11.6 / 14 & $3.7\pm0.5$ \\
56312.1718 & 119 & 9--33 & 11.5 & $1.3\pm0.1$ & $2.9\pm0.1$ & \nodata & 8.8 / 20 & $3.9\pm0.4$ \\
56316.2417 & 29 & 24--30 & 10.9 & $2.4\pm0.2$ & $2.3\pm0.1$ & \nodata & 21.5 / 17 & $8.1\pm0.9$ \\
56327.0731 & 25 & 16--22 & 25.7 & $8.6\pm0.4$ & $2.27\pm0.05$ & \nodata & 20.5 / 19 & $34\pm2$ \\
56333.1147 & 29 & 9--10 & 15.8 & $4.4\pm0.3$ & $2.34\pm0.09$ & \nodata & 15.5 / 15 & $16\pm1$ \\
\multirow{2}{*}{56335.0795} & \multirow{2}{*}{116} & \multirow{2}{*}{9--24} & \multirow{2}{*}{38.3} & $5.7\pm0.2$ & $2.52\pm0.03$ & \nodata & 25.9 / 19 & \nodata \\
 & & & & $6.4\pm0.3$ & $2.53\pm0.04$ & $0.33\pm0.09$ & 9.6 / 18 & $20.5\pm0.7$ \\
56340.1722 & 29 & 23--36 & 13.5 & $3.0\pm0.3$ & $2.4\pm0.1$ & \nodata & 18.7 / 14 & $10\pm1$ \\
56362.0826 & 29 & 15--21 & 20.1 & $6.0\pm0.4$ & $2.36\pm0.07$ & \nodata & 22.1 / 19 & $20.7\pm1.4$ \\
\multirow{2}{*}{56363.1066} & \multirow{2}{*}{56} & \multirow{2}{*}{23--33} & \multirow{2}{*}{29.1} & $5.8\pm0.2$ & $2.56\pm0.04$ & \nodata & 34.0 / 20 & \nodata \\
 & & & & $6.8\pm0.4$ & $2.59\pm0.06$ & $0.56\pm0.06$ & 18.6 / 19 & $20\pm1$ \\
\enddata

\tablenotetext{a}{Detection significance, computed according to Equation~(17) from \citet{li+ma-1983} using data above 200~GeV integrated over the exposure time.}
\tablenotetext{b}{Power-law model of the form $dN/dE=F_0\ (E/E')^{-\Gamma}$, $E'=300$~GeV, is fitted to each observation. A log-parabolic model fit of the form $dN/dE = F_0\ (E/E')^{-\alpha-\beta \log(E/E')}$ is shown for observations in which it provides a better description of the spectrum than the power-law model. The normalization constant, $F_0$, is in units of $10^{-10}$\,s$^{-1}$\,cm$^{-2}$\,TeV$^{-1}$.}		
\tablenotetext{c}{Quoted uncertainties are statistical.}

\end{deluxetable*} 

\begin{deluxetable*}{ccccccccc} 
\tabletypesize{\scriptsize}
\tablewidth{0pt}

\tablecaption{ Summary of the \veritas observations of Mrk\,421\ (January--March 2013) \label{tab:observations-veritas} }

\tablehead{
  \colhead{\multirow{2}{*}{\shortstack[c]{\, \\ \bf Start Time \\ \rm ( MJD ) }}} &
  \colhead{\multirow{2}{*}{\shortstack[c]{\, \\ \bf Exp. \\ \rm ( min. ) }}} &
  \colhead{\multirow{2}{*}{\shortstack[c]{\, \\ \bf Zenith Angle \\ \bf Range \rm ( $^{\circ}$ ) }}} &
  \colhead{\multirow{2}{*}{\shortstack[c]{\, \\ \bf $\sigma$\,\tablenotemark{a} }}} &
  \multicolumn{4}{c}{{\bf Model Fit\,\tablenotemark{b,c} }} &
  \colhead{\multirow{2}{*}{\shortstack[c]{\, \\ \bf Flux $>$200~GeV\,\tablenotemark{c}\\ \rm ( $10^{-11}$~s$^{-1}$\,cm$^{-2}$ ) }}} \\
  \cline{5-8} \\
  \colhead{} & \colhead{} & \colhead{} & \colhead{} &
  \colhead{$F_0$} &
  \colhead{$\Gamma$ or $\alpha$} &
  \colhead{$\beta$} &
  \colhead{$\chi^2/$d.o.f.} &
  \colhead{}
}

\startdata
\multirow{2}{*}{56302.3411} & \multirow{2}{*}{131} & \multirow{2}{*}{9--33} & \multirow{2}{*}{18.7}
       & $1.7\pm0.2$ & $3.2\pm0.2$ & \nodata     & 10.2 / 5 & \nodata  \\
 & & & & $1.8\pm0.2$ & $3.3\pm0.2$ & $1.1\pm0.7$ & 7.9 / 4  & $5.2\pm0.6$  \\
56307.3487 & 170 & 8--28 & 34.5 & $3.6\pm0.2$ & $3.1\pm0.1$ & \nodata & 15.7 / 9 & $12.2\pm0.7$ \\
56312.3762 & 197 & 6--31 & 32.6 & $1.5\pm0.1$ & $3.2\pm0.2$   & \nodata & 4.9 / 5  & $5.0\pm0.4$  \\
56329.2864 & 49  & 7--28 & 25.2 & $4.5\pm0.5$ & $2.7\pm0.2$   & \nodata & 8.2 / 5  & $17\pm2$     \\
56340.3344 & 89  & 7--23 & 22.7 & $2.5\pm0.2$ & $3.2\pm0.2$   & \nodata & 5.1 / 5  & $8.5\pm0.8$  \\
\multirow{2}{*}{56356.2352} & \multirow{2}{*}{43} & \multirow{2}{*}{6--21} & \multirow{2}{*}{33.6}
       & $6.0\pm0.5$ & $3.0\pm0.1$ & \nodata     & 10.4 / 8 & \nodata \\
 & & & & $6.3\pm0.5$ & $3.0\pm0.1$ & $0.6\pm0.4$ & 8.2 / 7  & $20\pm2$ \\
\multirow{2}{*}{56363.2355} & \multirow{2}{*}{127} & \multirow{2}{*}{7--17} & \multirow{2}{*}{39.2}
       & $3.5\pm0.2$ & $3.5\pm0.1$ & \nodata     & 16.7 / 6 & \nodata \\
 & & & & $4.1\pm0.3$ & $3.9\pm0.1$ & $1.9\pm0.4$ & 5.6 / 5  & $10.9\pm0.8$ \\
\multirow{2}{*}{56368.1885} & \multirow{2}{*}{123} & \multirow{2}{*}{9--26} & \multirow{2}{*}{40.4}
       & $4.3\pm0.3$ & $3.2\pm0.1$ & \nodata     & 9.2 / 9  & \nodata \\
 & & & & $4.3\pm0.3$ & $3.2\pm0.1$ & $0.7\pm0.3$ & 4.6 / 8  & $13\pm1$     \\
\enddata

\tablenotetext{a}{Detection significance, computed according to Equation~(17) from \citet{li+ma-1983} using data above 200~GeV integrated over the exposure time.}
\tablenotetext{b}{Power-law model of the form $dN/dE=F_0\ (E/E')^{-\Gamma}$, $E'=300$~GeV, is fitted to each observation. A log-parabolic model fit of the form $dN/dE = F_0\ (E/E')^{-\alpha-\beta \log(E/E')}$ is shown for observations in which it provides a better description of the spectrum than the power-law model. The normalization constant, $F_0$, is in units of $10^{-10}$~s$^{-1}$\,cm$^{-2}$\,TeV$^{-1}$.}	
\tablenotetext{c}{Quoted uncertainties are statistical.}

\end{deluxetable*} 


\subsection{\em MAGIC} 

\label{sec:data-magic}

\magic is a system of two 17-m diameter imaging air-Cherenkov telescopes (IACTs) located at the Roque de los Muchachos Observatory on La~Palma, one of the Canary Islands (28$^{\circ}$46\arcmin\,N, 17$^{\circ}$53.4\arcmin\,W at 2231~m above sea level). The hardware was substantially upgraded during 2011 and 2012 \citep{aleksic+2016a}, which yielded a performance characterized by a sensitivity of $\simeq$0.7\% of the Crab Nebula flux to detect a point-like source above 200~GeV at 5\,$\sigma$ in 50~hours of observation. Equivalently, a 1-hour integration yields a detection of a source with approximately 5\% Crab flux. The angular resolution is $\lesssim$0.07$^{\circ}$ (68\% containment, $>200$~GeV), and the energy resolution is 16\%. The systematic uncertainties in the spectral measurements for a Crab-like point-source were estimated to be 11\% in the normalization factor (at $\simeq$200~GeV) and 0.15 in the power-law slope. The systematic uncertainty in the absolute energy determination is estimated to be 15\%. Further details about the performance of the MAGIC telescopes after the hardware upgrade in 2011--2012 can be found in \citet{aleksic+2016b}.

After data-quality selection, Mrk\,421\ was observed with \magic for a total of 10.8~h between 2013~January~8 and 2013~March~18. Most of these observations (8~h in total) were strictly simultaneous with the {\em NuSTAR}\ observations. They were performed in the ``false-source tracking'' mode \citep{fomin+1994}, where the target source position has an offset of 0.4$^{\circ}$ from the camera center, so that both signal and background data are taken simultaneously. Those data were analyzed following the standard procedure described in \citet{aleksic+2016b}, using the MAGIC Analysis and Reconstruction Software (MARS; \citealt{moralejo+2009}). The analysis cuts to extract $\gamma$-ray signals from the hadronic background were optimized on the Crab Nebula data and dedicated Monte Carlo simulations of $\gamma$-ray induced showers.

The significance of the source detection, calculated using Equation~(17) from \citet{li+ma-1983}, varied between 8.3\,$\sigma$ (MJD 56310) and 38.3\,$\sigma$ (MJD~56335). Observed intranight variability is not statistically significant, so we used data integrated over complete observations for the spectral analysis. Spectra were modeled with a power-law function with a normalization energy of 300~GeV. The normalization energy was chosen to be 300 GeV for both \magic and \veritas, in order to facilitate a direct comparison of the VHE spectral results. For observations in which the spectrum is not well described by a power-law model (MJD~56302, 56335 and 56363), we additionally fit a log-parabolic model.  A summary of the observations and the spectral modeling is given in Table~\ref{tab:observations-magic}. All uncertainties quoted in the table and in the rest of the paper are statistical only.

\subsection{\em VERITAS} 

\label{sec:data-veritas}

The Very Energetic Radiation Imaging Telescope Array System (\veritas$\!\!$) is an array of four 12-m diameter IACTs located in southern Arizona \citep{weekes+2002,holder+2006} designed to detect emission from astrophysical objects in the energy range from $\sim$100~GeV to greater than 30~TeV. A source with 1\% of the Crab Nebula flux can be detected in $\simeq$25 hours of observations; equivalently, a source with approximately 5\% Crab flux can be detected in a 1-hour integration. \veritas has an energy resolution of $\simeq$15\% and an angular resolution (68\% containment) of $\sim0.1^{\circ}$ per event at 1~TeV. The uncertainty on the \veritas energy calibration is approximately 20\%. The systematic uncertainty on reconstructed spectral indices is estimated at 0.2, independent of the source spectral index, according to studies of \citet{madhavan-2013}. Details of the sensitivity of the system after the recent hardware upgrade can be found on the \veritas website\footnote{\url{http://veritas.sao.arizona.edu/specifications}}.

\veritas observations of Mrk\,421\ were carried out under good weather conditions during the period of the {\em NuSTAR}\ campaign, resulting in a total, quality-selected exposure time of 15.5~h during the period 2013~Jan~10 to 2013~Mar~17, almost all of which is strictly simultaneous with {\em NuSTAR}\ exposures. These observations were taken at 0.5$^{\circ}$ offset in each of four cardinal directions from the position of Mrk\,421\ to enable simultaneous background estimation using the ``false-source tracking'' method \citep{fomin+1994}. Detected events are parametrized by the principal moments of the elliptical images of the Cherenkov shower in each camera \citep{hillas-1985}. Cosmic-ray background rejection is carried out by discarding events based on a set of selection cuts that have been optimized {\it a priori} using \veritas observations of the blazar 1ES\,1218$+$304 (photon index~3.0) and the Crab Nebula (photon index~2.5). The results were verified using two independent analysis packages \citep{cogan-2008,daniel-2008}.

The significance of the source detection was calculated using Equation~(17) from \citet{li+ma-1983}, and was found to vary between 18.7\,$\sigma$ (on MJD~56302) and 40.4\,$\sigma$ (on MJD~56368). No significant intranight variability was detected. Since the observations spanned several hours during each night, the energy threshold varied due to the range of zenith angles observed. Fluxes are therefore quoted at a commonly reached minimum energy of 200~GeV. We modeled the spectra with a power-law function with a normalization energy of 300~GeV. For four observations (MJD~56302, 56356, 56363 and 56368), the spectrum is better described with a log-parabolic model, while for the other observations this model does not provide a significantly better fit than the simpler power-law model. A summary of the observations, VHE flux and spectral parameters with their statistical uncertainties is given in Table~\ref{tab:observations-veritas}.

\begin{figure*} 
\begin{center}
\includegraphics[width=2\columnwidth]{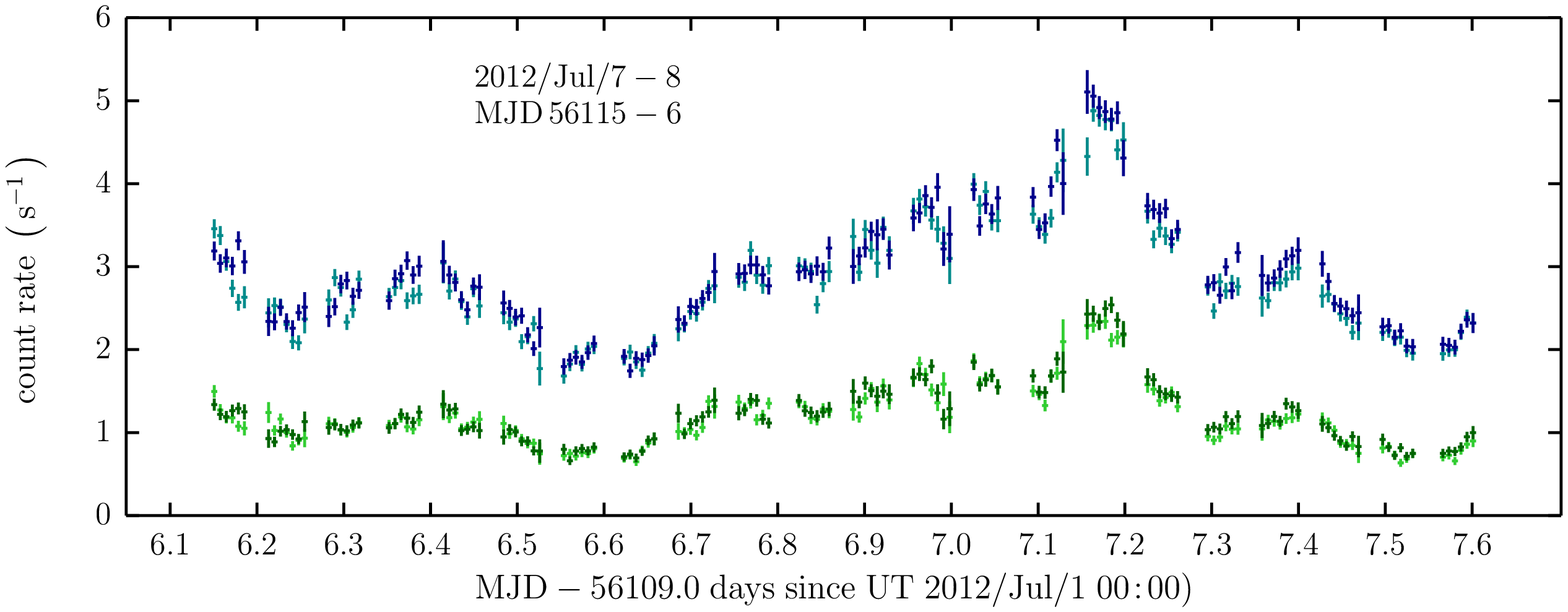}
\includegraphics[width=2\columnwidth]{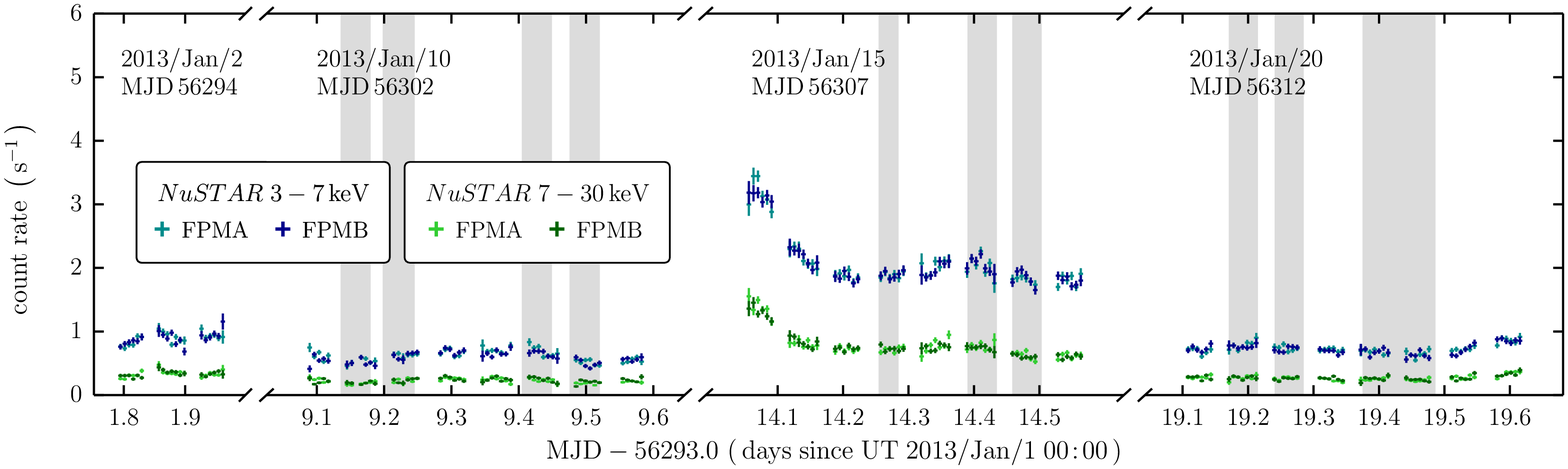}
\includegraphics[width=2\columnwidth]{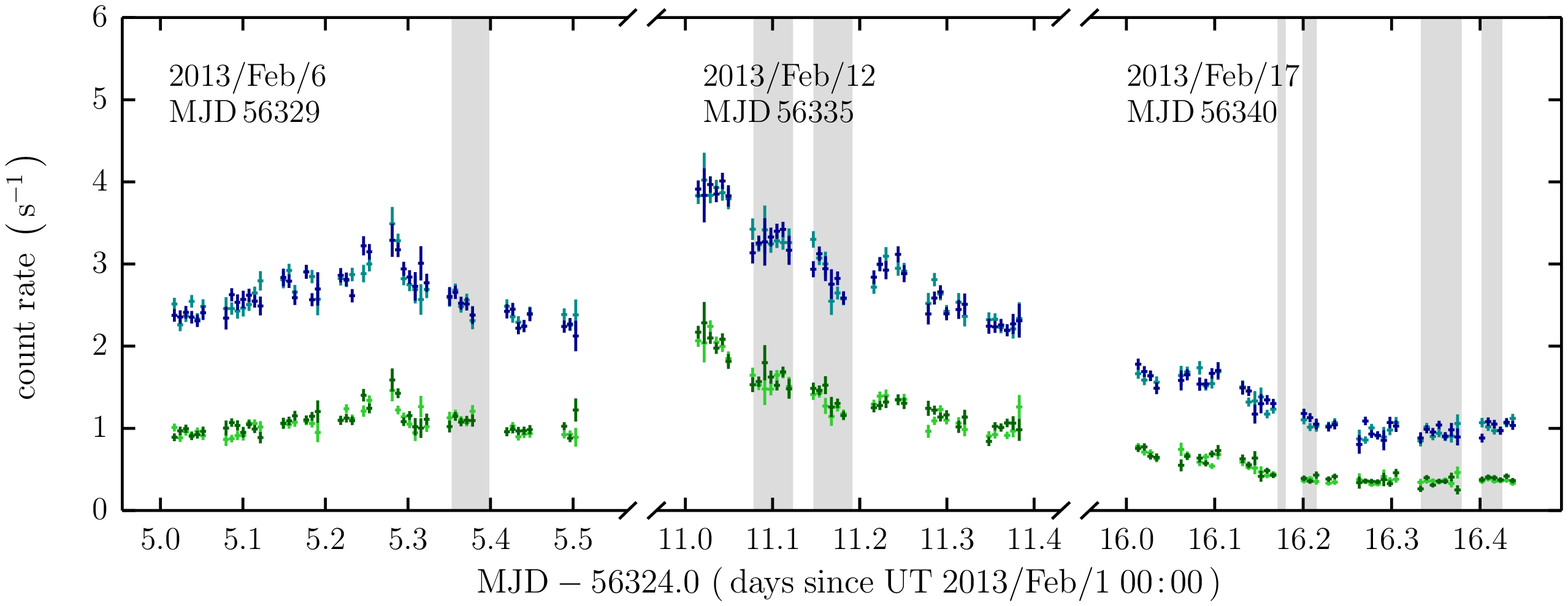}
\includegraphics[width=2\columnwidth]{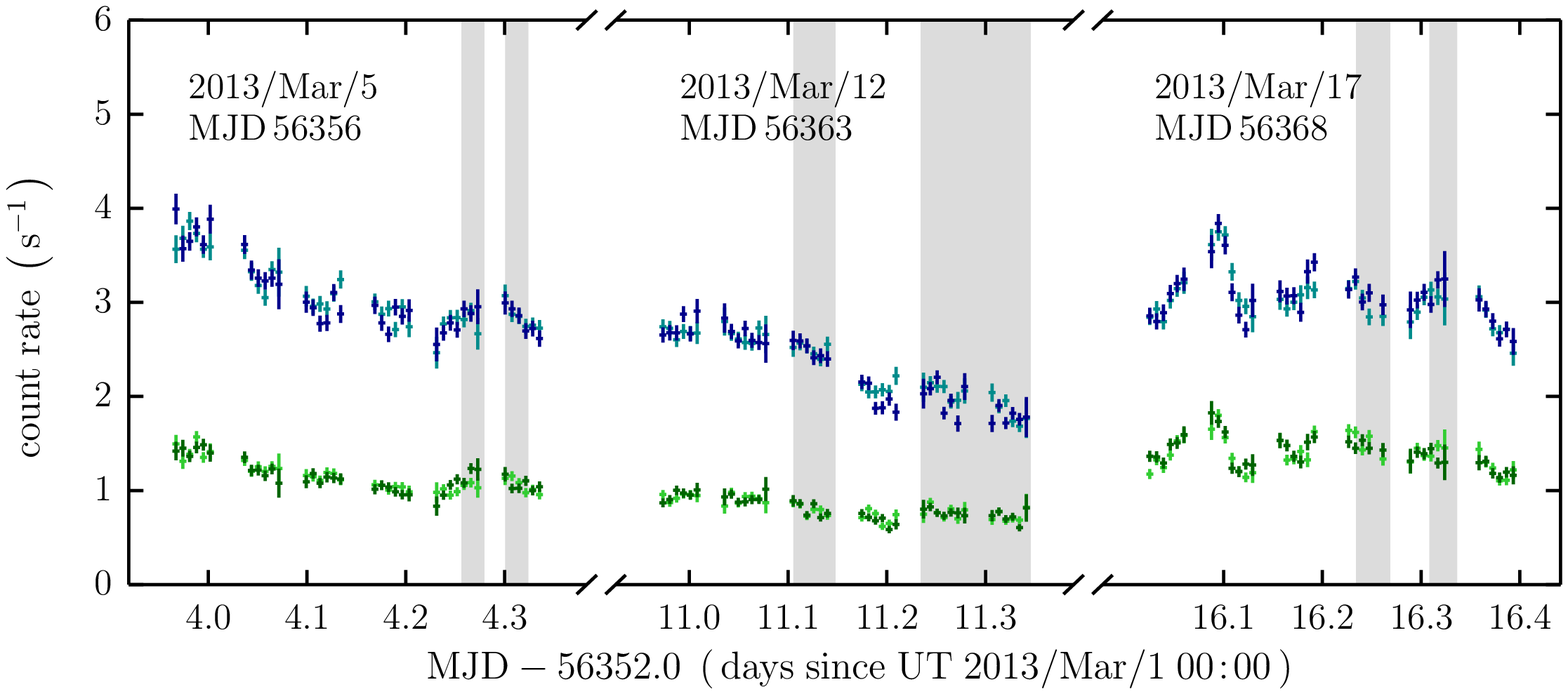}
\caption{ Count rates for {\em NuSTAR}\ in the 3--7 keV (blue) and 7--30~keV (green) bands. The legend given in the second panel from the top applies to all panels; for both bands FPMA count rates are plotted with a lighter color. The data are binned into 10-minute bins. The time axis of each row starts with the first day of the month and the UTC and MJD dates are printed above the light curves for each particular observation. Note that the data shown in the top panel represent two contiguous observations (broken up near MJD~56115.15). The intervals shaded in grey show times for which simultaneous data from \magic and \veritas are presented in this paper. Both the horizontal and the vertical scales are equal in all panels. \label{fig:nustar_count_rates}}
\end{center}
\end{figure*} 

\section{Characterization of the Hard X-ray Spectrum of Mrk\,421\ with {\em NuSTAR}} 

\label{sec:nustar}

\subsection{Flux and Hardness Ratio Variability} 

\label{sec:nustar-variability}

\begin{figure} 
\centering
\includegraphics[width=\columnwidth]{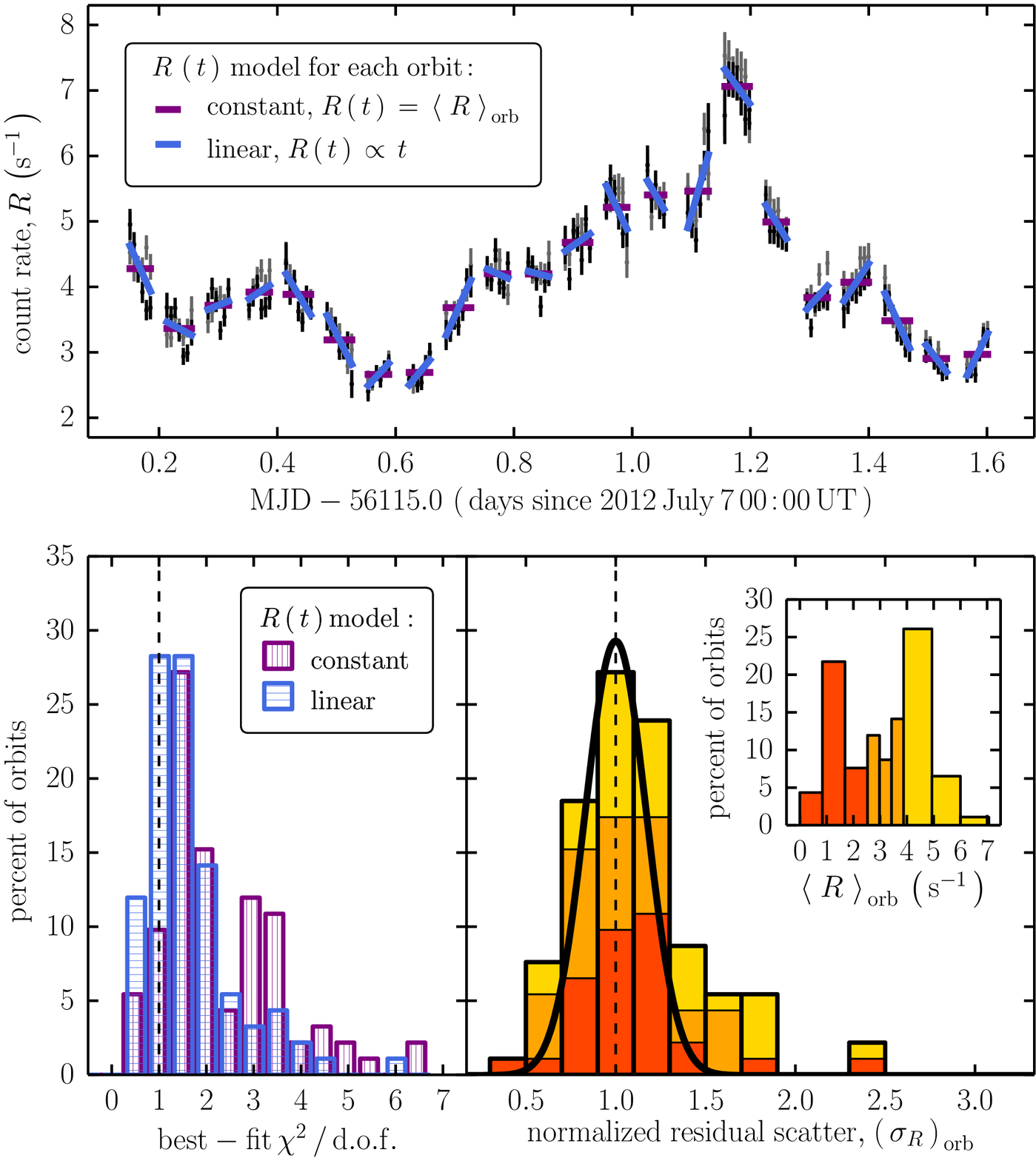}
\caption{ {\em Upper Panel:} Light curve of the July 2012 observation (FPMA in black, FPMB in grey) shown as an example for the count rate modeling; the two models fitted to each orbit of data are represented by purple (constant model) and blue lines (linear model). {\em Lower Panels:} The colored histograms on the left-hand side (colors matching the upper panel) show the distributions of reduced $\chi^2$ for the two models fitted to every {\em NuSTAR}\ orbit up to the end of March 2013. The number of degrees of freedom (d.o.f.) in each fit varies slightly due to the different duration of the orbits, but it is typically around 10. The right panel shows the distribution of the residual scatter after subtraction of the best-fit linear trend from the observed count rates in each orbit, in units of the median rate uncertainty within the orbit, ${\left( \sigma_R \right)}_{\rm orb}$. The colors reflect the mean count rate of the orbit: the lowest-rate third in red, the mid-rate third in orange and the highest-rate third in yellow, distributed as shown in the inset. The residual scatter distribution is slightly skewed to values greater than unity, indicating that $\lesssim$20\% of orbits show excess variability on suborbital timescales. A Gaussian of approximately matching width is overplotted with a thick black line simply to highlight the asymmetry of the observed distribution. }
\label{fig:nustar_var}
\end{figure} 

\begin{figure} 
\begin{center}
\includegraphics[width=\columnwidth]{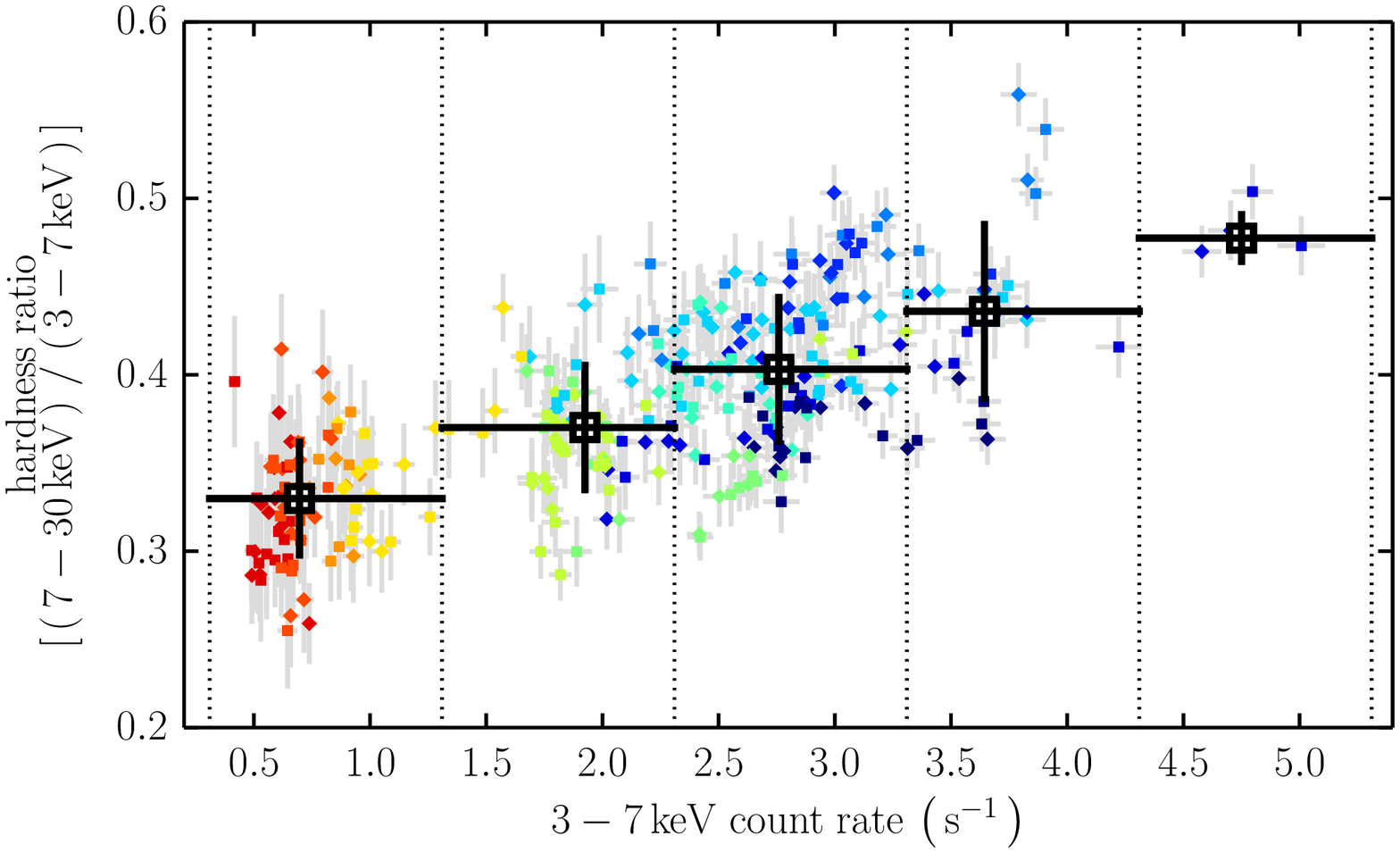}
\caption{ Hardness ratio (defined as the ratio of the number of counts in the 7--30~keV band to that in the 3--7~keV band) as a function of the count rate in individual 30-minute bins of {\em NuSTAR}\ data is shown with colored symbols: FPMA plotted with squares, and FPMB with diamonds. The colors distingush different observations and match those in Figure~\ref{fig:nustar_full_obs_spectra}. The thick black error bars and symbols show median count rate and hardness ratio in bins (1 ct\,s$^{-1}$ width), delimited by the vertical dotted lines. The vertical error bars denote standard deviation within each bin. \label{fig:nustar_hr}}
\end{center}
\end{figure} 

Figure~\ref{fig:nustar_count_rates} shows the background-subtracted X-ray light curves extracted from the {\em NuSTAR}\ observations of Mrk\,421 listed in Table~\ref{tab:observations-nustar}. The split into subbands at 7~keV is based on the spectral analysis and justified in later sections. The count rate above 30~keV is dominated by the background on short timescales and is therefore not shown here. The differences in count rates between observations, and the range covered in each particular observation, are entirely dominated by the intrinsic variability of the target. For example, the calibration observation taken in July 2012 (the top panel of Figure~\ref{fig:nustar_count_rates}) includes a possible ``flare'' in which the count rate increased by a factor of 2.5 over a 12-hour period and dropped by almost a factor of 2 in 3~h. Several observations in 2013 have shown steadily decreasing count rate over the course of $\simeq$12~h. We did not observe any sharp increases followed by exponential decay typical of flaring events, although we cannot exclude the possibility that the observed count rate decreases are due to such events. All of the observed increases in the count rate (e.g., 2012 July~7 and~8, 2013 February~6, as well as 2013 March~5 and~17 on a shorter timescale) appear rather symmetric with respect to subsequent decreases. The campaign observations up to the end of March 2013 have predominantly covered relatively low-flux states of Mrk\,421, even though the lowest and the highest observed fluxes span almost an order of magnitude.

The observed count rates are not consistent with a constant flux during any of the observations. However, the dominant variations in the count rate can be described as smooth on a timescale of several hours. If a simple exponential decay fit, $R(t)\propto e^{-t/\tau_{\rm \tiny var}}$, is performed on the observations that show significant downward trends (2012 July~8, 2013 January~15, February~12 and~17, March~5 and~12), the typical decay timescale ($\tau_{\rm var}$) is found to be between 6 and 12~h. These rough fits are not meant to describe the light curves fully, but only to provide an estimate of the timescale on which the flux changes significantly. For the remainder of the paper we use $\tau_{\rm var}=9\pm3$~hours as our best estimate of this timescale.

In order to characterize the variability on shorter timescales ($\Delta t\ll\tau_{\rm var}$), we consider the data in individual {\em NuSTAR}\ orbits as this represents a natural, albeit still arbitrary, way of partitioning the data. The {\em NuSTAR}\ orbits are approximately 90~minutes long and contain roughly 50~minutes of source exposure. We treat each orbit independently and fit two simple light-curve models to the observed count rates: a constant flux during the orbit, $R(t)=\,$constant, and a linear trend in time, i.e.,~$R(t)\propto t$. The top panel of Figure~\ref{fig:nustar_var} provides an example of both models fitted to the July 2012 data binned into 10-minute time bins, so that each orbit is divided into 5--7 bins per focal-plane module.\footnote{Because these data were taken in the calibration phase of the mission in suboptimal pointing conditions, a systematic uncertainty of 4\% was added to the light curve to reflect the total uncertainty in the true count rate.} The lower panel of Figure~\ref{fig:nustar_var} shows the results of this fitting procedure performed on all 88 orbits. 

We find that the flux during the majority of orbits is better described by a linear trend than by a constant-flux model. Linear trends account for most of the orbit-to-orbit variability and approximate smooth variations on super-orbital timescales of $\tau_{\rm var}\approx9$~hours. In 10-minute bins, for example, the variability amplitude typically does not exceed the observed count rate uncertainty of 3\%. Based on the mildly overpopulated tail of the reduced-$\chi^2\!$ distribution for the linear-trend fits, we estimate that up to 20\% of orbits show excess variance beyond the simple linear trend. Subtracting that trend and comparing the residual scatter to the median rate uncertainty within each orbit, ${\left( \sigma_R \right)}_{\rm orb}$, gives a distribution slightly skewed towards values greater than unity (see lower right panel of Figure~\ref{fig:nustar_var}). This is consistent with intrinsic suborbital variability on a $\sim$10-minute timescale in $\lesssim$20\% of orbits, while for the majority of the observations the short-timescale variability can be constrained to have a $\lesssim$5\% amplitude. These results are independent of the exact choice of the bin size, and hold for any subhour binning. Based on a separate analysis of the low- and high-flux data alone, we do not find significant evidence for a change in variability characteristics with flux.

The hardness of the spectrum, defined here as the ratio of count rates in the hard 7--30~keV and in the soft 3--7~keV bands, changes over the course of the observations. In Figure~\ref{fig:nustar_hr} we show the general trend of the spectrum hardening when the count rate is higher. Although the observed range of hardness ratios is relatively large at any specific count rate, the overall trend is clearly present in the binned data shown with thick black lines. There are no apparent circular patterns observed in the count rate versus hardness ratio plane, as previously seen in soft X-ray observations during bright flaring periods (e.g.,~\citealt{takahashi+1996,ravasio+2004,tramacere+2009}). We note, however, that the circular patterns might not be observable in the {\em NuSTAR}\ data presented here simply because the observations predominantly covered instances of declining flux, whereas the patterns arise from differences in the rising and the declining phases of a flare. The apparent symmetric features observed on 2012~July~7/8, 2013~February~6 and 2013~March~17 do not display enough contrast in flux and hardness ratio to show well-defined circular patterns.

\begin{turnpage} 

\begin{deluxetable*}{c cc cccc ccc cccc} 
\tabletypesize{\scriptsize}
\tablewidth{0pt}
\tablecolumns{11}

\tablecaption{ Models fitted to the {\em NuSTAR}\ spectra of each observation \label{tab:full_obs_models} }

\tablehead{
  \colhead{\multirow{2}{*}{\shortstack[c]{\, \\ \bf Start Time \\ \rm [ MJD ]\,}}} &
  \multicolumn{2}{c}{\bf Power Law} &
  \multicolumn{4}{c}{\bf Broken Power Law} &
  \multicolumn{3}{c}{{\bf Log-parabola} ($E_{*}=5$~keV)} &
  \multicolumn{4}{c}{{\bf Time-averaged Flux}\tablenotemark{b}} \\
  \cline{2-3} \cline{4-7} \cline{8-10} \cline{11-14} \\
  \colhead{} &
  \colhead{$\Gamma$} &
  \colhead{$\chi^2$/d.o.f.} &
  \colhead{$\Gamma_1$} &
  \colhead{$E_b$\tablenotemark{a}} &
  \colhead{$\Gamma_2$} &
  \colhead{$\chi^2$/d.o.f.} &
  \colhead{$\alpha$} &
  \colhead{$\beta$} &
  \colhead{$\chi^2$/d.o.f.} &
  \colhead{$3-7$~keV} &
  \colhead{$2-10$~keV} &
  \colhead{$7-30$~keV} \\
}

\startdata

56115.1353 & $2.82\pm0.01$\tablenotemark{c} & 1085/949 & $2.74\pm0.02$ & $7.0_{-0.6}^{+0.8}$ & $2.92\pm0.03$ & 922/947 & $2.76\pm0.01$ & $0.21\pm0.03$ & 906/948 & $9.47\pm0.03$ & $18.8\pm0.1$ & $6.48\pm0.03$ \\

56116.0732 & $2.87\pm0.01$\tablenotemark{c} & 1126/833 & $2.78\pm0.03$ & $7.0_{-0.7}^{+1.0}$ & $2.98\pm0.04$ & 978/831 & $2.99\pm0.01$ & $0.24\pm0.04$ & 967/832 & $11.08\pm0.03$ & $22.1\pm0.1$ & $7.16\pm0.04$ \\

56294.7778 & $3.10\pm0.04$ & 415/390 & $3.19_{-0.05}^{+0.07}$ & $7.6_{-1.4}^{+1.1}$ & $2.9\pm0.1$ & 399/388 & $3.16\pm0.05$ & $-0.2\pm0.1$ & 403/389 & $2.87\pm0.05$ & $6.4\pm0.1$ & $1.57\pm0.04$ \\

56302.0533 & $3.07\pm0.03$ & 512/506 & $3.08\pm0.04$ & $7.5$ (f) & $3.04\pm0.07$ & 511/505 & $3.07\pm0.04$ & $0.0\pm0.1$ & 512/505 & $1.91\pm0.02$ & $4.10\pm0.05$ & $1.05\pm0.01$ \\

56307.0386 & $3.02\pm0.01$ & 865/710 & $2.89\pm0.03$ & $6.4_{-1.3}^{+0.8}$ & $3.13_{-0.04}^{+0.06}$ & 742/708 & $2.92\pm0.02$ & $0.28\pm0.05$ & 741/710 & $6.43\pm0.03$ & $13.0\pm0.1$ & $3.55\pm0.02$ \\

56312.0980  & $3.05\pm0.02$ & 571/543 & $2.9_{-0.4}^{+0.1}$ & $4.6_{-0.9}^{+1.3}$ & $3.09_{-0.03}^{+0.04}$ & 568/541 & $3.03\pm0.03$ & $0.1\pm0.1$ & 572/542 & $2.26\pm0.02$ & $4.71\pm0.05$ & $1.22\pm0.01$ \\

56329.0116 & $2.93\pm0.01$ & 925/724 & $2.80\pm0.02$ & $7.5_{-0.4}^{+0.5}$ & $3.13\pm0.04$ & 709/722 & $2.82\pm0.02$ & $0.39\pm0.05$ & 709/723 & $8.27\pm0.05$ & $16.3\pm0.1$ & $4.94\pm0.04$ \\

56335.0106 & $2.73\pm0.01$ & 839/742 & $2.66\pm0.02$ & $10.2_{-1.1}^{+1.5}$ & $2.92_{-0.06}^{+0.8}$ & 738/740 & $2.64\pm0.02$ & $0.21\pm0.05$ & 742/741 & $9.25\pm0.05$ & $18.1\pm0.1$ & $7.00\pm0.06$ \\

56339.9828 & $3.02\pm0.02$ & 577/559 & $2.95\pm0.03$ & $7.5$ (f) & $3.09\pm0.05$ & 559/558 & $2.96\pm0.03$ & $0.14\pm0.09$ & 558/558 & $3.83\pm0.03$ & $7.9\pm0.1$ & $2.14\pm0.03$ \\

56355.9631 & $3.01\pm0.01$ & 823/701 & $2.91_{-0.04}^{+0.03}$ & $6.3_{-0.6}^{+0.9}$ & $3.09\pm0.04$ & 751/699 & $2.94\pm0.02$ & $0.22\pm0.05$ & 751/700 & $9.74\pm0.04$ & $19.9\pm0.1$ & $5.45\pm0.05$ \\ 

56362.9690 & $3.10\pm0.01$ & 640/640 & $3.01_{-0.15}^{+0.05}$ & $5.8_{-1.6}^{+3.1}$ & $3.16_{-0.03}^{+0.16}$ & 603/638 & $3.04\pm0.02$ & $0.19\pm0.06$ & 600/639 & $7.49\pm0.05$ & $15.6\pm0.1$ & $3.80\pm0.04$ \\

56368.0210 & $2.75\pm0.01$ & 848/760 & $2.67\pm0.02$ & $8.8_{-1.3}^{+0.9}$ & $2.91\pm0.05$ & 756/758 & $2.68\pm0.02$ & $0.24\pm0.05$ & 749/759 & $9.51\pm0.05$ & $18.6\pm0.1$ & $6.93\pm0.05$ \\

\enddata

\tablenotetext{a}{Energy (in keV) at which the model sharply changes slope from $\Gamma_1$ to $\Gamma_2$. For some observations this parameter is unbound, so we fix it to 7.5~keV and mark with (f).}
\tablenotetext{b}{Flux calculated from the best-fit model, in units of $10^{-11}$~erg~s$^{-1}$~cm$^{-2}$. Note that the 2--10~keV band requires some extrapolation below the {\em NuSTAR}\ bandpass, but we provide it here for easier comparison with the literature.}
\tablenotetext{c}{Formal statistical uncertainty is 0.008, however, the {\em NuSTAR}\ bandpass calibration is limited to 0.01 \citep{madsen+2015}, so we round up the values assuming this lowest uncertainty limit for these cases.}

\end{deluxetable*} 

\end{turnpage} 

\subsection{Observation-averaged Spectroscopy} 

\label{sec:nustar-fullobs}

\begin{figure*} 
\centering
\includegraphics[width=0.43\textwidth]{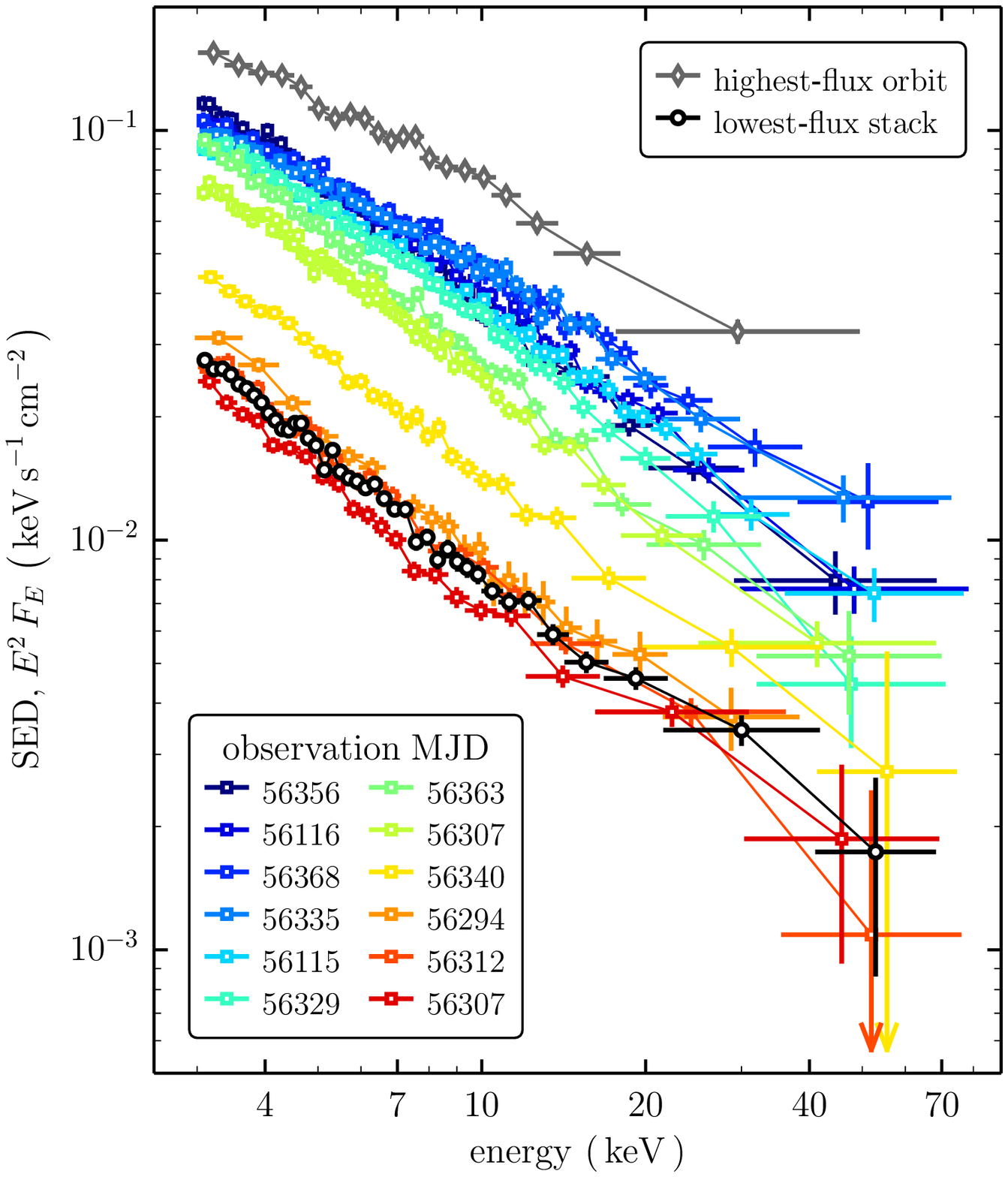}
\includegraphics[width=0.43\textwidth]{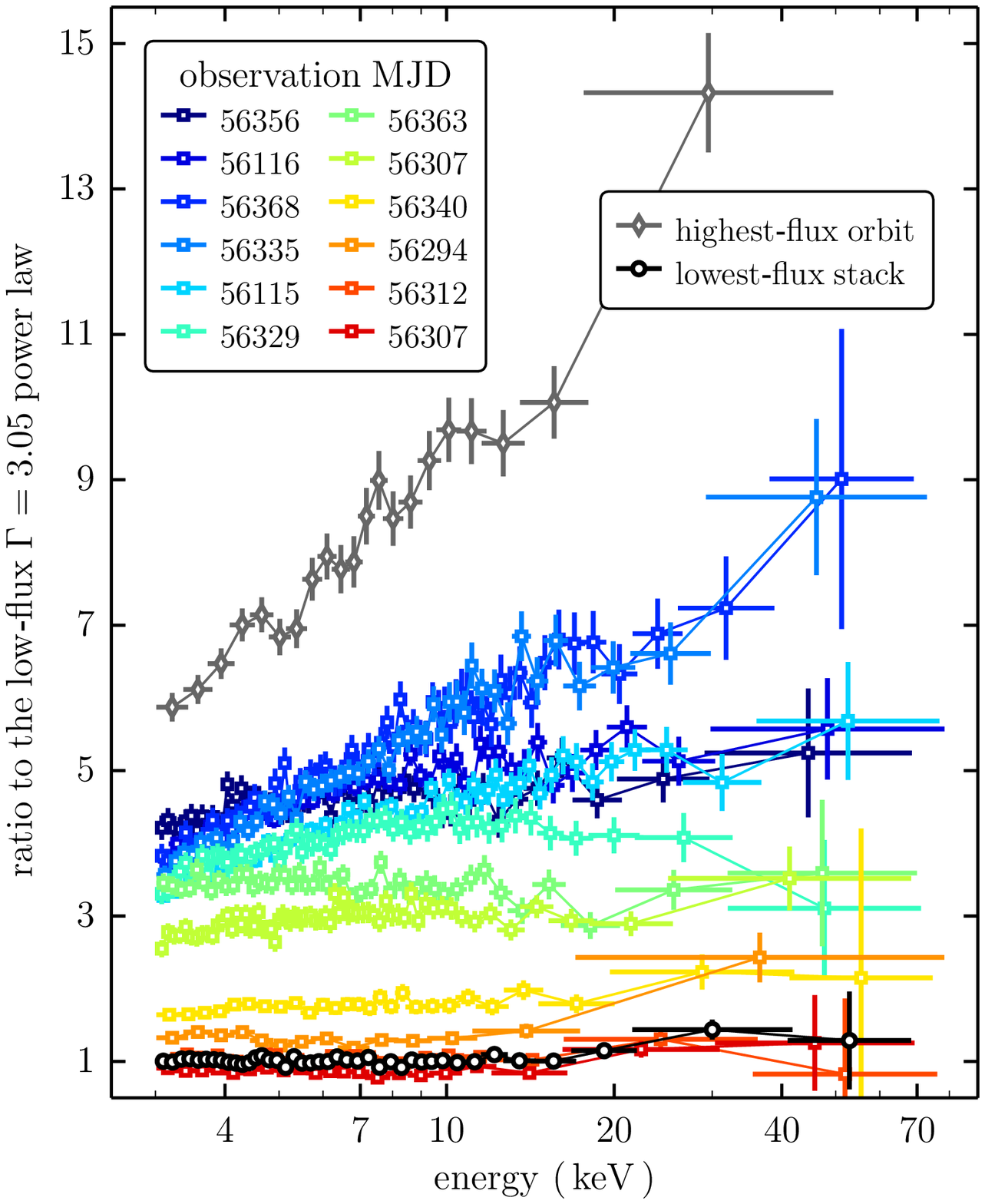}
\caption{ Unfolded {\em NuSTAR}\ spectra of Mrk\,421\ in each of the 12 observations. Colors are arranged by the 3--7~keV flux. Also shown are spectra of the single highest-flux orbit (grey symbols) and the stack of three lowest-flux observations (black symbols). Modules FPMA and FPMB have been combined for clearer display and the bin mid-points for each spectrum are shown connected with lines of the same color to guide the eye. The left panel shows the unfolded spectra in the $\nu F_{\nu}$ representation, while the right panel shows the same spectra plotted as a ratio to the best-fit model for the lowest-flux stacked spectrum (a power law with $\Gamma=3.05$). Note that the vertical scale is logarithmic in the left, and linear in the right panel. }
\label{fig:nustar_full_obs_spectra}
\end{figure*}

We first model the {\em NuSTAR}\ spectra for each observation separately, before examining the clear intra-observation spectral variability in \S\,\ref{sec:nustar-timesep} (see Figures~\ref{fig:nustar_var} and \ref{fig:nustar_hr}, and \citealt{balokovic+2013-granada}). All observation-averaged spectra are shown in Figure~\ref{fig:nustar_full_obs_spectra}. For spectral analysis we use spectra grouped to a minimum of 20 photons per bin and perform the modeling in \xspec (version 12.8.0). The simplest model for a featureless blazar spectrum is a power-law function:
\begin{equation}
F(E)\propto E^{-\Gamma},
\end{equation}
where $\Gamma$ is the photon index. The \xspec model is formulated as \texttt{phabs(zpow)}, where the \texttt{phabs} component accounts for the Galactic absorption with fixed hydrogen column density of $N_{\rm H}=1.92\times10^{20}$~cm$^{-2}$ \citep{kalberla+2005}. We first fit each of the 12 observation-averaged spectra with a power-law model and find that this model fits the observations with lower mean flux better than the ones where the mean flux is high (see Table~\ref{tab:full_obs_models}). This is likely due to the fact that the higher-flux spectra are somewhat more curved than lower-flux ones, although the curvature is not immediately obvious to the eye, i.e.,\,in Figure~\ref{fig:nustar_full_obs_spectra}. The fits confirm that the effective photon index decreases with increasing flux, as suggested by the observed harder-when-brighter behavior shown in Figure~\ref{fig:nustar_hr}.

The fitting results imply that a power-law model with $\Gamma\approx3$ describes the data well for observations with the lowest flux observed in the campaign. The poorer fit of the power-law model for the high-flux observation data may be due to intrinsic curvature, or it may be simply an effect of superposition of different curved or broken-power-law spectra. The latter effect can certainly be expected to be present since the hardness does vary with the flux and the source exhibited significant variability during most of the observations (see Figure~\ref{fig:nustar_count_rates}). We address this issue by examining spectra on a shorter timescale in Section~\ref{sec:nustar-timesep}. In order to better characterize the observation-averaged spectra, we replace the power-law model with two other simple models that allow extra degrees of freedom. The first one is a broken power-law model, \texttt{bknpow}:
\begin{eqnarray}
\nonumber F(E) & \propto & E^{-\Gamma_1}, E<E_b \mbox{;} \\
F(E) & \propto & E^{-\Gamma_2}, E>E_b.
\label{eqn:bknpow}
\end{eqnarray}
This model provides better fits to highest-flux spectra. However, the broken power-law form is degenerate at low flux and degrades into the simpler power-law shape discussed above for observations of low mean flux (i.e.,\ the photon indices converge to a single value and $E_b$ becomes unconstrained). Both photon indices depend on the flux, but dependence of $\Gamma_2$ seems to be weaker. The break energy seems to be largely independent of flux and relatively poorly constrained to the range roughly between 5 and 10~keV.

The third \xspec model we use is a simple log-parabolic shape, \texttt{logpar}:
\begin{equation}
f(E) \propto {\left( E/E_{*} \right)}^{-\alpha-\beta \log \left( E/E_{*} \right)}.
\label{eqn:logpar}
\end{equation}
In this model, $\alpha$ and $\beta$ are free parameters, while $E_{*}$ is the so-called pivot energy (fixed parameter) at which $\alpha$ equals the local power-law photon index. The $\beta$ parameter describes deviation of the spectral slope away from $E_{*}$. In our analysis we fix the value of $E_{*}$ to 5~keV, so that $\alpha$ closely approximates the photon index in the 3--7~keV band. This model fits all observations well and also hints at spectral trends outlined earlier. The log-parabolic model does not provide statistically better fits than the broken power-law model; in most cases they fit equally well (see Table~\ref{tab:full_obs_models}). However, the log-parabolic is often used for modeling blazar spectra in the literature, and does not contain a rather unphysical sharp break in the spectrum. All relevant parameters of the fits to the observation-averaged spectra are summarized in Table~\ref{tab:full_obs_models}. We note that for the {\em NuSTAR}\ observation on 2013~January~2 (MJD~56294), the best-fit parabolic model has marginally significant negative curvature, $\beta=-0.2\pm0.1$. As this is the shortest of all {\em NuSTAR}\ observations and the same effect is not observed in any of the other spectra, while negative curvature is a physically possible scenario, it is likely that this anomalous result arises from the fact that the high-energy background is not sufficiently well sampled in such a short exposure.

\begin{deluxetable}{cccc} 
\tabletypesize{\scriptsize}
\tablewidth{0pt}

\tablecaption{ Best-fit linear relations parametrized in Equation~(\ref{eq:trends}) for describing the change of spectral parameters with the X-ray flux. The uncertainties are quoted at the 1\,$\sigma$ level. \label{tab:nustar_flux_trends} }

\tablehead{
  \colhead{Model} &
  \colhead{Parameter, $X$} &
  \colhead{Slope, $s$} &
  \colhead{Zero Point, $X_0$} \\
}

\startdata

{\tt pow} & $\Gamma$ & $-0.32\pm0.07$ & $2.88\pm0.02$ \\
{\tt bknpow} & $\Gamma_1$ & $-0.46\pm0.07$ & $2.80\pm0.02$ \\
{\tt bknpow} & $E_b$\,\tablenotemark{a} & $1\pm9$ & $7\pm2$ \\
{\tt bknpow} & $\Gamma_2$ & $-0.11\pm0.05$ & $3.00\pm0.01$ \\
{\tt logpar} & $\alpha$ & $-0.42\pm0.07$ & $2.82\pm0.02$ \\
{\tt logpar} & $\beta$ & $0.22\pm0.06$ & $0.24\pm0.01$ \\

\enddata

\tablenotetext{a}{Due to poor constraint on this parameter, it is kept fixed at 7~keV while quantifying the trends in the $\Gamma_1$ and $\Gamma_2$ parameters.}

\end{deluxetable} 

\subsection{Time-resolved Spectroscopy} 

\label{sec:nustar-timesep}

We next consider spectra integrated over time intervals shorter than the complete {\em NuSTAR}\ observations. Separating the data into individual orbits represents the most natural although still arbitrary way of partitioning. Any particular orbit has a smaller spread in flux compared to a complete observation, since variability amplitude is significantly lower -- we have established in \S\,\ref{sec:nustar-variability} that the dominant flux variations occur on a super-orbital timescale of $\tau_{\rm var}\approx9$~hours. The shorter exposures significantly reduce the statistical quality, but still allow for a basic spectral analysis, such as the one presented in the preceding section, to be performed on spectra from single orbits. The average orbit exposure is 2.8~ks, and the total number of source counts per orbit is between 2,000 and 20,000 per focal-plane module. 

As with the observation-averaged spectra, we fit power-law, broken power-law and log-parabolic models using \xspec$\!\!$. We again find that the broken power-law model parameter $E_b$ (the break energy) is poorly constrained in general, so we fix it at 7~keV for the remainder of this analysis. Choosing a different value in the interval between 5~and 10~keV does not significantly alter any results; however, break energies outside of that interval cause one of the photon indices to become poorly constrained in a considerable number of orbits. Similarly, the curvature parameter of the log-parabolic model ($\beta$) is poorly constrained for the lowest-flux data, likely due to both the lack of intrinsic curvature and relatively poor photon statistics. In a similar manner to the observation-averaged spectral modeling, the log-parabolic model does not necessarily provide statistically better fits than the broken power-law model. We use it because it provides a smooth spectrum over the modeled energy range, and in order to facilitate comparison to other work in the literature. 

\begin{figure} 
\begin{center}
\includegraphics[width=0.99\columnwidth]{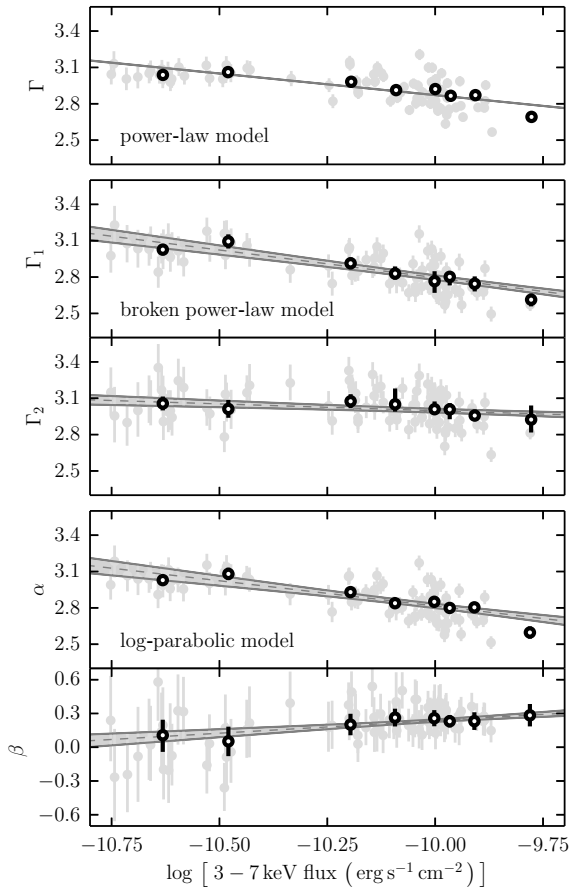}
\caption{ Trends in the hard X-ray spectral parameters as functions of flux for three simple spectral models revealed by time- and flux-resolved analyses of the {\em NuSTAR}\ data. The values fitted to high-S/N stacked spectra separated by flux (see text for an explanation) are shown with black-lined empty circles. A linear function is fitted to each of the trends and the uncertainty region is shaded in grey. Parameters of the fitted linear trends are given in Table~\ref{tab:nustar_flux_trends}. The filled grey circles in the background show spectral modeling results for spectra of 88 individual orbits.}
\label{fig:nustar_flux_trends}
\end{center}
\end{figure}

With less smearing over different spectral states of the source, the spectral variability is more clearly revealed by this analysis. As shown with the grey data points in Figure~\ref{fig:nustar_flux_trends} (one for each {\em NuSTAR}\ orbit), for any of the three models statistically significant spectral changes occur as the X-ray flux varies. The spectrum becomes harder as the flux increases. Most of the change happens below $\simeq$7~keV, as shown by the substantial variations in the parameters $\Gamma_1$ and $\alpha$, compared to the lower-amplitude variations in $\Gamma_2$ and $\beta$. In all cases the trends are consistent with the well-established harder-when-brighter behavior, also evident in the more basic representation using hardness ratios in Figure~\ref{fig:nustar_hr}. Since for orbits with the lowest count rate the uncertainties on the spectral parameters are relatively large, in the following section we verify that the same spectral variability trends persist for data with higher signal-to-noise ratio (S/N).

\begin{figure} 
\begin{center}
\includegraphics[width=\columnwidth]{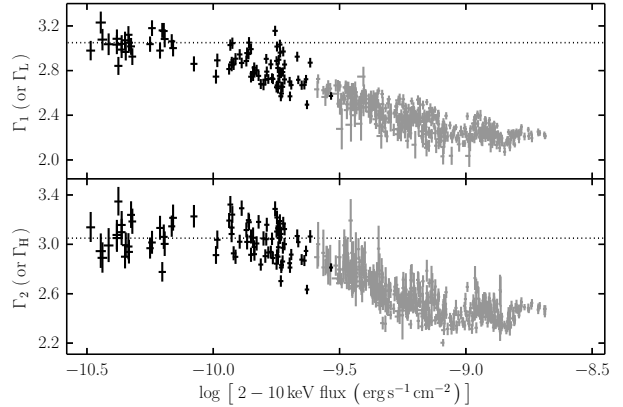}
\caption{ Comparison of the spectral trends revealed in our data (black symbols) with the ones published by~\citet{giebels+2007} (grey symbols). For this set of fits to the {\em NuSTAR}\ data the break energy ($E_b$) was kept fixed at 7~keV. The uncertainties are given at the 68\% confidence level in order to match the previous results. Note the smoothness of the trends covering nearly two orders of magnitude and the apparent saturation effects at each end. The dotted lines are median photon indices for 2--10~keV flux below $10^{-10}$~erg\,s$^{-1}$\,cm$^{-2}$, representing the apparent low-flux saturation values.}
\label{fig:comp_giebels}
\end{center}
\end{figure} 

\subsection{Flux-resolved Spectroscopy} 

\label{sec:nustar-fluxsep}

In order to verify that the spectral parameter trends we identify in the time-resolved spectra are not spuriously produced by relatively low photon statistics at the low-flux and high-energy ends, we proceed to examine stacked single-orbit spectra of similar flux. Stacking provides the highest possible S/N in well-defined flux bins, and allows us to use the data up to 70~keV -- where the signal is fainter by a factor of a few than the {\em NuSTAR}\ detector background. We combine spectra for each focal-plane module separately due to intrinsic differences in response matrices. Spectra from both modules are fitted simultaneously in \xspec$\!\!$, just like the observation-averaged and the single-orbit ones. Note that this procedure implicitly assumes that the source behaves self-consistently, in the sense that a particular flux level corresponds to a unique spectral shape within the data-taking time interval. The validity of this assumption is further discussed in Section~\ref{sec:discussion}.

We first stack the spectra of three complete observations (2013 January 2, 10 and 20), since Mrk\,421\ displayed a nearly-constant low flux during all three (see Figure~\ref{fig:nustar_count_rates}). The combined spectrum is very similar to the spectra from any of the constituent observations, but has significantly higher S/N. It can be statistically well described as a simple power law with $\Gamma=3.05\pm0.02$ from 3 to 70~keV ($\chi_{r}^{2}=1.05$). For completeness, modeling with a broken power-law model gives the break energy at $6\pm3$~keV, and low- and high-energy photon indices both formally consistent with the $\Gamma$ value found for the simpler power-law model. Furthermore, the curvature parameter of the log-parabolic model is consistent with zero ($\beta=0.01\pm0.04$), and $\alpha\approx\Gamma$. These fitting results lead us to conclude that in the lowest-flux state observed in 2013 the hard X-ray spectrum follows a steep  power law with $\Gamma\approx3$. Extrapolating below 3~keV for the sake of comparison with the literature, we derive a 2--10~keV flux of $(3.5\pm0.2)\times10^{-11}$~erg~s$^{-1}$~cm$^{-2}$ for the $\Gamma=3$ power-law model normalized to the lowest observed orbital flux (orbit \#6 of the 2013 January 20 observation). A state of such low X-ray flux has not yet been described in the published literature,\footnote{To the best of our knowledge, the lowest published 2--10~keV flux thus far was $(4.1\pm0.2)\times10^{-11}$~erg~s$^{-1}$~cm$^{-2}$ (1997 May; \citealt{massaro+2004}).} making the results of this analysis new and unique.

We combine the spectra of all 88 orbits according to their 3--7~keV flux in order to obtain higher S/N for the flux states covered in the data set. The choice of flux bins shown with black symbols in Figure~\ref{fig:nustar_flux_trends} is such that relatively uniform uncertainty in spectral parameters is achieved across the flux range; this condition requires stacking of $\sim$10 orbits of data on the faint end, while a single orbit is sufficient at the bright end. The results, however, are largely independent of the exact choice of which orbits to combine into a particular flux bin. Fitting the stacked spectra with the same simple spectral models as before reveals spectral trends much more clearly than for observation-averaged or time-separated spectra, as shown by black symbols in Figure~\ref{fig:nustar_flux_trends}. The spectra of the lowest-flux stack and the highest-flux orbit are displayed in Figure~\ref{fig:nustar_full_obs_spectra} for comparison with the observation-averaged spectra. The analysis performed here describes the spectral changes happening between those two extremes as a smooth function of the X-ray flux.

For each of the parameters of the power-law, broken power-law and log-parabolic models we parametrize their dependence on the X-ray flux as
\begin{equation}
\label{eq:trends}
X \left( F_{3-7~\mbox{\small{keV}}} \right) = s \log \left( F_{3-7~\mbox{\small{keV}}} / F_0 \right) + X_0,
\end{equation}
where $X$ stands for any of the spectral parameters ($\Gamma$, $\Gamma_1$, $E_b$, $\Gamma_2$, $\alpha$, $\beta$), $s$ is the slope of the relation, $F_{3-7~\mbox{\tiny{keV}}}$ is the 3--7~keV flux, $F_0$ is a reference flux in erg\,s$^{-1}$ (chosen to be the median flux of our dataset, $\log F_0 =-10.1$) and $X_0$ is the vertical offset (parameter value at the reference flux). We find that in all cases this linear function adequately describes the general trends. Since we find that the break energy of the broken power-law model ($E_b$) is independent of flux within its large uncertainties, we keep it fixed at 7~keV while fitting for the trends in the $\Gamma_1$ and $\Gamma_2$ parameters. For $\Gamma_2$, the high-energy photon index of the broken power-law model, the best-fit slope is small, but different from zero at the 2\,$\sigma$ level. For the rest of the spectral parameters, the trends are statistically more significant. Figure~\ref{fig:nustar_flux_trends} shows the best-fit $X \left( F_{3-7~\mbox{\tiny{keV}}} \right)$ relations superimposed on the time- and flux-resolved fitting results, clearly matching the trends that the former analysis hinted at. We list the best-fit linear trend parameters $s$ and $X_0$, with their 1\,$\sigma$ uncertainties, in Table~\ref{tab:nustar_flux_trends}.

Finally, we briefly return to the broken power-law model fits only to make a comparison to the previously observed spectral variability during flares. The \rxte $2\!-\!20$~keV data analysed by \citet{giebels+2007} overlap in the 2--10~keV flux only for the highest-flux single-orbit data presented here, and extends almost a decade above that. These authors showed that the break energy is essentially independent of flux and $\langle E_b \rangle = 5.9\pm1.1$~keV (68\% confidence interval), which is consistent with the median value of approximately 7~keV found from our nondegenerate fits of the broken power-law model. The photon indices were found to vary with flux up to approximately $10^{-9}$~erg~s$^{-1}$~cm$^{-2}$, above which they saturate at $\Gamma_1\approx2.2$ and $\Gamma\approx2.5$. The data presented here smoothly connect to those trends (see Figure~\ref{fig:comp_giebels}), extending them towards the faint end. Whereas the low-energy photon index ($\Gamma_1$ or $\Gamma_{\mbox{\tiny L}}$) continues to increase with decreasing flux, reaching $\Gamma\approx3$ at $\lesssim10^{-10}$~erg~s$^{-1}$~cm$^{-2}$, the high-energy one ($\Gamma_2$ or $\Gamma_{\mbox{\tiny H}}$) essentially levels off to the same $\Gamma\approx3$ at a factor of a few higher flux. A naive extrapolation of the \citet{giebels+2007} trends is therefore not supported by the new {\em NuSTAR}\ observations. Our analysis reveals a clear low-flux saturation effect that none of the previous studies could have constrained due to lack of sensitivity.

\begin{figure} 
\begin{center}
\includegraphics[width=\columnwidth]{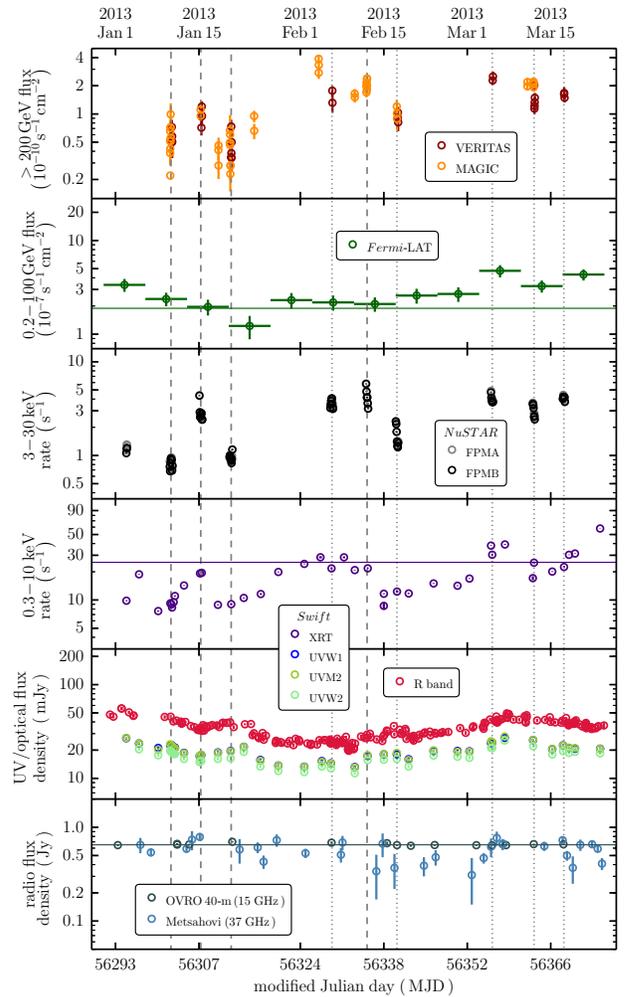}
\caption{ Light curves for Mrk\,421\ from \magic$\!\!$, \veritas (both above 200~GeV, binned in $\sim$30-minute intervals), {\em Fermi}\,-LAT\ (0.2--100~GeV, binned weekly), {\em NuSTAR}\ (3--30~keV, binned by orbit), \swiftxrt (0.3--10~keV, complete observations), \swiftuvot (UVW1, UVM2 and UVW2 bands, complete observations), ground-based optical observatories (R~band, intranight cadence), OVRO and Mets{\"a}hovi (15 and 37~GHz, both with 3--4-day cadence). The host-galaxy contribution in the R~band has been subtracted out according to \citet{nilsson+2007}. The dynamic range in all panels is 40. Vertical and horizontal error bars show statistical uncertainties and the bin width, respectively, although some of the error bars are too small to be visible in this plot. The vertical lines mark midpoints of the coordinated {\em NuSTAR}\ and VHE observations: dashed lines mark the epochs for which we discuss SED snapshots in \S\,\ref{sec:results-sed}, while the rest are shown with dotted lines. The horizontal lines in some panels show the long-term median values (see text for details).}
\label{fig:lc-big}
\end{center}
\end{figure} 

\section{Results from the Multiwavelength Observations} 

\label{sec:results}

\subsection{Multiwavelength Variability} 

\label{sec:results-mwvariability}

The majority of observations performed in January through March 2013 were coordinated between the participating observatories so as to maximize the strictly simultaneous overlap in the X-ray and VHE bands. In particular, nine 10--12-hour long observations performed by {\em NuSTAR}\ were accompanied with \swift pointings at the beginning, middle and the end, and the ground-based Cherenkov-telescope arrays \magic and \veritas covered roughly half of the {\em NuSTAR}\ exposure each. Approximately 50~h of simultaneous observations with {\em NuSTAR}\ and either \magic or \veritas resulted in total exposure of 23.5~h (the remainder being lost due to poor weather conditions and quality cuts). Figure~\ref{fig:lc-big} shows the multiwavelength light curves and highlights the dates of {\em NuSTAR}\ observations taken simultaneously with \magic and \veritas observations with vertical lines. A zoomed-in view of the VHE $\gamma$-ray observation times is shown overlaid on the expanded {\em NuSTAR}\ light curves in Figure~\ref{fig:nustar_count_rates}. 

The VHE flux varied between approximately 0.1 and 2 Crab units, reaching substantially lower and higher than the long-term average of $0.446\pm0.008$~Crab \citep{acciari+2014}, which is considered typical for a nonflaring state of Mrk\,421\ \citep{aleksic+2015b}. In Figure~\ref{fig:lc-big} we show typical fluxes for {\em Fermi}\,-LAT, \swiftxrt and OVRO bands, represented by medians of the long-term monitoring data that are publicly available. The {\em Fermi}\,-LAT\ light curve reveals elevated flux with respect to the median, as do the optical and UV data when compared to historical values. Modest soft X-ray flux is apparent from \swiftxrt data in comparison with the long-baseline median (\citealt{stroh+falcone-2013,balokovic+2013-granada}) and the intense flaring episodes covered in the literature (e.g.,\,\citealt{acciari+2009,tramacere+2009,aleksic+2012}). Count rates of $\lesssim10$~s$^{-1}$ in the 0.3--10~keV band are up to a factor of $\simeq2$ lower than those observed in quiescent periods during multiwavelength campaigns in 2009 \citep{aleksic+2015b} and 2010~March \citep{aleksic+2015c}. The radio flux was only slightly elevated above the values that have remained steady for the past 30~years, apart from the exceptional radio flare observed in October 2012 (see \S\,\ref{sec:flare2012} for more details).

\begin{figure} 
\centering
\includegraphics[width=\columnwidth]{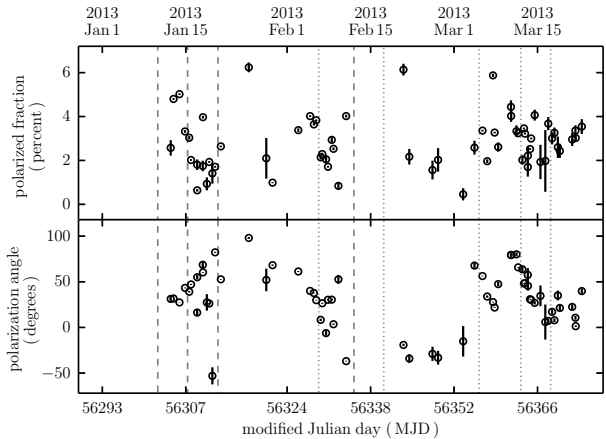}
\caption{Optical polarization of Mrk\,421 between 2013 January and March. The degree of polarization is shown in the upper panel and the position angle of polarization is shown in the lower panel.  Measurement uncertainties are based on photon statistics and are often smaller than the data points plotted. As in Figure~\ref{fig:lc-big}, the vertical lines mark midpoints of the coordinated {\em NuSTAR} and VHE observations: dashed lines mark the epochs for which we discuss SED snapshots in \S\,\ref{sec:results-sed}, while the rest are shown with dotted lines. }
\label{fig:optpol}
\end{figure}

Remarkably well-correlated flux variability in the X-ray and VHE bands on a timescale of about a week is already apparent from Figure~\ref{fig:lc-big}, and will be discussed in more detail in \S\,\ref{sec:results-correlations-xgammaray}. The fluxes in the UV and {\em Fermi}\,-LAT\ bands (to the extent allowed by the limited photon statistics) are consistent with a slow increase in flux between January and March, but do not show a clear short-term flux correlation. Further details regarding these bands are presented in \S\,\ref{sec:results-corelations-uvoptical}. The activity observed in the first three months of 2013 can be generally described as low. Note in particular that on January~10 and 20, Mrk\,421\ showed a remarkably low X-ray and VHE flux in comparison to the historical X-ray and VHE fluxes reported in \citet{stroh+falcone-2013} and \citet{acciari+2014}, respectively. Optical polarization, shown in Figure~\ref{fig:optpol}, showed random and statistically significant variations around the average polarized fraction of 3\%, and the polarization angle also varied significantly without any obvious coherent structure.

A general trend observed in the 2013 campaign is a gradual rise in broadband emission between January and March by a factor of\ $\lesssim$10, depending on the band. This was followed by an intense flaring period in April 2013 (not shown in Figure~\ref{fig:lc-big}), rivaling the brightest flares ever observed for Mrk\,421\ \citep{april-flare-atel1,april-flare-atel2,april-flare-atel3,pian+2014}. Analysis of the campaign data from the flaring period and more detailed analysis of the multiwavelength variability properties will be presented in separate publications. In the following sections, we focus on quantifying short-timescale and time-averaged correlations between different spectral bands and on the basic modeling of the Mrk\,421\ SED in the low-activity state that has not previously been characterized in any detail, except very recently in \citet{aleksic+2015b}.

The variability across the electromagnetic spectrum can be described using the fractional variability distribution. Fractional variability, $F_{\rm var}$, is mathematically defined in \citet{vaughan+2003}, and its uncertainty is calculated following the prescription from \citet{poutanen+2008}, as described in \citet{aleksic+2015a}. It can be intuitively understood as a measure of the variability amplitude, with uncertainty primarily driven by the uncertainty in the flux measurements and the number of measurements performed. While the systematic uncertainties on the absolute flux measurements\footnote{Estimated to be 20\% in the VHE band, and around 10\% in the optical, X-ray and GeV bands -- see \S\,\ref{sec:obs+data} for details.} do not directly add to the uncertainty in $F_{\rm var}$, it is important to stress that different observing sampling and, more importantly, different instrument sensitivity, do influence $F_{\rm var}$ and its uncertainty: a densely sampled light curve with very small temporal bins and small error bars might allow us to see flux variations that are hidden otherwise, and hence we might obtain a larger $F_{\rm var}$. Some practical issues of its application in the context of multiwavelength campaigns are elaborated in \citet{aleksic+2014,aleksic+2015b,aleksic+2015c}.

\begin{figure} 
\begin{center}
\includegraphics[width=0.98\columnwidth]{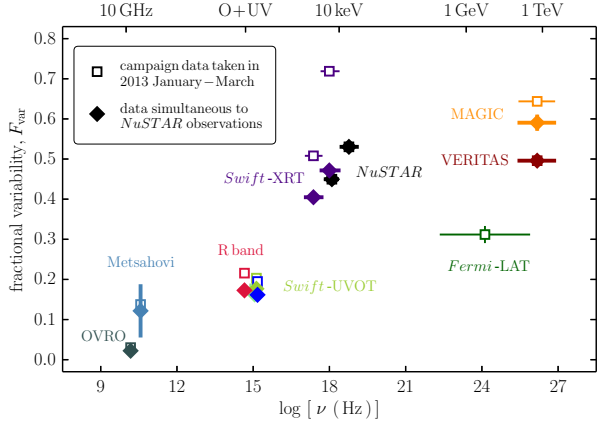}
\caption{ Fractional variability amplitude, $F_{\rm var}$, as a function of frequency for the period 2013 January--March. The vertical error bars depict the statistical uncertainty, while the horizontal error bars show the energy band covered (with the markers placed in the center of the segments). We show $F_{\rm var}$ computed in two ways: using the complete light curves acquired in the campaign (empty symbols), and using only the data taken within narrow windows centered on the coordinated {\em NuSTAR}\ and VHE observations (filled symbols). Note that the points overlap where only the coordinated observations are available. The {\em Fermi}\,-LAT\ point is based on the weekly-binned light curve shown in Figure~\ref{fig:lc-big}. }
\label{fig:fvar}
\end{center}
\end{figure}

In this paper we explore two cases, as shown in Figure~\ref{fig:fvar}. First, we use the full January--March dataset reported in Figure~\ref{fig:lc-big} (which has different cadence and different number of observations in each band), and second, we use only data collected simultaneously, in narrow windows centered on observations coordinated between {\em NuSTAR}\ and VHE telescopes. In the latter case the fluxes are averaged over the complete {\em NuSTAR}, \swift and VHE observations, effectively smoothing over any variability on shorter timescales. The optical and radio fluxes are taken from single measurements closest in time to the coordinated observations. In the former case, however, we sample shorter timescales and during a longer time span, which allows us to detect somewhat higher variability, as one can infer by comparing the \swift and \magic observations reported by the open/filled markers in Figure~\ref{fig:fvar}. $F_{\rm var}$ for {\em Fermi}\,-LAT\ is calculated from the weekly-binned light curve shown in Figure~\ref{fig:lc-big}; the relatively low GeV~$\gamma$-ray flux observed by {\em Fermi}\,-LAT\ precludes us from using significantly shorter time bins, or dividing the {\em Fermi}\,-LAT\ band into subbands as we do for \swiftxrt and {\em NuSTAR}.  Figure~\ref{fig:fvar} shows that $F_{\rm var}$ determined from our campaign rises significantly from the radio towards the X-ray band (consistent with \citealt{giebels+2007}), decreases over the {\em Fermi}\,-LAT\ band (consistent with \citealt{abdo+2011}), and then rises again in the VHE band. This double-bump structure relates to the two bumps in the broadband SED shape of Mrk\,421\, and has been recently reported for both low activity \citep{aleksic+2015b} and high activity \citep{aleksic+2015c}. The less variable energy bands (radio, optical/UV and GeV $\gamma$-ray bands) relate to the rising segments of the SED bumps, while the most variable energy bands (X-rays and VHE $\gamma$-ray bands) relate to the falling segments of the SED bumps.

\subsection{Correlations between Spectral Bands} 

\label{sec:results-correlations}

\subsubsection{X-ray versus VHE $\gamma$-ray Band} 

\label{sec:results-correlations-xgammaray}

\begin{figure*} 
\begin{center}
\includegraphics[width=0.97\textwidth]{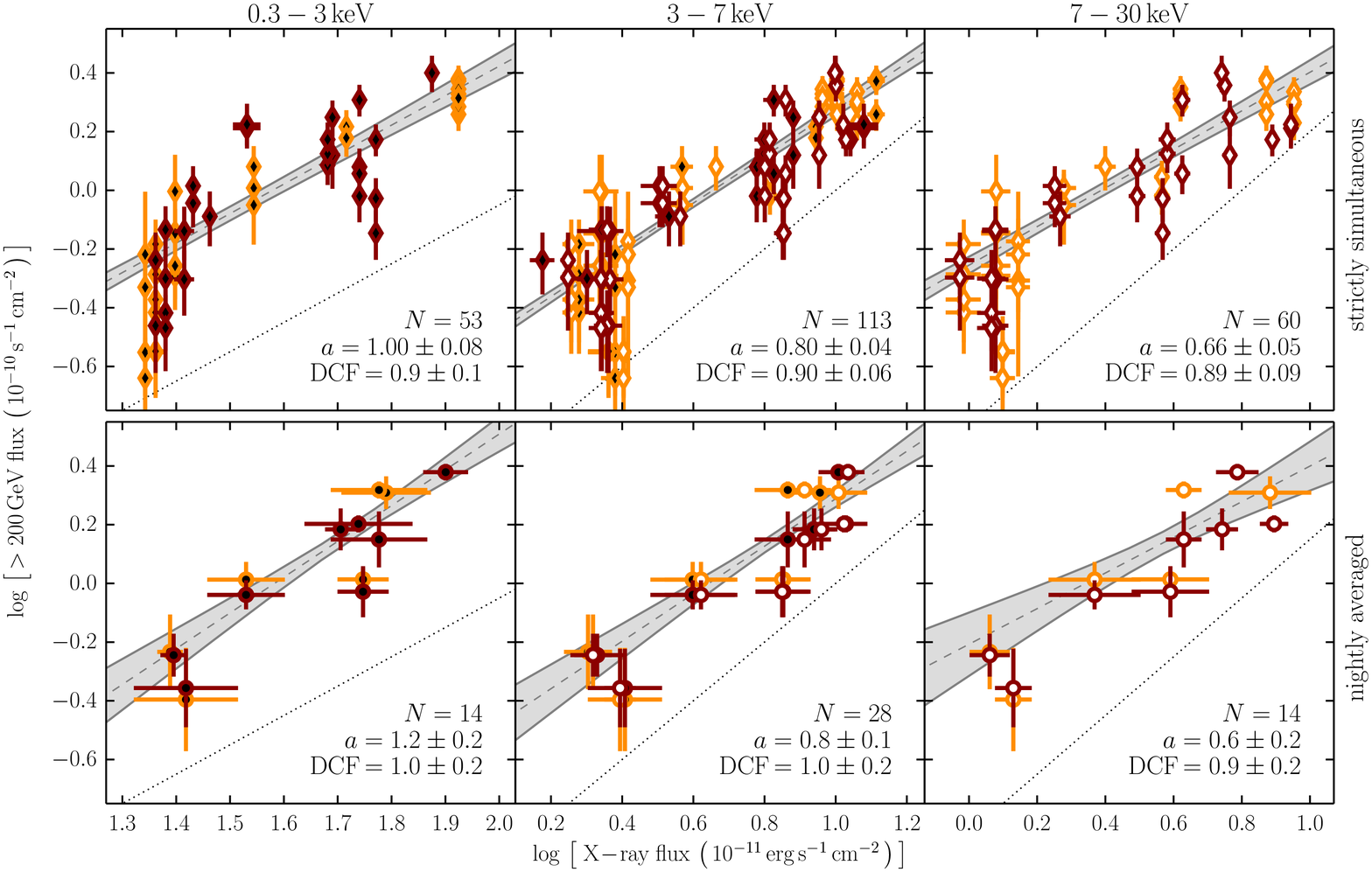}
\caption{ Flux--flux correlation between the X-ray and VHE ($>$200~GeV) flux in three different X-ray bands: \swiftxrt 0.3--3~keV in the left panel, \swiftxrt and {\em NuSTAR}\ 3--7~keV in the middle, and {\em NuSTAR}\ 7--30~keV in the right panel. \swiftxrt and {\em NuSTAR}\ measurements are shown with black-filled and white-filled symbols, respectively. Orange symbols mark \magic measurements, while dark red mark \veritas$\!\!$. In the upper panels we show only the data taken essentially simultaneously (within 1.5~hours). The lower panels show data averaged over the nights of simultaneous observations with X-ray and VHE instruments. The $N$, $a$ and DCF values given in each panel are the number of data points considered, the slope of the log--log relation, and the discrete correlation function. The best fit linear relation (dashed grey line) and its uncertainty region are shown with grey shading. The thin dotted line of slope unity is shown in all panels for comparison. }
\label{fig:Xray-VHE_fluxflux}
\end{center}
\end{figure*} 

The existence of a correlation between the X-ray and VHE fluxes is well established on certain timescales and in certain activity states of Mrk\,421: claims of correlated variability stem from long-term monitoring of fluxes in these bands that include high-activity states \citep{bartoli+2011,acciari+2014}, as well as observations of particular flaring events which probe correlated variability on timescales as short as 1~hour \citep{giebels+2007,fossati+2008,acciari+2009}. Detection of such a correlation in a low state was reported for the first time in \citet{aleksic+2015b}, using the X-ray (\swiftxrt$\!\!$, \rxtepca$\!\!$) and VHE (MAGIC, \veritas$\!\!$) data obtained during the 4.5-month multiwavelength campaign in 2009, when Mrk\,421\ did not show any flaring activity and varied around its typical \swiftxrt 0.3--10~keV count rates of $\sim$25~s$^{-1}$ and VHE flux of 0.5~Crab units. In this section, we confirm the flux--flux correlation in the X-ray and VHE bands with higher confidence, during a period of even lower activity. We also study the characteristics of such a correlation in different X-ray bands using the strictly simultaneous \swift$\!\!$, {\em NuSTAR}, \magic and \veritas observations.\footnote{The \magic and \veritas observations reported in this paper were performed after the extensive hardware upgrades performed on these two facilities in 2011 and 2012. They are therefore much more sensitive than the ones performed in 2009, which allows for a significant detection of lower flux in a single night of observation.} We summarize the results in Figure~\ref{fig:Xray-VHE_fluxflux} for three nonoverlapping X-ray bands. The flux in each band was calculated from the best-fit broadband model (power-law, or log-parabolic where needed; see \S\,\ref{sec:nustar}).

In the following, we use the discrete correlation function (DCF) and the associated uncertainty as defined in \citet{edelson+krolik-1988}. We carried out the correlation analysis on two timescales: a\ $\sim$1-hour timescale, using strictly simultaneous observations, and a\ $\lesssim$1-day timescale, using data averaged over complete 6--10-hour observations (i.e.,~averaged over one night of observations with each of the VHE observatories). Figure~\ref{fig:nustar_count_rates} shows the exact overlap of the {\em NuSTAR}\ and VHE observations. For both timescales and for all three X-ray bands we find significant correlations between fluxes in log--log space: DCF$\gtrsim0.9$ in all cases, with typical uncertainty of 0.1--0.2. The DCF is therefore inconsistent with zero with a minimum significance of 3.5\,$\sigma$ (nightly-averaged fluxes, 7--30~keV band) and maximum significance of 15\,$\sigma$ (simultaneous data, 3--7~keV band). As a sanity check, we also compute Pearson's correlation coefficients and find $>5$\,$\sigma$ significance in all cases. The strongest correlation, at 14\,$\sigma$, is again found for the 3--7~keV band and strictly simultaneous data shown in the upper middle panel of Figure~\ref{fig:Xray-VHE_fluxflux}. Note that we compute the correlation coefficients and DCF values using the logarithm of flux: because of greater dynamic range, the true flux--flux correlations are even more significant.

The similar slope of the $\log F_{\rm X-ray} - \log F_{\rm VHE}$ correlation on both timescales may indicate that the correlation is mainly driven by flux variability on a timescale of several days, i.e., between different observations, rather than within single observations spanning several hours. The statistical significance of the correlation on the $\sim$daily timescale is lower, due both to the smaller number of data points and to the fact that flux variance is larger because of the presence of strong variability on shorter timescales. For a chosen X-ray band, the best-fit slopes of the relation ($a$; listed in Figure~\ref{fig:Xray-VHE_fluxflux}) are statistically consistent with a single value. This is in good agreement with our finding that the dominant X-ray flux variability timescale is $\tau_{\rm var}\approx$9~hours (see \S\,\ref{sec:nustar-variability}). It could also be indicative of a lag between the bands which is longer than the binning of the data taken strictly simultaneously, however, such an analysis is outside of the scope of this paper. Results of \citet{aleksic+2015b} point to absence of any lags between the X-ray and VHE bands in a nonflaring state of Mrk\,421\ in 2009.

An interesting result stems from our ability to broaden the search for the correlation over a very wide band in X-rays, enabled by the simultaneous \swift and {\em NuSTAR}\ coverage. As shown in the upper three panels of Figure~\ref{fig:Xray-VHE_fluxflux}, the slope of the relation systematically shifts from $1.00\pm0.08$ for the soft 0.3--3~keV band, to $0.80\pm0.04$ for the 3--7~keV band, and to $0.66\pm0.05$ for the hard 7--30~keV band. The same behavior is seen in the nightly-averaged data, with somewhat lower significance. The persistence of this trend on both timescales and in all observatory combinations counters the possibility of a systematic bias related to those choices. We interpret it as an indication that the soft X-ray band scales more directly with the VHE flux (which is dominated by {\it soft} $\gamma$-ray photons on the low-energy end of the VHE band) due to the emission being produced by the same population of relativistic electrons. The greater relative increase in the hard X-ray flux with respect to the soft band is consistent both with the spectral hardening already revealed by our analysis (see \S\,\ref{sec:nustar-fluxsep}) and the fractional variability distribution determined from our data (Figure~\ref{fig:fvar}). Our interpretation would imply that the hard X-ray band scales more directly with the higher-energy VHE flux (e.g., $>1$~TeV), which we cannot quantify well with the current data.

We find no significant correlation of the simultaneously observed spectral slopes in the hard X-ray and VHE bands. Remarkably, on two dates when the observed X-ray flux was lowest, January~10 and 20 (MJD 56302 and 56312, respectively), steep spectra with $\Gamma\approx3$ were observed by {\em NuSTAR}\ and both VHE observatories. Other simultaneous observations yield $\Gamma>2.6$ in the 3--30~keV band and $\Gamma>2.4$ in the VHE band, with an average photon index of approximately 3 in both bands. In comparison to the previously published results, we note that the observed steepness of the X-ray and VHE spectra presented here is atypical for Mrk\,421. In a more typical low state, such as that observed in the 2009 campaign \citep{abdo+2011}, a photon index of $\simeq2.5$ has been observed in both bands. Here we compare our {\em NuSTAR}\ spectral slopes to that of the \rxte spectrum (2--20~keV) integrated over several months. Care should be taken in comparing with previously published results, because direct slope measurements in the 3--30~keV band were not available before, especially on short timescales. While the simultaneously observed steep slopes add support to the connection between X-ray and VHE bands, higher-quality data for the quiescent states are clearly needed in order to quantify it further.

\begin{figure} 
\begin{center}
\includegraphics[width=0.90\columnwidth]{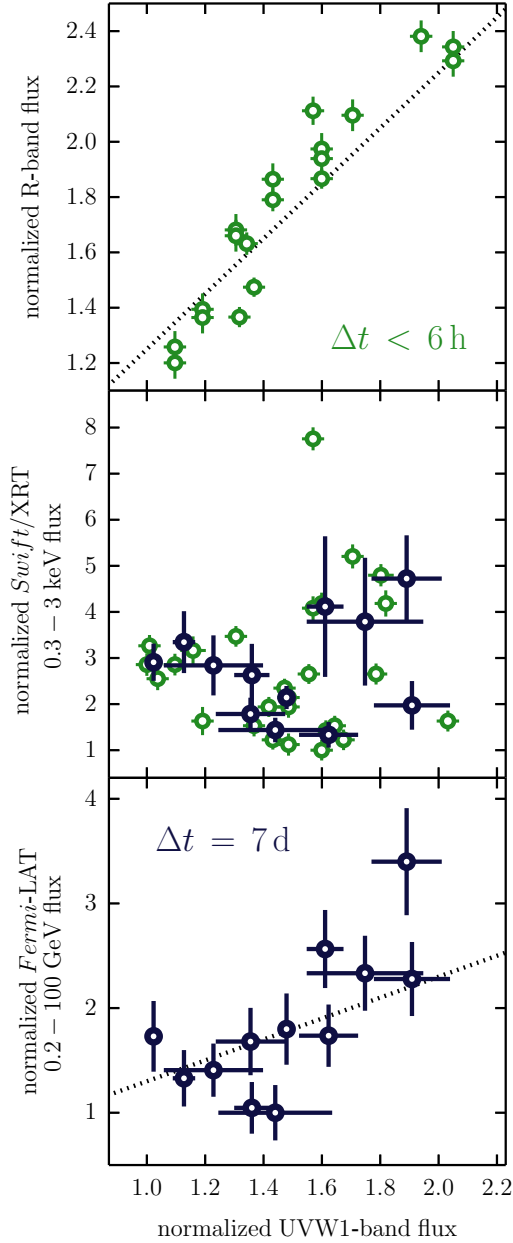}
\caption{ Flux-flux correlations between the UV band (\swiftuvot$\!\!$ filter UVW1) and the optical (R~band; top), soft X-ray (\swiftxrt 0.3--3~keV; middle) and GeV $\gamma$-ray ({\em Fermi}\,-LAT\ 0.2--100~GeV; bottom) bands. The green data points are based on flux measurements that are coincident within 6~hours. The blue data points are derived by averaging over 7-day intervals, i.e.,~integration times used for determination of the flux in the {\em Fermi}\,-LAT\ band. In the top and bottom panels we overplot the slope of one-to-one proportionality (not a fit) with a black dotted line. Note that the vertical scale is linear in all panels, but has a different dynamic range in each one.}
\label{fig:uv_fluxflux}
\end{center}
\end{figure} 

\subsubsection{UV/optical Versus Other Bands} 

\label{sec:results-corelations-uvoptical}

Despite the low flux observed in the X-ray and $\gamma$-ray bands, the range of the UV/optical flux was higher than in some flaring episodes reported in the past (e.g.,\,\citealt{aleksic+2012}). In this section we present flux correlation analyses with respect to the UV band, as represented by measurements using \swiftuvot$\!\!$. The choice of band UVW1 ($\lambda_{\rm eff}=2120$\,\AA) for this work is arbitrary; results do not change for either of the other two filters, as all of them sample the flux on the opposite side of the extremely variable synchrotron SED peak from the X-ray band. In Figure~\ref{fig:uv_fluxflux} we show the correlations between the UV and optical, soft X-ray and GeV $\gamma$-ray bands, each normalized to the lowest flux observed in the 2013 campaign. As in previous sections, we use the DCF and the associated uncertainty to quantify the correlation significance.

A strong correlation is expected between the UV and optical fluxes, and is confirmed by the data presented here. Previous work hinted at a possible correlation of the optical flux and the X-ray flux, but over a very narrow dynamic range and with low significance \citep{lichti+2008}. The states of Mrk\,421\ observed in early 2013 are not consistent with that result, indicating perhaps that a physically different regime was probed. We examine two different timescales in more detail here: for the UV and X-ray measurements taken within 6~hours of each other the DCF is $0.2\pm0.2$, while for weekly-averaged values it is $0.3\pm0.4$, i.e., consistent with zero in both cases. Note that the X-ray data require averaging in the latter case, and that the uncertainty in flux is dominated by intrinsic variability. Given the established difference in the variability characteristics (see Figure~\ref{fig:fvar}) the lack of a significant correlation, especially on the shorter timescale, is not unexpected.

The most interesting correlation in terms of constraints on physical models is the one between the UV (and hence the optical, given their essentially 1-to-1 correspondence) and the {\em Fermi}\,-LAT\ band (0.2--100~GeV), shown in the bottom panel of Figure~\ref{fig:uv_fluxflux}. In the framework of the SSC model, emission in these bands is due to electrons of roughly the same energy. Since the {\em Fermi}\,-LAT\ light curve had to be derived in $\approx$1-week bins due to the low photon counts (see Section~\ref{sec:data-fermi}), we average the UV flux over the same time periods in order to cross-correlate them. Overall, the DCF is $0.8\pm0.3$, revealing a possible correlation with a $2.7$\,$\sigma$ significance. In order to examine its robustness against contributions from the outlying data points, we perform the following test: we first remove the highest-flux {\em Fermi}\,-LAT\ data point and find that it does not change the DCF, then we remove the lowest-flux \swiftuvot data point, which increases the DCF to $0.9\pm0.4$ but lowers the significance. We therefore estimate the correlation significance to be $\simeq$2.5\,$\sigma$ based on the data presented here. Existence of a real correlation between UV and $\gamma$-ray bands cannot be confirmed with the current data, but this may be possible with the Mrk\,421\ observations at higher $\gamma$-ray flux, such as those taken during our multiwavelength campaign in April 2013.

\begin{figure} 
\centering
\includegraphics[width=0.96\columnwidth]{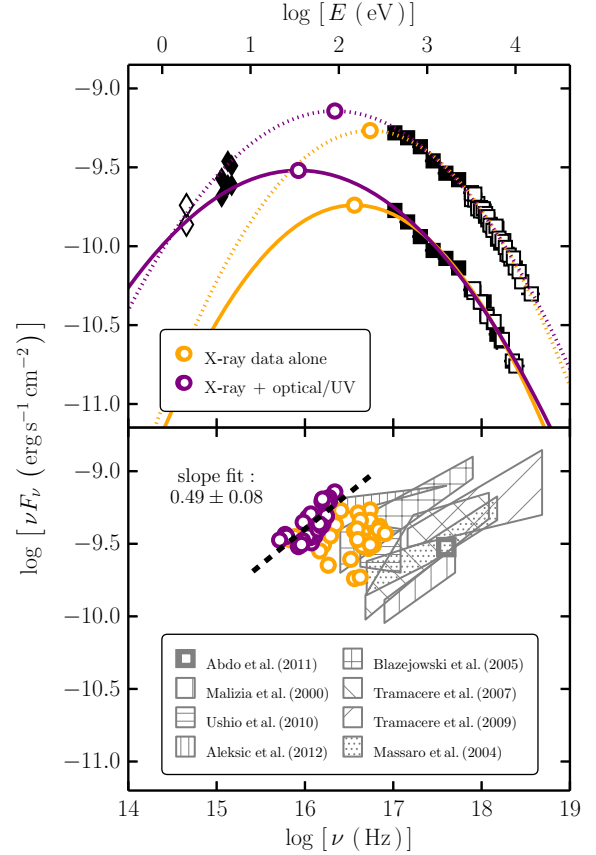}
\caption{ {\it Upper panel:} Examples of approximate localization of the synchrotron SED component peak for two orbits of simultaneous observations with \swift and {\em NuSTAR}. The \swift data are shown as black filled symbols (diamonds for the UVOT and squares for the XRT) and the {\em NuSTAR}\ data are shown as black empty squares. Empty diamonds represent R-band data. For each of the epochs we show a log-parabolic curve fit to X-ray data alone (yellow) and all data (purple). For each curve, we mark the SED peak with an empty circle of matching color. {\it Lower panel:} Results of the SED peak localization based on data from strictly simultaneous \swift and {\em NuSTAR}\ orbits. The colored data points show $\nu_{\rm syn.\,peak}$ and $(\nu F_{\nu})_{\rm syn.\,peak}$, i.e.,~the frequency of the SED peak and the flux at the peak. The assumption of the log-parabolic model connecting the UV/optical and X-ray data (purple empty circles) reveals a proportionality between $\log (\nu F_{\nu})_{\rm syn.\,peak}$ and $\log \nu_{\rm syn.\,peak}$; the dashed black line shows a linear fit best describing that relation. The other method (using only X-ray data) does not show a similar relation. In comparison with the observations published previously, shown here with different hatched grey regions, in 2013 January--March we observed a state in which the peak occurred at atypically low energy and high flux. }
\label{fig:synchpeak}
\end{figure} 

\subsection{The Peak of the Synchrotron SED Component} 

\label{sec:results-synchpeak}

Previous work on modeling the Mrk\,421 SED established that the lower-energy peak of the SED, likely arising from synchrotron processes, is usually located at frequencies between $\sim10^{17}$ and $\sim10^{18}$~Hz. The peak itself is therefore often directly observable in the \swiftxrt band, as in other similar HBL sources (e.g., \citealt{tramacere+2007a} and \citealt{furniss+2015}). Its location in frequency space can be estimated from the UV/optical and X-ray data, using a reasonable smooth interpolation or extrapolation model (e.g.,\,\citealt{massaro+2004,blazejowski+2005,tramacere+2009,ushio+2010}). For the analysis presented here, we use 30 pairs of \swiftxrt and {\em NuSTAR}\ spectra assembled from data taken simultaneously, together with UV data taken within the same \swift observation and R-band data taken within 24~hours. Since optical variability is significantly lower, especially on short timescales, the nonsimultaneity of the optical flux measurements is not a serious concern. We employ the two most commonly used methods from the literature to localize the synchrotron SED peak. In the top panel of Figure~\ref{fig:synchpeak} we show examples of both methods applied to two representative sets of data. The two methods are: i) fitting a log-parabolic model to the X-ray data alone, using \xspec model \texttt{logpar} described by Equation~(\ref{eqn:logpar}), and extrapolating to lower energies; ii) fitting a log-parabolic model to both optical/UV and X-ray data.

The X-ray-based extrapolation underestimates the UV/optical flux by more than an order of magnitude in nearly all cases. The peak frequencies predicted by this method uniformly cover the frequency range from $10^{16}$ to $10^{17}$~Hz and a factor of $\simeq3$ in peak flux. Interpretation of this simplistic parametrization of the SED would imply that it should consist of two superimposed components in order to match the observed UV/optical flux. However, log-parabolic fits which additionally include the R-band and UVOT fluxes provide a simpler solution that matches the data well on both sides of the SED peak. This is demonstrated by the two examples shown in Figure~\ref{fig:synchpeak}. Both methods are somewhat sensitive to the systematic uncertainties in the cross-normalization between the instruments, and to the exact values of the line-of-sight column density and extinction corrections. We conservatively estimate that the combination of these effects results in a factor of $\simeq$2 uncertainty in the synchrotron peak frequency ($\nu_{\rm syn.\,peak}$), which dominates any statistical uncertainty from the fits. For this reason we do not show the uncertainties for individual $\nu_{\rm syn.\,peak}$ estimates in Figure~\ref{fig:synchpeak}.

Both methods consistently show the peak at an atypically low frequency ($\nu_{\rm syn.\,peak} < 10^{17}$~Hz), with peak flux comparable to high-activity states (see lower panel of Figure~\ref{fig:synchpeak} and references listed there). The scatter is found to be larger for the fits using only the X-ray data, which can be easily understood since the curvature is subtle in all but the lowest-energy bins of the \swiftxrt band and the parabola has no constraint at energies below the peak. The optical/UV data provide the leverage to constrain the parabolic curves significantly better. We find an interesting trend using the second method: the flux at the SED peak is approximately proportional to the square root of the peak frequency. This is highlighted with a linear fit shown in the lower panel of Figure~\ref{fig:synchpeak}. The best-fit slope of the relation, $\log (\nu F_{\nu})_{\rm syn.\,peak} \propto b \log \left( \nu_{\rm syn.\,peak} \right)$, is $b=0.49\pm0.08\approx0.5$. The dynamic range over which the relation holds, assuming that this parametrization is valid at all, is unknown. It is clear from the comparison with the peak localization taken from the literature (listed in the lower panel of Figure~\ref{fig:synchpeak}) that the relation is not universal, although its slope is broadly consistent with the slopes of previously identified relations of the same kind. Given the fact that the simple method used to derive it is purely phenomenological and sensitive to systematic uncertainties, we refrain from quantifying and interpreting this correlation further.

For all observations presented here $\nu_{\rm syn.\,peak}<10^{17}$~Hz, essentially independent of the choice of model. If only a single peak is assumed to exist, our data unambigously imply that the synchrotron peak frequency was well below the \swiftxrt band ($<$0.3~keV) during this period of very low X-ray activity. This is atypical for Mrk\,421\ based on the previously published data. We found that, for the states of lowest X-ray activity, our data show a peak frequency as low as $\nu_{\rm syn.\,peak}\lesssim10^{16}$~Hz, estimated with a (symmetric) log-parabolic function fitted to both optical/UV and X-ray (\swiftxrt and {\em NuSTAR}) data. A peak frequency as low as $\sim10^{16}$~Hz has been reported in \citet{ushio+2010}, but only when using a specific model that led to a very asymmetric parameterization of the synchrotron bump. When using a symmetric function (comparable to a log-parabola), the fit to the same observational data reported in \citet{ushio+2010} led to a peak frequency of $\sim10^{17}$~Hz, which is typical for Mrk\,421.

Blazars are classified by the frequency of the peak of the synchrotron emission as LBLs, IBLs and HBLs (for low-, intermediate- and high-$\nu_{\rm syn.\,peak}$ BL\,Lacs) if $\nu_{\rm syn.\,peak}$ is below $10^{14}$~Hz, in the range $10^{14}-10^{15}$~Hz, or above $10^{15}$~Hz, respectively. The data presented here show that Mrk\,421, which is one of the archetypal TeV HBLs, with a synchrotron peak position well above that of the typical HBL, changed its broadband emission in such a way that it almost became an IBL. This effect may also happen to other HBLs that have not been as extensively observed as Mrk\,421, and suggests that the SED classification may denote a temporary characteristic of blazars, rather than a permanent one.

\begin{figure*} 
\begin{center}
\includegraphics[width=0.48\textwidth]{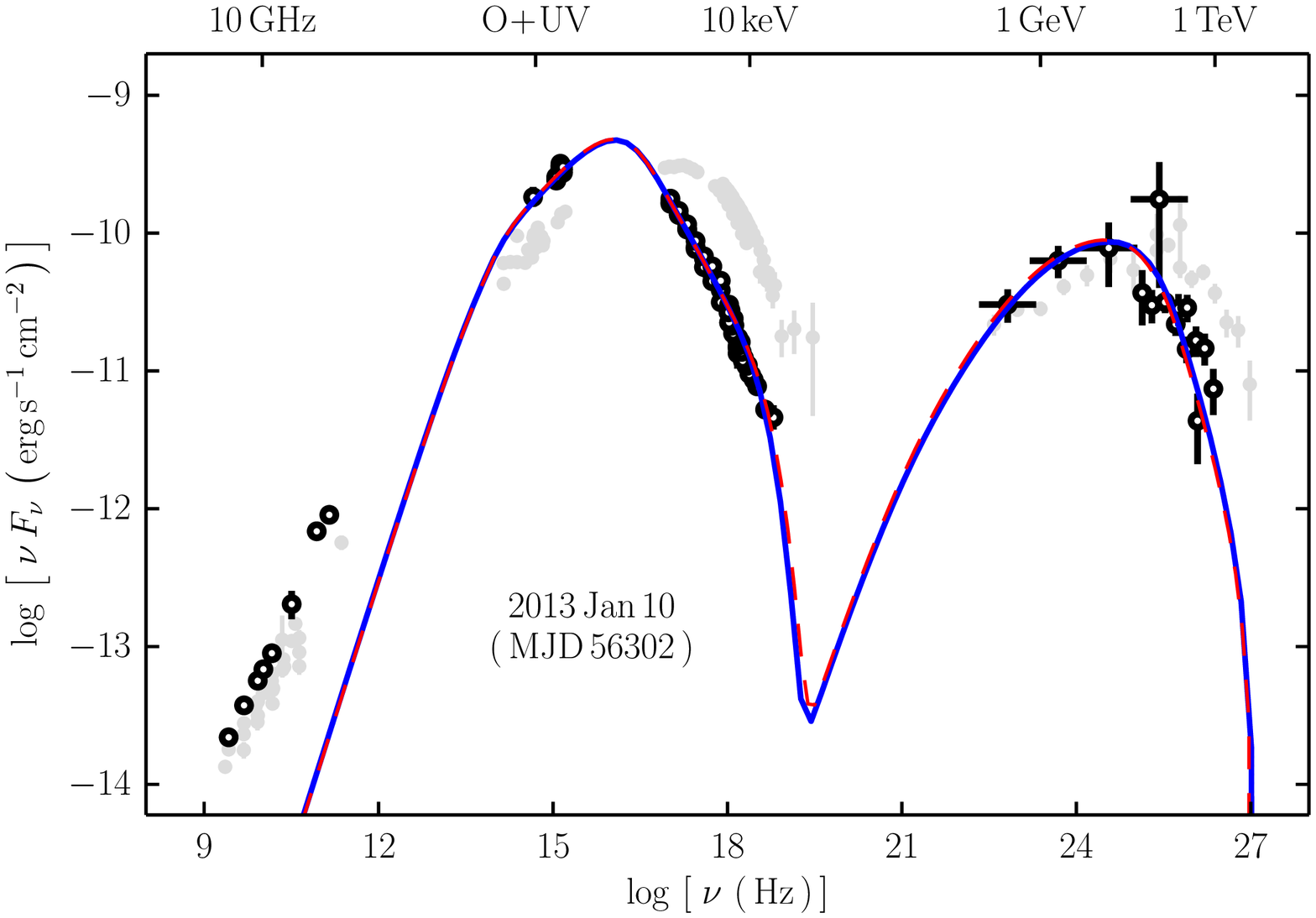}
\includegraphics[width=0.48\textwidth]{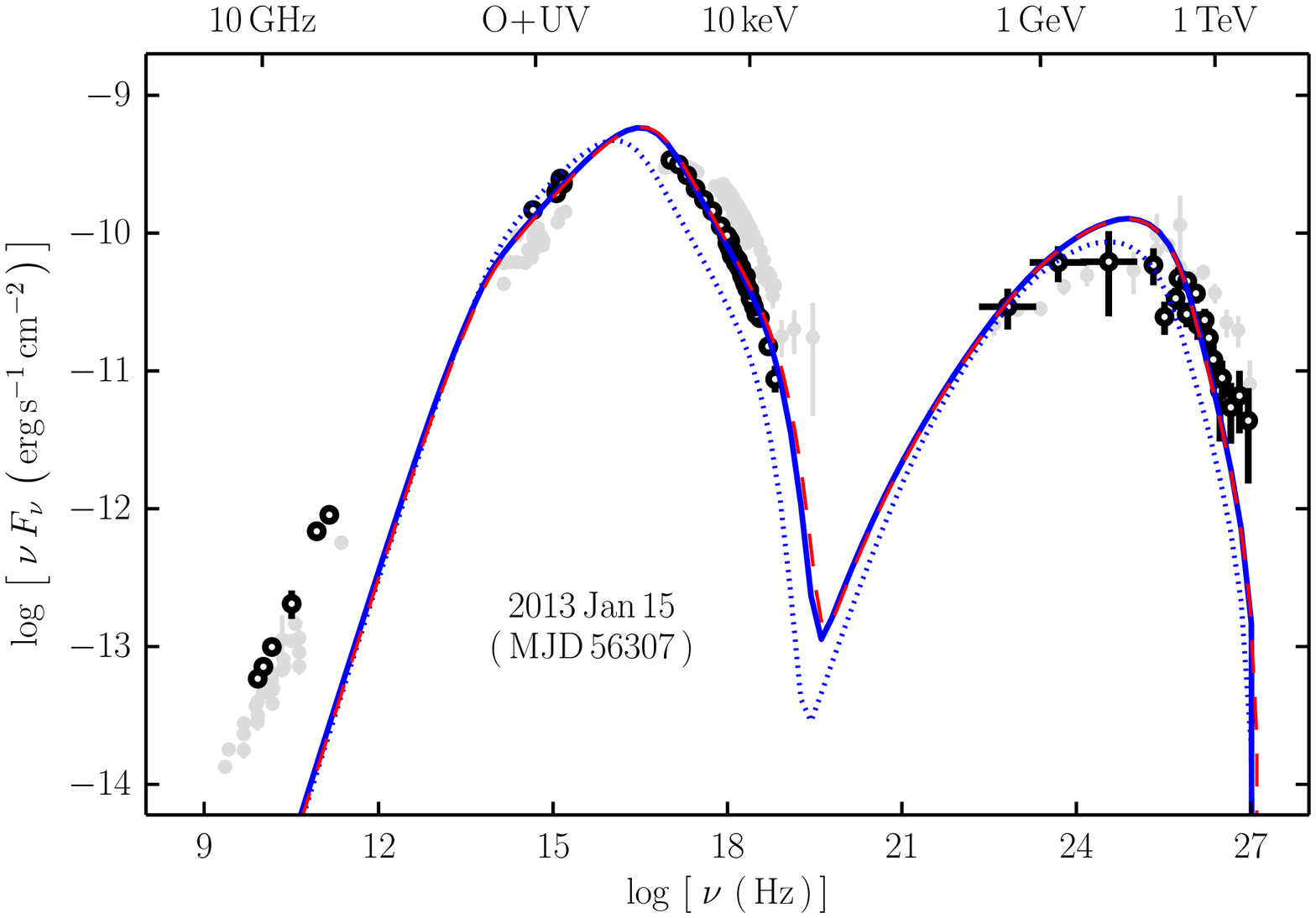}
\includegraphics[width=0.48\textwidth]{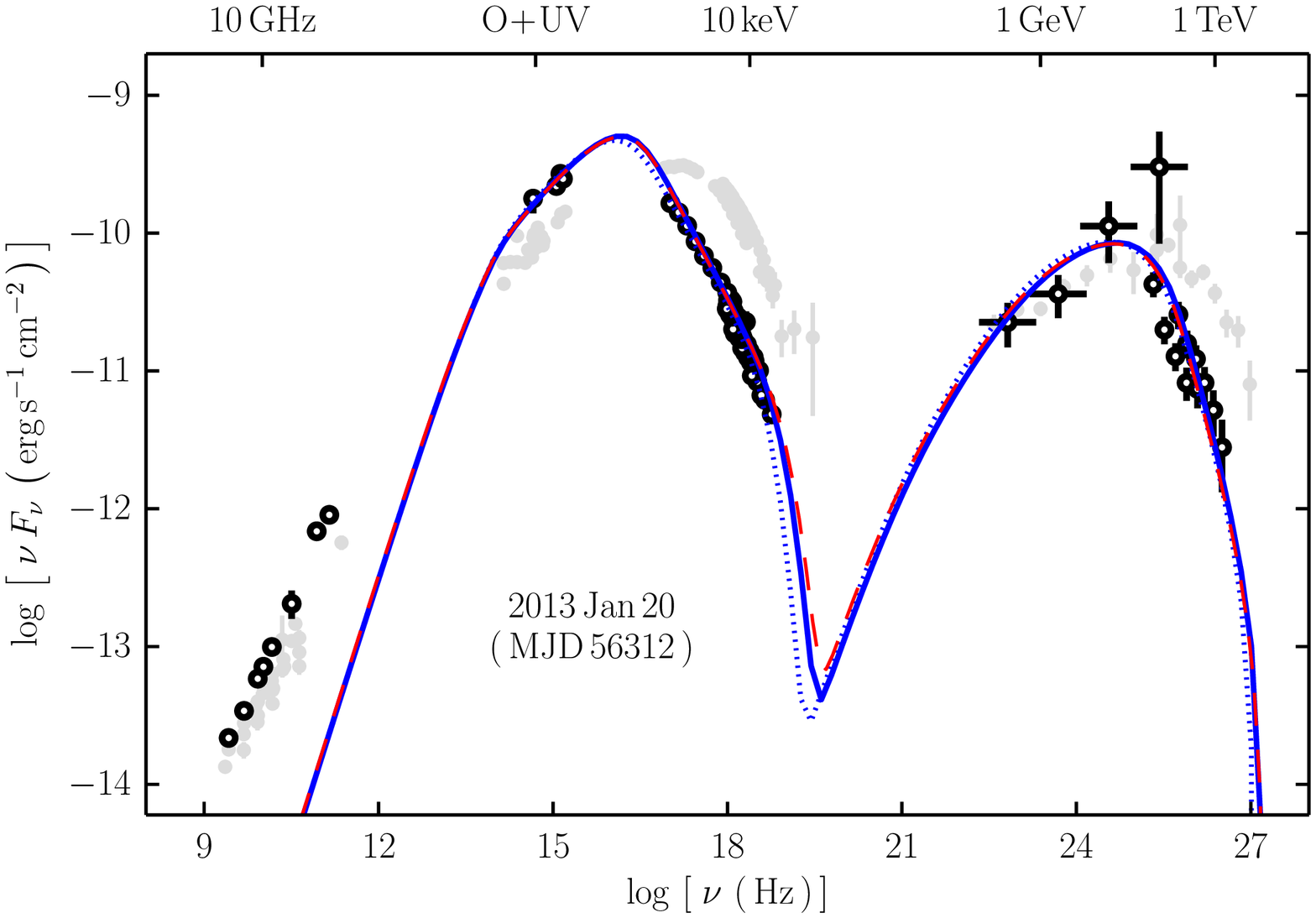}
\includegraphics[width=0.48\textwidth]{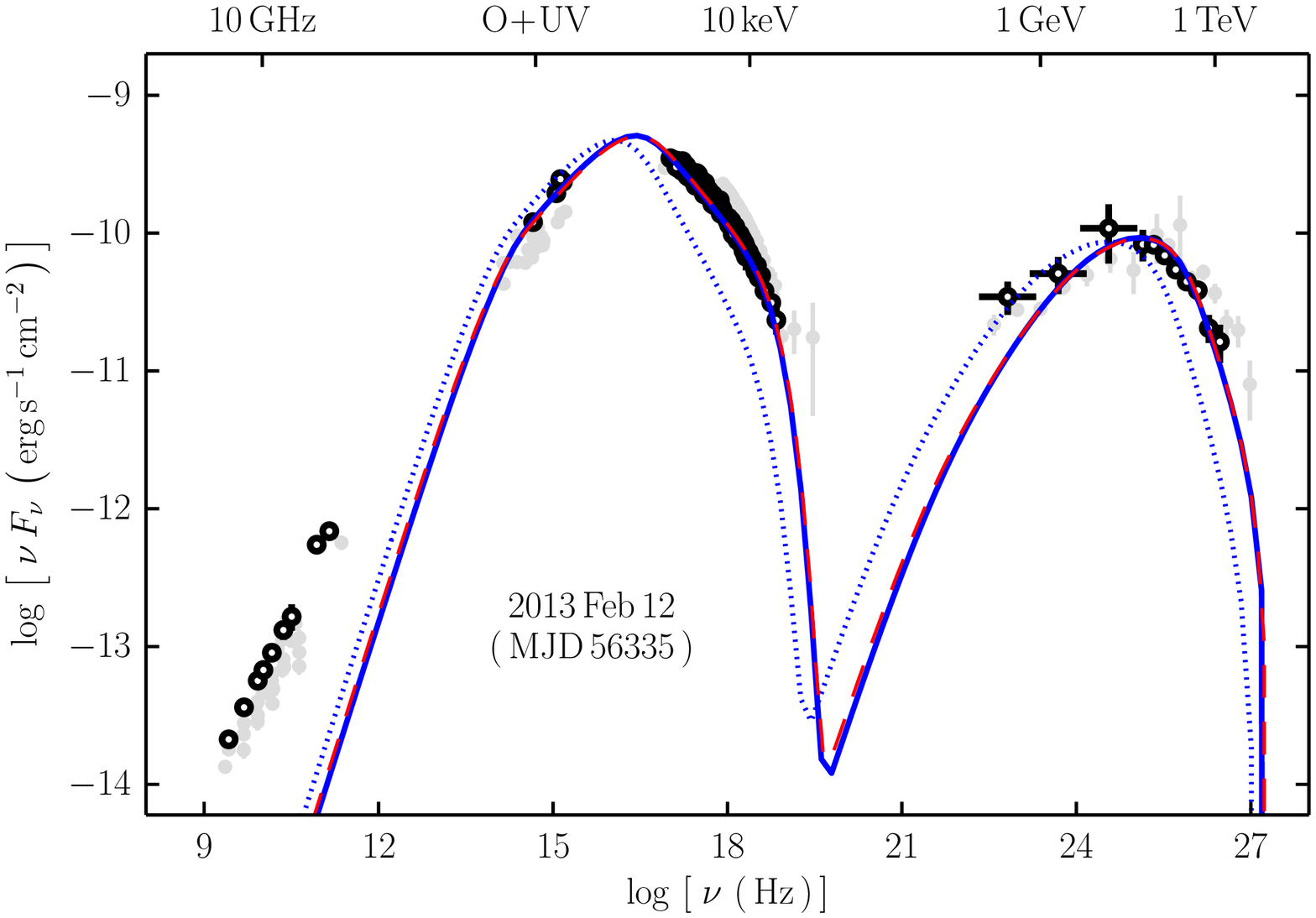}
\caption{ SED snapshots for four selected epochs during the campaign, assembled using simultaneous data from \swiftuvot$\!\!$, \swiftxrt$\!\!$, {\em NuSTAR}, {\em Fermi}\,-LAT, \magic$\!\!$, and \veritas$\!\!$. Most of the data were acquired over a period shorter than 12~hours in each case; the exceptions are the {\em Fermi}\,-LAT\ data and part of the radio data, which were accumulated over roughly one-week time intervals. The two left panels show low-state SEDs, while the two on the right show elevated states (not flaring, but among the highest presented in this paper). The grey symbols in the background of each panel show the SED of Mrk\,421\ from~\citet{abdo+2011} averaged over a quiescent 4.5-month period. The solid blue lines show a simple one-zone SSC model discussed in \S\,\ref{sec:discussion-ssc}. To aid comparison, the model curve from the first panel is reproduced in the other panels with a blue dotted line. The dashed red lines show SED models with a time-averaged electron distribution discussed in \S\,\ref{sec:discussion-ssc} for comparison with previously published results.}
\label{fig:seds}
\end{center}
\end{figure*} 

\subsection{Broadband SED at Different Epochs} 

\label{sec:results-sed}

For a better understanding of the empirically observed correlations, we need to consider the complete broadband spectrum. SED snapshots for four selected epochs (marked with dashed vertical lines in Figure~\ref{fig:lc-big}) are shown in Figure~\ref{fig:seds}. They were selected to show a state of exceptionally low X-ray and VHE flux (January~10 and~20; see \S\,\ref{sec:results-correlations-xgammaray}), in contrast to higher, though not flaring, states (January~15, February~12, for example). In all SED plots we also show data accumulated over 4.5 months in the 2009 multiwavelength campaign \citep{abdo+2011}, which is currently the best-characterized quiescent broadband SED available for Mrk\,421\ in the literature. For the two epochs of very low X-ray flux, we show for the first time states in which both the synchrotron and inverse-Compton SED peaks are shifted to {\em lower} energies by almost an order of magnitude compared to the typical quiescent SED. The accessibility of the low-activity state shows the large scientific potential brought by the improvements in the X-ray and VHE instrumentation in the last several years with the launch of {\em NuSTAR}\ and upgrades to the \magic and \veritas telescopes.

We note that the empirical SEDs from the 2013 campaign shown here represent $\lesssim$12 hours of observation in the X-ray and VHE bands, rather than integrations over a time period of weeks or even months. We match the simultaneous UV, X-ray and VHE data to optical data taken within at most 2~days, radio data taken within at most 2~weeks, and {\em Fermi} data integrated over time intervals of 6--10~days centered on the time of the coordinated X-ray and VHE observations. Mrk\,421\ is a point-like and unresolved source for single-dish radio instruments, which means that the data shown in Figure~\ref{fig:seds} include emission from spatial scales larger than the jet itself and therefore should be considered as upper limits for the SSC models of jet emission. We further discuss the SED in the context of the SSC model in \S\,\ref{sec:discussion-ssc} and \S\,\ref{sec:discussion-ssc2}.

\subsection{Brief Summary of the Flaring Activity in 2012} 

\label{sec:flare2012}

\begin{figure} 
\begin{center}
\includegraphics[width=\columnwidth]{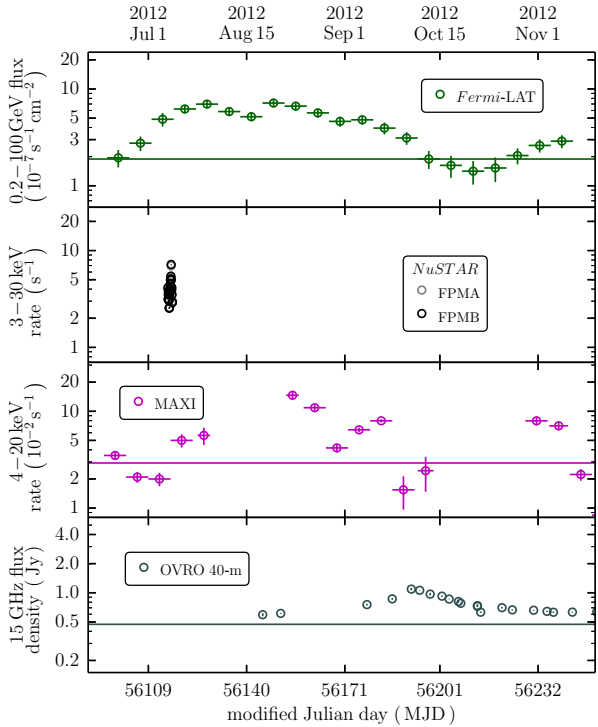}
\caption{ Light curves for Mrk\,421\ around the $\gamma$-ray flare detected by {\em Fermi}\,-LAT\ in 2012, from {\em Fermi}\,-LAT\ (0.2--100~GeV, binned weekly), {\em NuSTAR}\ (3--30~keV, binned by orbit; see the top panel of Figure~\ref{fig:nustar_count_rates} for greater time resolution), MAXI (4--20~keV, weekly bins) and OVRO (15~GHz, $\sim$weekly--daily cadence). Vertical and horizontal error bars show statistical uncertainties and the bin width, respectively, although some of the error bars are too small to be visible in this plot. The colored horizontal lines show the long-term median flux calculated from publicly available monitoring data. The dynamic range in all panels is 40, as in Figure~\ref{fig:lc-big}, so that the two figures are directly comparable. }
\label{fig:lc-2012}
\end{center}
\end{figure} 

In addition to the coordinated multiwavelength campaign conducted in 2013, Mrk\,421\ was observed independently with several instruments in July--September 2012, including {\em NuSTAR}, {\em Fermi} and OVRO. In July~2012, the flux in the {\em Fermi}\,-LAT\ band increased above the median level and peaked twice over the following two months (see Figure~\ref{fig:lc-2012}). The first peak was reported on July~16 by the {\em Fermi}\,-LAT\ collaboration~\citep{fermi-flare-2012} and, within the same day, by the ARGO-YBJ collaboration~\citep{argo-ybj-flare-2012}. The daily flux seen by {\em Fermi}\,-LAT\ increased to $(1.4\pm0.2)\times10^{-6}$~s$^{-1}$cm$^{-2}$, a factor of $\simeq$8 above the average flux reported in the second {\em Fermi}\,-LAT\ catalog (2FGL; \citealt{nolan+2012}). Light curves from several observatories monitoring Mrk\,421\ in July--September 2012 are shown in Figure~\ref{fig:lc-2012} in order to provide a timeline for this flaring event.

An observation of Mrk\,421\ was performed by {\em NuSTAR}\ on 2012 July 7 and 8, shortly before the start of flaring activity in the $\gamma$-ray band. The observation\footnote{The observation consists of two contiguous segments, sequence IDs 10002015001 and 10002016001; see Table~\ref{tab:observations-nustar}.} was not originally intended for scientific usage, as the pointing was suboptimal at this early point in the mission (less than a month after launch). However, it represents both the longest and the most variable {\em NuSTAR}\ observation considered in this paper (see Figure~\ref{fig:nustar_count_rates}), and thus represents an important part of the {\em NuSTAR}\ data presented here. The available X-ray and $\gamma$-ray data are clearly too sparse to allow for associations to be inferred between any specific features in the light curves. There is indication from the MAXI public monitoring data\footnote{\url{http://maxi.riken.jp/top/}} that the X-ray flux in the 4--20~keV band increased further after the {\em NuSTAR}\ observation, peaking between 2 and 5 weeks later (see Figure~\ref{fig:lc-2012}). The {\em NuSTAR}\ observation therefore makes it possible to investigate the hard X-ray spectrum of Mrk\,421\ at the time when its $\gamma$-ray activity was rapidly increasing.

A unique feature of this event is the well-defined rapid radio flare observed at 15~GHz from OVRO \citep{ovro-flare-2012,hovatta+2015}. Approximately 70~days after the first peak of the $\gamma$-ray flare Mrk\,421\ reached a flux density of $(1.11\pm0.03)$~Jy, approximately 2.5 times its median flux density. Note that in Figure~\ref{fig:lc-2012} we show the light curves on a logarithmic scale with a fixed dynamic range in order to facilitate direct comparison with Figure~\ref{fig:lc-big}; on a linear plot both the $\gamma$-ray and the radio flare appear strikingly peaked. Based on the statistical properties of {\em Fermi}\,-LAT\ and OVRO 15~GHz light curves, \citet{maxMoerbeck+2014} have shown that the $\gamma$-ray and the radio flare are likely causally related. During most of the flaring activity in 2012 Mrk\,421\ was very close to the Sun on the sky, which resulted in relatively poor multiwavelength coverage. We therefore do not attempt a more comprehensive analysis of the sparse data available for this epoch. \citet{hovatta+2015} present the radio and $\gamma$-ray data, as well as a physical model for the flaring activity, both for the 2012 flare and the flare observed during the multiwavelength campaign in April~2013. As the latter event has been covered with numerous instruments (e.g., \citealt{april-flare-atel1,april-flare-atel2,april-flare-atel3,pian+2014}), we defer a comprehensive analysis of this period to a separate publication.

\section{Discussion} 

\label{sec:discussion}

\subsection{Spectral Variability\\in the X-ray Band} 

\label{sec:discussion-xray}

The good temporal coverage of the {\em NuSTAR}\ data reveals a typical variability timescale of $\tau_{\rm var}\approx9\pm3$~hours. Significant variability is clearly detectable even in the low-flux states, which is the case for several epochs in early 2013. We find no evidence for strong intrahour variability; on timescales as short as $\sim$10 minutes the variability amplitude is $\lesssim$5\%, approximately an order of magnitude lower than the typical flux change over a 10-hour observation. This can be inferred from the fact that in the light curves shown in Figure~\ref{fig:nustar_count_rates} only a small fraction of adjacent bins differ in count rate by more than 3\,$\sigma$. We summarize this more formally in Figure~\ref{fig:nustar_var}. In contrast to some previous studies (e.g.,~\citealt{takahashi+1996}), we observed no clear time-dependent circular patterns in the count rate--hardness plane. The reason for this may be that most of the observations seem to have covered periods of decreasing flux, with no well-defined flare-like events except for a few ``mini-flares'' of modest $\lesssim40$\% amplitude.

We find that the X-ray spectrum of Mrk\,421\ cannot generally be described as a simple power law, but that instead it gradually steepens between 0.3 and $\sim$70~keV. For most of the \swift and {\em NuSTAR}\ observations in the 2013 January--March period, we find that the spectra in both bands are better described when a curvature term is added to the basic power law, as in the log-parabolic model available in \xspec$\!$. Using this model, we find significant curvature at the highest observed fluxes---still notably lower than in any flaring states---gradually vanishing as flux decreases (see Figure~\ref{fig:nustar_flux_trends}). This has a simple explanation because the X-ray band samples the Mrk\,421\ SED close to the synchrotron peak: when the X-ray flux is low, and the SED peak shifts to lower energy (away from the {\em NuSTAR}\ band), the hard X-ray spectra can be well described by a power law. This behavior is consistent with the steady increase of fractional variability with energy through the X-ray band, as shown in Figure~\ref{fig:fvar}. The high sensitivity of {\em NuSTAR}\ reveals that the hard X-ray spectrum does not exhibit an exponential cutoff, and it is well described by a power law with a photon index $\Gamma\approx3$,  even during the epochs related to the lowest X-ray fluxes. The {\em NuSTAR}\ data also show no signature of spectral hardening up to $\sim80$~keV, meaning that the onset of the inverse-Compton bump must be at even higher energies.

\subsection{Correlated Variability in the X-ray\\and VHE Spectral Bands} 

\label{sec:discussion-XrayVHE}

The data gathered in the 2013 multiwavelength campaign contribute some
unique details to the rich library of blazar phenomena revealed by
Mrk\,421. The object is highly variable on a wide range of timescales
and fluxes, with the fractional variability amplitude highest at the
high-energy ends of the synchrotron and inverse-Compton SED bumps (see
Figure~\ref{fig:fvar}). The well-matched coverage in the X-ray and VHE
bands reveals that the steep spectral slope observed in the X-ray band
at very low flux occurs simultaneously in time with an atypically
steep slope observed in the VHE band. For the first time we observed a simultaneous shift of both the synchrotron and the inverse-Compton SED peaks to lower energies in comparison to the typical quiescent state (see Figure~\ref{fig:seds}), constrained primarily by the X-ray and VHE data. The measurements in those bands do not support the existence of high-energy cutoffs up to $\simeq80$~keV and $\simeq1$~TeV. All of this indicates that the energies of radiating particles must be very high (up to $\gamma\sim10^6$; see \S\,\ref{sec:discussion-ssc} below) even when the source is in such a low state.

In \S~\ref{sec:results-correlations-xgammaray} we have shown that the X-ray and VHE fluxes are correlated at $>$3\,$\sigma$ significance. Parametrizing the correlation as $\log \left( F_{\rm X-ray} \right) = a \log \left( F_{\rm VHE} \right) + b$, the correlation is found to be approximately linear ($a\approx$1) both on half-hour and half-day timescales. This is consistent with most previous results considering similar spectral bands: $a=1.7\pm0.3$ \citep{tanihata+2004}, $a\approx1$ (\citealt{fossati+2008}, for averaged and nonflaring periods; also \citealt{aleksic+2015b}), $a=1$ (\citealt{acciari+2014}; assumed linear) etc. We emphasize the importance of distinguishing between i) a correlation of count rates versus a correlation of fluxes, since the conversion between them is nonlinear due to spectral variability, and ii) a general correlation versus a correlation associated with isolated flares, since those could potentially be produced by different physical mechanisms (as argued by, e.g.,\,\citealt{katarzynski+2005}). Indeed, for isolated flaring periods \citet{fossati+2008} and \citet{giebels+2007} find $a\approx2$ and $a=2.9\pm0.6$, respectively. Care should be taken with direct comparison of the results in the literature, since the chosen spectral bands are not always the same, and we have shown in \S\,\ref{sec:results-correlations-xgammaray} that the slope does depend on the band choice as a consequence of the spectral variability.

We note that in the simplest, one-zone SSC model, one expects a close correlation between the X-ray and VHE fluxes. However, if the scattering takes place exclusively in the Thomson regime, the inverse-Compton flux should obey a quadratic ($a=2$) relationship, since increasing both the number of electrons and the seed photon flux results in a quadratic increase in the scattering rate. Since we detect a linear relationship, this would argue that the scattering cross-section is diminished, possibly because the scattering takes place in the less-efficient Klein-Nishina regime. For example, a quadratic relation has recently been observed in a similar HBL object, Mrk\,501 \citep{furniss+2015}. It has been shown previously (e.g.,~\citealt{katarzynski+2005}), that this implication is valid only if the normalization of the entire electron distribution is changed to produce flux variations. For changes in other parameters of the electron distribution, or in physical conditions within the emission region, this is no longer strictly correct. The linearity of the flux--flux correlation itself does not uniquely indicate Klein-Nishina effects; we therefore combine the broadband SED modeling and variability properties in the following section, in order to further investigate this issue.

\begin{deluxetable}{ccccc} 

\tablecolumns{5}
\tablewidth{0pc}
\tablecaption{Model parameters for the equilibrium SSC model for four selected epochs \label{tab:sed-parameters} }
\scriptsize

\tablehead{
  \colhead{Parameter} &
  \colhead{Jan 10} &
  \colhead{Jan 15} &
  \colhead{Jan 20} &
  \colhead{Feb 12}
}

\startdata

$\elgamma_{\rm min}$ ( $10^4$ ) & 2.2 & 3.5 & 2.4 & 5.0 \\
$\elgamma_{\rm max}$ ( $10^5$ ) & 4.0 & 4.8 & 5.8 & 6.8 \\
$q$ & 3.5 & 3.5 & 3.6 & 3.0 \\
$\eta$ & 35 & 35 & 35 & 35 \\
$B$ ( G ) & 0.17 & 0.25 & 0.16 & 0.10 \\
$\Gamma$ & 25 & 25 & 25 & 25 \\
$R$ ( $10^{16}$ cm ) & 0.9 & 0.55 & 1.0 & 1.4 \\
$\theta$ ( $^{\circ}$ ) & 2.29 & 2.29 & 2.29 & 2.29 \\
$\tau$ ( h ) & 3.4 & 2.1 & 3.8 & 5.3 \\
$L_e$ ( $10^{43}$ erg~s$^{-1}$ ) & 3.0 & 2.6 & 3.0 & 3.3 \\
$\epsilon = L_B/L_e$ & 0.18 & 0.25 & 0.20 & 0.14 \\

\enddata

\tablecomments{The electron energy distribution parameters listed here refer to the electrons injected into the emission region, and the equilibrium distribution is calculated self-consistently within the model, as described in \S\,\ref{sec:discussion-ssc}. Model SED curves are shown in Figure~\ref{fig:seds}.}

\end{deluxetable} 

\subsection{Interpretation within the Framework\\of a Single-zone SSC Model} 

\label{sec:discussion-ssc}

In the framework of an SSC model, if the peak energies of the
synchrotron and inverse-Compton SED are resolved, then along with
constraints from temporal variability and an estimate of the bulk
Doppler factor of the emitting material, fairly general and robust
estimates can be made of the characteristic particle energies, the
magnetic field strength, and the overall size of the emitting region.
An estimate for the characteristic electron Lorentz factor ($\elgamma_c$, measured in the
comoving frame of the emitting plasma) is given roughly by the square
root of the ratio of the energies of the synchrotron and inverse-Compton peaks.
The radiation at the peaks is dominated by electrons of intermediate
energy, where the Klein-Nishina reduction in the scattering cross-section is not expected to be significant.
From the SEDs shown in Figure~\ref{fig:seds}, we estimate
$\nu_{\rm syn.\,peak} \simeq 10^{16}$\,Hz and $\nu_{\rm IC\,peak} \simeq 10^{25}$\,Hz.
It then follows that $\elgamma_c \sim (\nu_{\rm IC\,peak}/\nu_{\rm syn.\,peak})^{1/2} \sim
3\times 10^4$. Assuming a bulk Doppler factor of $\delta \sim 25$, we
can also estimate the magnetic field strength in the plasma comoving frame:
$B = 4\pi \nu_{\rm syn.\,peak} m_e c/(3e\delta\elgamma_c^2)\simeq0.1$~G.
Finally, an upper limit for the size of the emitting
region in the comoving frame is given by the observed variability time
scale, $\tau_{\rm var}\approx9$\,h, and the bulk Doppler factor $\delta$:
$R \lesssim c\tau_{\rm var}\delta \simeq 2\times 10^{16}$~cm.

We can constrain the properties of the jet emission region more
precisely by directly modeling the multiwavelength SED data shown in
Figure~\ref{fig:seds} with a standard one-zone SSC model.
Specifically, we apply an equilibrium version of the SSC model from \cite{bottcher+chiang-2002}.  This model has
already been used to represent Mrk\,421\ in two different states
\citep{acciari+2009}.  In this model, the emission originates from a
spherical region with radius $R$, containing relativistic electrons which
propagate down the jet with a bulk Lorentz factor $\Gamma$.
In order to decrease the number of free parameters, we assume a 
value $\Gamma = 25$ with the jet axis aligned near the line of sight with the critical angle 
$\theta=1/\Gamma=0.04$~rad~$=2.29^{\circ}$, which makes the Doppler factor equal to the jet Lorentz factor ($\delta=\Gamma$). This simplifying choice is often used in the literature when direct measurements are not available (see, e.g., \citealt{abdo+2011} and the discussion therein). A Doppler factor of 25 is higher than the value inferred from VLBA measurements of the blob movement by \citet{Piner10}. This is a common situation in VHE blazars, often referred to as ``bulk Lorentz factor crisis'', and requires that the radio and VHE emission are produced in regions with different Lorentz factors \citep{2003ApJ...594L..27G,2005A&A...432..401G,2006ApJ...640..185H}.
High Doppler factors ($\gtrsim$10) are required to explain previously reported rapid variations in the VHE band \citep{gaidos+1996,celotti+1998,galante+2011}, and are typically used in theoretical scenarios to describe the broadband emission of VHE blazars.
Relativistic leptons are injected according to a power-law distribution $dn/d\gamma \propto
\gamma^{-q}$ between $\gamma_{\rm min}$ and $\gamma_{\rm max}$.  These
particles lose energy through synchrotron and inverse-Compton
radiation, leading to an equilibrium between particle injection,
radiative cooling and particle escape.  The particle escape is
characterized with an escape efficiency factor $\eta$, defined so that
$\tau_{\rm esc}=\eta R/c$ is the escape time.  This results in a particle
distribution which propagates along the jet with power $L_e$.
Synchrotron emission results from the interaction of particles with a
magnetic field $B$, generating a Poynting-flux luminosity of $L_B$.
$L_e$ and $L_B$ allow the calculation of the equipartition parameter
$L_B/L_e$.  Various other blazars have been represented with this
model, with the resulting model parameters summarized in
\cite{aliu+2013}. In application to the broadband data, the intrinsic source VHE flux from the SSC model is absorbed by the \citet{franceschini+2008} model describing the extragalactic photon field. In Table~\ref{tab:sed-parameters}, we list the
relevant model parameters that reproduce the observed SED of Mrk\,421\ for
the four selected epochs in 2013. 

Since the injected particle distribution in our SSC model follows a single
power law, the observed spectral shapes in the GeV and VHE bands imply
certain constraints on the model parameters. In the 0.1--100~GeV
band, the observed spectra have photon indices in the range $\Gamma \sim
1.6-1.7$, while, by contrast, the VHE spectra have photon indices of
$\Gamma \sim 2.3-3.5$ (see Tables~\ref{tab:observations-magic}
and~\ref{tab:observations-veritas}). These indices imply spectral
breaks of $\Delta\Gamma \sim$0.6--1.9, which are moderately to
significantly larger than the ``cooling'' break of $\Gamma = 0.5$ that
arises from incomplete (or ``weak'') synchrotron cooling of an
injected power-law distribution of electrons.  In the strong-cooling
regime, i.e., where the synchrotron cooling timescale is shorter than
the particle-escape time, the cooled electron distribution has a break
at the lower bound of the injected power law, $\elgamma_b =
\elgamma_{\rm min}$, and has power-law shapes $dn/d\elgamma \propto
\elgamma^{-2}$ for $\elgamma < \elgamma_b$ and $dn/d\elgamma \propto
\elgamma^{-(q+1)}$ for $\elgamma > \elgamma_b$.  For the parameters
shown in Table~\ref{tab:sed-parameters}, this particle distribution
implies a synchrotron spectrum with a peak at $\nu_{\rm syn.\,peak} \simeq
1$--$2\times 10^{16}$~Hz, and spectral shapes $F_\nu \propto
\nu^{-1/2}$ ($dN/dE \propto E^{-1.5}$) for $\nu < \nu_{\rm syn.\,peak}$ and
$F_\nu \propto \nu^{-q/2}$ ($dN/dE \propto E^{-(2.6-3.2)}$) for $\nu >
\nu_{\rm syn.\,peak}$ (cf. Table~\ref{tab:full_obs_models} and
Figure~\ref{fig:seds}). For $\elgamma_{\rm min} = 3\times 10^4$ and
$B = 0.2$~G, a synchrotron cooling timescale of $\tau_{\rm syn} =
4\times10^5$~s is obtained in the comoving frame of the emitting
plasma; this is slightly larger than the nominal (i.e., in the absence
of any scattering) escape time of $\tau_{\rm esc,\,nom} = R/c =
3\times10^5$~s. The escape efficiency factor, $\eta = 35$, ensures that the cooled
electron distribution extends to sufficiently low energies to model
both the optical/UV points, and the {\em Fermi}\,-LAT\ data down to 0.1~GeV.
The Larmor radius of the lowest-energy electrons in the modeled
magnetic field is small enough that the electrons have sufficient time
to cool within the emission region before escaping.

Past SED modeling of HBL-type blazars has often used SSC calculations
that have electron distributions which are assumed to persist in the
specified state for the entire duration of the observation.  For
example, for a given variability timescale, a single, time-averaged,
multiply broken power-law electron distribution is used by
\cite{abdo+2011} to model the multiwavelength data obtained for Mrk\,421\
over a 4.5-month period in early 2009.  By contrast, the SED
calculations we have performed in this paper attempt to model specific
flaring or quiescent periods for which most of the data (optical,
X-ray, and VHE) were obtained within 12-hour intervals.  Our modeling
assumes that an initial power-law electron spectrum is injected into
the emission region, and we compute the resulting quasi-equilibrium
particle distribution for those epochs given the radiative and
particle escape timescales.  Since the 2009 observations could
contain a large number of similarly short flaring and quiescent
episodes with a range of physical properties, it would be
inappropriate to attempt to model those data with the procedure we
have used here.  However, as we have indicated above, it should be
possible to obtain equivalent time-averaged SED models that have
multiply broken power-law electron distributions and which we can
compare directly to the \cite{abdo+2011} results.

\newcommand{\elgammata}{{\langle\elgamma\rangle}}

\begin{deluxetable}{ccccc} 

\tablecolumns{5}
\tablewidth{0pc}
\tablecaption{Model parameters for the snapshot SSC model for four selected epochs \label{tab:sed-snapshot-parameters} }
\scriptsize

\tablehead{
  \colhead{Parameter} &
  \colhead{Jan 10} &
  \colhead{Jan 15} &
  \colhead{Jan 20} &
  \colhead{Feb 12}
}

\startdata

$\elgamma_{\rm min}$ ( $10^3$ ) & 2.0 & 1.3 & 1.7 & 3.0 \\
$\elgamma_{\rm brk}$ ( $10^4$ ) & 2.5 & 3.7 & 3.0 & 5.2 \\
$\elgamma_{\rm max}$ ( $10^5$ ) & 4.0 & 4.8 & 5.8 & 6.8 \\
$p_l$ & 2.0 & 2.0 & 2.0 & 2.0 \\
$p_h$ & 4.5 & 4.5 & 4.6 & 4.0  \\
$B$ ( G ) & 0.21 & 0.28 & 0.19 & 0.10 \\
$\Gamma$ & 25 & 25 & 25 & 25 \\
$R$ ( $10^{16}$ cm ) & 0.93 & 0.60 & 1.04 & 1.69 \\
$\theta$ ( $^{\circ}$ ) & 2.29 & 2.29 & 2.29 & 2.29 \\
$\tau$ ( h ) & 3.4 & 2.2 & 3.9 & 6.3 \\
$L_e$ ( $10^{43}$ erg~s$^{-1}$ ) & 3.3 & 2.9 & 3.3 & 4.1 \\
$\epsilon=L_B/L_e$ & 0.51 & 0.47 & 0.56 & 0.33 \\

\enddata

\tablecomments{The electron energy distribution parameters listed here refer to the distribution directly responsible for the SSC emission. This simplified model is described in \S\,\ref{sec:discussion-ssc} and used for comparison with the literature. Model SED curves are shown in Figure~\ref{fig:seds}.}

\end{deluxetable} 

We have performed such modeling, and in
Table~\ref{tab:sed-snapshot-parameters}, we give the parameters for
the same four selected epochs appearing in
Table~\ref{tab:sed-parameters}. The SEDs produced by the two models
can be matched very well, as shown in Figure~\ref{fig:seds} with the
blue and dashed red lines. As we noted above, the equivalent
time-averaged electron distributions can be represented via a broken
power law with a break at $\elgammata_{\rm brk} \simeq \elgamma_{\rm min}$
and index $p_l=2.0$ below $\elgammata_{\rm brk}$ and index
$p_h=q+1$ above the break.

As \cite{abdo+2011} demonstrate and as we discuss above, for data that
resolve the shapes of both the synchrotron and self-Compton components
of the SED, the model parameters in these sorts of leptonic
time-averged models are largely determined once either the variability
timescale or the Doppler factor is constrained or set. Therefore,
when comparing the current model parameters to those of
\citet{abdo+2011}, we consider just their $\delta=21$ results. In
terms of the shape of the underlying particle distributions, the value
of $p_l=2.0$ we find is comparable to their value of $p_1=2.2$,
and our values of $p_h = 4.0$--$4.6$ are similar to their high-energy
index of $p_3=4.7$. While this does not uniquely imply
that the same energy loss mechanisms and acceleration processes are at work
in both cases, the consistency is encouraging.
The \citet{abdo+2011} modeling does require an additional
medium-energy power-law component which is dictated by their generally
broader SED peaks (see the grey points in Figure~\ref{fig:seds}). In the
context of the quasi-equilibrium modeling, this would arise from a
distribution of physical parameters in shorter flaring and
quiescent episodes that are averaged over the 4.5-month observation time.

Several physical parameters in the current modeling do differ
substantially from those of \cite{abdo+2011}. The characteristic
electron Lorentz factors are about an order of magnitude lower
($\elgammata_{\rm brk} = 2.5$--$5.2 \times 10^4$ versus
$\gamma_{\rm brk2} = 3.9 \times 10^5$), while the inferred emitting region radius
is about a factor of 3--10 smaller ($R = 0.6$--$1.7\times 10^{16}$~cm
versus $5.2\times 10^{16}$~cm), the inferred magnetic field is
substantially higher ($B = 0.10$--$0.28$ versus $0.038$), and the
resulting jet powers differ by a factor of 3--4. As a consequence,
the current modeling yields equipartition parameters that are much
closer to unity, in the range $\epsilon = 0.33$--$0.56$ versus
$\epsilon = 0.1$ for the \citet{abdo+2011} result. Given the overall
similarity in the size and shape of the synchrotron and SSC components
among all five datasets (i.e., the four 2013 epochs and the 2009 data
shown in Figure~\ref{fig:seds}), the differences in model parameters can
be understood as being driven mostly by the combination of the
order-of-magnitude larger characteristic electron Lorentz factor and
the order-of-magnitude higher peak synchrotron frequency required by
the \citet{abdo+2011} data and modeling. Since $\nu_{\rm syn} \propto
B\gamma^2$, the order-of-magnitude higher magnetic fields in the
current modeling are readily understood, and those in turn largely
account for the equipartition parameters being substantially closer to
unity. We note that other authors, such as \citet{aleksic+2015c}, have
inferred SSC model parameters that are below equipartition by much
more than an order-of-magnitude. However, \citet{aleksic+2015c}
considered a flaring state of Mrk\,421\ that had much higher synchroton
peak frequencies, as well as substanially higher fluxes at all
wavelengths. Accordingly, their much larger disparity in the
partitioning of the jet power compared to the current results is not
surprising and roughly fits in with the preceding discussion.

Studying broadband emission of Mrk\,421\ at different epochs, \citet{mankuzhiyil+2011}
found that there were no substantial shifts in the location of the peaks
of the synchrotron and the inverse-Compton bumps. They concluded that the variability
in the blazar emission was dominated by changes in the parameters related
to the environment, namely, the emission-region size, the Lorentz (Doppler) factor,
and the magnetic field. The observational results presented here, with
substantially broader energy coverage and better instrumental sensitivity
due to the advent of new $\gamma$-ray and X-ray instruments, differ from those
presented in \citet{mankuzhiyil+2011}. We show that, besides changes in the
magnetic field, the distortions in the broadband emission of Mrk\,421\ also require
changes in the electron energy distribution, which may be due to
variations in the mechanism accelerating the electrons to high energies.

Having modeled the broadband SEDs with
single-zone SSC calculations, we can test the hypothesis that the
VHE emission occurs in the Klein-Nishina regime.
The SED modeling yields injected electron Lorentz factors in the
range $\sim 3\times 10^4$ to $\sim 6 \times 10^{5}$. Assuming that
the target synchrotron photons for inverse-Compton scattering have energies
around the synchrotron peak at $\nu_{\rm syn.\,peak} \sim 10^{16}$~Hz, 
the parameter governing the transition between Thomson and Klein-Nishina regimes is
$4h\nu_{\rm syn.\,peak}\elgamma/ m_e c^2 $ \citep{blumenthal+gould-1970}, 
which in the observer frame becomes $4h\nu_{\rm syn.\,peak}\elgamma/ \delta m_e c^2 $.

When considering photons from the synchrotron peak position ($E=h\nu_{\rm syn.\,peak}\sim$40~eV;
i.e., about one order of magnitude lower than the typical position of the 
synchrotron peak in Mrk\,421), we obtain $4h\nu_{\rm syn.\,peak}\elgamma/\delta m_e c^2 \simeq 0.4-8$, 
indicating that the inverse-Compton scattering of photons with energy $h\nu_{\rm syn.\,peak}$ 
takes place, at least partially, in the Klein-Nishina regime.
The X-ray energies probed with \swiftxrt are roughly one order of magnitude above
$h\nu_{\rm syn.\,peak}$, far above the range where Thomson scattering is relevant,
and consistent with the linear ($a\simeq1$) relationship between the soft X-ray and VHE flux.

\subsection{Toward a multi-zone emission scenario } 

\label{sec:discussion-ssc2}

The electrons responsible for the broadband emission of Mrk\,421\ lose energy mostly due to synchrotron cooling,
as one can infer from the dominance of the synchrotron bump over the inverse-Compton bump shown in the SEDs
from Figure~\ref{fig:seds}. Note that the inefficiency of cooling via the Compton channel is independently implied from the observed
slope of the X-ray--VHE flux correlation. The observed variability timescale
(measured in a stationary observer's frame) due to synchrotron cooling alone is given by 
$\tau_{\rm syn} = 1.2 \times 10^{3} {B}^{-3/2} E^{-1/2} \delta^{-1/2}$~s,
where $E$ is photon energy in keV, and $B$ is comoving frame magnetic field strength in G.
Taking $E\approx10$~keV as the energy typical for the {\em NuSTAR}\ band, assuming $B \approx 0.2$~G,
as found from our SED modeling, and $\delta=25$ as before, we arrive at $\tau_{\rm syn}$ of
$\sim 10^3$ seconds.\footnote{ Note that the longer synchrotron cooling timescale
discussed in \S\,\ref{sec:discussion-ssc} refers to emission at much lower energies,
below the synchrotron peak of the SED.} This is more than an order of magnitude shorter than the
 variability timescale $\tau_{\rm var}\approx9$~hours
measured in the observer's frame, as we can derive from  the {\em NuSTAR}\ light curves.
Since the synchrotron cooling timescales are so short, this requires that the electron
acceleration must be happening locally, very close to where the emission takes place.

Considering the disparity between the variability timescale
and the synchrotron cooling timescale, along with the similarity of
the increases and decreases in flux during the {\em NuSTAR}\ observations
(Figure~\ref{fig:nustar_count_rates}), it seems unlikely that the
output is dominated by a single shocked region as a site of particle
acceleration, such as is often argued to be the case in flaring
episodes. Instead, we can interpret the flux changes as a geometrical
effect due to a spatially extended region containing multiple particle-acceleration
zones contributing comparably to the overall SED.
Observation of variability due to geometrical effects of a spatially extended region
would lack sharp flux increases in the X-ray band which might result from sudden particle-acceleration events,
because the sharp flux increases and decreases from the different regions (even if partially connected)
would not occur at exactly the same time. In this scenario, the shortest variability timescales, comparable
to the electron cooling timescales, would be produced only when a single region dominates the overall emission,
which is expected to occur during flaring episodes, but not during the relatively low activity reported in this paper.
As described in \S\,\ref{sec:nustar-variability}, the observed increases appear at least as smooth
and as slow as the observed decreases, in consistency with this picture.

One may argue that the X-ray flux variability reported in
\S\,\ref{sec:nustar-variability} is not due to acceleration/cooling of
electrons, but rather produced by variations in the parameters related
to the environment (e.g., $B$, $R$) or the Dopper factor $\delta$
(e.g., due to a change in the viewing angle). In that case, the smooth
and relatively slow changes observed in the {\em NuSTAR}\ light curves
would not be related to the short electron cooling timescales derived
above, but rather to the variations in the above-mentioned
parameters. However, such a theoretical scenario is strongly disfavored by the fractional variability as a function of energy reported in Figure~\ref{fig:fvar}, as well as by the lack of correlation between optical and X-ray fluxes reported in Figure~\ref{fig:uv_fluxflux}, while there is a correlation between optical and GeV $\gamma$-ray fluxes, as well as X-ray and VHE fluxes, reported in Figures~\ref{fig:uv_fluxflux} and \ref{fig:Xray-VHE_fluxflux}, respectively. The only possibility for the parameters $R$, $B$ or $\delta$ to dominate the measured flux variations would be to have, at least, two distinct emission regions, one dominating the optical and GeV $\gamma$-ray bands, and the other dominating the X-ray and VHE bands. Therefore, despite the success of the one-zone SSC scenario in describing the broadband SED (see \S\,\ref{sec:discussion-ssc}), we argue that the observed multiwavelength variability and correlations
point towards an emission region composed of several distinct zones, 
and dominated by changes in the electron energy distribution. The increase in the fractional variability with energy for both SED
bumps, and the harder-when-brighter trend that is clearly observed in the X-ray spectra measured with {\em NuSTAR}\ (which is the
segment of the broadband SED reconstructed with the highest accuracy), indicate that the changes in the electron energy
distribution are generally chromatic\footnote{In the sense of larger relative
increase at higher energies.}, with strongest variability in the highest-energy electrons. However, the saturation of the X-ray spectral shape at the lowest and highest X-ray fluxes (see \S\,\ref{sec:nustar-fluxsep} and Figure~\ref{fig:comp_giebels}) suggests that at the times of lowest and highest activity, the variations in the electron energy distribution become achromatic, at least for those electron energies responsible for the X-ray emission. It is possible that at those times the variability is not dominated by acceleration and cooling of the electrons, but rather by variations in the physical parameters of the environment in which particle acceleration occurs. For the periods of very low activity, a possibility would be that the radiation is being produced within a larger region by particles accelerated by Fermi~II processes (e.g., stochastic acceleration on magnetic turbulence), as suggested, for instance, by \citet{massaro+2004} and \citet{ushio+2009}.

The magnetic field implies a size constraint for the acceleration zones, since electrons cannot attain
energies corresponding to a gyroradius significantly larger than the
characteristic size of a zone. The {\em NuSTAR}\ data imply no cutoff in
the synchrotron SED up to $\sim80$\,keV, so we can estimate the electron gyroradius
$R_{\rm G}$ corresponding to that photon energy using $B = 0.2$\,G and the maximal $\elgamma \sim 10^{6}$.
Since $R_{\rm G} = \elgamma m_{\rm e} c^2 e^{-1} B^{-1}$, we have $R_{\rm G}
\lesssim 10^{11}$~cm, which is much smaller than the inferred emission-region size of $10^{16}$~cm.
Given the large difference of five orders of magnitude between the gyroradius for the highest-energy electrons
and the size of the overall emitting region, the electrons cannot travel far from their acceleration site without
losing a substantial fraction of their energy, and hence the particle acceleration and the emission need to be essentially co-spatial.
We therefore conclude that the set of physical parameters discussed
here offers a self-consistent picture in which the observed properties
of Mrk\,421\ in a nonflaring state are consistent with compact zones of particle acceleration
distributed within a significantly larger volume that produces the total emission.
While detailed characterization of the acceleration process is outside the scope of the paper,
one possible scenario involves magnetic reconnection and ``mini-jets'' formed within a larger emission volume,
as suggested, for instance, by \citet{nalewajko+2011}, and developed further by \citet{nalewajko+2015}.
For a recent summary of arguments in favor of magnetic reconnection for powering blazar jets, the reader is referred to \citet{sironi+2015}.

Regardless of the exact acceleration mechanism, emitting regions composed of multiple zones,
e.g, as in the model proposed by \citet{2014ApJ...780...87M}, would be consistent with other behavior observed in blazars,
such as the increase in the degree of polarization of the synchrotron radiation when the polarization electric
vector rotates, or curvature in the SED arising from non-uniform particle acceleration and energy losses.
In a low-activity state, where no single zone dominates the output, the addition of polarization vectors
from individual zones would result in a low overall level of polarization with random fluctuations in both the polarization degree and angle.
Our optical polarization measurements, shown in Figure~\ref{fig:optpol}, are consistent with that prediction.
While multizone scenarios have previously been considered for flaring states
(e.g., \citealt{massaro+2004,ushio+2009,cao+wang-2013,aleksic+2015c}),
it has usually been assumed that the quiescent state can be well described by
a simpler single-zone SSC model (e.g., \citealt{abdo+2011}). The observations presented here,
however, show that, even in this state of very low activity, the emission region may have a more
complex structure than previously assumed.

\section{Summary and Conclusions} 

\label{sec:summary}

We have observed the blazar Mrk\,421\ in an intensive multiwavelength campaign in 2013,
including GASP-WEBT, \swift$\!\!$, {\em Fermi}\,-LAT, \magic$\!\!$, \veritas$\!\!$, and,
for the first time, the new high-sensitivity hard X-ray instrument {\em NuSTAR}.
In this paper we present part of the data from the campaign between the 
beginning of January and the end of March 2013, with the focus on the 
unprecedented coverage of the X-ray part of the broadband spectrum.  
Another successful aspect of the campaign is the achieved goal of strictly 
simultaneous observations in the X-ray and VHE $\gamma$-ray bands, 
in order to constrain the correlated variability. During the data-taking 
period presented in this work, Mrk\,421\ exhibited relatively low activity, 
including the lowest-flux state ever investigated with high temporal 
and broadband spectral coverage. 

The rich data set yields a number of important empirical results:

\begin{itemize}

  \item During the first three months of 2013, the X-ray and VHE $\gamma$-ray activity of Mrk\,421\ was among the lowest ever observed. 

  \item {\em NuSTAR}\ performed half-day long observations of Mrk\,421\, which showed that this source varies predominantly on timescales of several hours, with  multiple instances of exponentially varying flux on timescales of 6--12~hours. Mrk\,421\ also exhibited smaller-amplitude, intrahour variations at the $\lesssim\,5$\,\% level.  However, only $\lesssim\,20$\,\% of the
    X-ray data show any appreciable intrahour variability. Within the
    dynamic range of our observations, we find no differences in the
    variability pattern or timescales between the lower and higher
    flux states.
    
  \item We find a systematic model-independent hardening of the X-ray spectrum with
    increasing X-ray flux. As the X-ray activity decreases, the curvature in the X-ray spectrum decreases and the spectral shape becomes softer. At 2--10~keV fluxes $\lesssim 10^{-10}$~erg\,s$^{-1}$\,cm$^{-2}$, the spectral curvature completely disappears, and the spectral shape saturates into a steep $\Gamma\approx3$ power law, with no evidence for an
    exponential cutoff or additional hard components up to $\simeq80$~keV.

\item For two epochs of extremely low X-ray and VHE flux, in a regime not previously reported in the literature, 
    we observed atypically steep spectral slopes with $\Gamma\approx3$ in both X-ray and VHE bands.
    Using a simple steady-state one-zone SSC scenario, we find that in these two epochs the peaks of both the
    synchrotron and inverse-Compton components of the SED shifted
    towards lower frequencies by more than an order of magnitude
    compared to their positions in the typical low states of Mrk\,421\ observed previously.
    The peak of the synchrotron bump of Mrk\,421\ shifted from $\sim0.5-1$~keV to $\sim0.04$~keV, which implies that HBLs can move towards becoming IBLs, leading to the conclusion that the SED classification based on the peak of the synchrotron bump may denote only a temporary rather than permanent characteristic of blazars. 

\item A clear double-bump structure is found in the fractional variability distribution, computed from radio to VHE $\gamma$-ray energies. 
This double-bump structure relates to the two peaks in the broadband SED shape of Mrk\,421, and has been recently reported (with less resolution) for both low-activity \citep{aleksic+2015b} and high-activity states \citep{aleksic+2015c}. The less variable energy bands (radio, optical/UV and GeV $\gamma$-rays) relate to the segments of the SED rising up towards the peaks as a function of photon energy, while the most variable energy bands (X-rays and VHE $\gamma$-rays) sample the SED above the peaks, where it steeply declines with photon energy.

 \item We find a tight X-ray--VHE flux correlation in three nonoverlapping X-ray bands between
    0.3 and 30~keV, with significantly different scaling.
     These results are consistent with an SSC scenario in which the X-ray and VHE
    radiation are produced by the same relativistic electrons, and the
    scattering of X-ray photons to VHE energies ($\sim$TeV) occurs in the less-efficient Klein-Nishina regime.
    From broadband SED modeling with a single-zone SSC model for four epochs, and assuming a constant Doppler factor of 25, we infer a
    magnetic field $B \sim 0.2$~G and electron Lorentz factors
    as large as $\elgamma \gtrsim 6 \times 10^{5}$.  These
    parameter values, which are typical for describing the broadband SED of HBLs,  further support the claim that,
    in the context of the SSC model, the
    inverse-Compton scattering responsible for the VHE emission takes place 
    in the Klein-Nishina regime.

 \item There is tentative evidence for an optical/UV--GeV flux
    correlation, which is consistent with the emission in these two bands being
    produced by the same lower-energy electrons within the SSC framework. 

\item No correlation is found
    between fluxes in the optical/UV and the soft X-ray bands on either short or long timescales. However, we do find that a simple parametrization
    of the SED around the synchrotron peak with a log-parabolic function leads to a correlation between the peak flux and the frequency at which it occurs over a limited frequency range.

\item The reported increase in the fractional variability with energy (for each of the two SED bumps), and the hardening of the X-ray spectra with increasing flux, suggest that the variability in the emission of Mrk\,421\ is produced by chromatic changes in the electron energy distribution, with the highest-energy electrons varying the most. The saturation of the X-ray spectral shape at the extremely high and low X-ray fluxes indicates that, for these periods of outstanding activity, the flux variability is instead dominated by other processes that lead to achromatic variations in the X-ray emission.

\item The lifetimes of relativistic electrons due
    to synchrotron losses are estimated to be $\tau_{\rm syn}\lesssim10^3$~seconds, which are substantially shorter than the $\sim3\times10^4$~seconds that dominate the large-amplitude variations in the {\em NuSTAR}\ light curves. Together with the fractional variability distribution and the multiwavelength correlations observed in this campaign, this observation suggests that the broadband emission of Mrk\,421\ during low activity is produced by multiple emission regions.

  \item The electron cooling times of $\tau_{\rm syn}\lesssim10^3$\,seconds are also shorter than the emission-region crossing
    time ($\gtrsim10^4$\, seconds), which points towards {\it in situ} electron acceleration.
    While particle acceleration in shocks is not excluded by our data,
    the gyroradii of the most energetic electrons (those radiating in the
    upper part of the {\em NuSTAR}\ band, or the upper part of the VHE
    band) is $\lesssim 10^{11}$~cm, which is shorter than the cooling
    (energy-loss) timescales inferred from our modeling. This is suggestive of
    an electron-acceleration process occurring in relatively compact zones within a larger emission volume.

\end{itemize}

\acknowledgments 

We thank the anonymous referee for constructive suggestions that helped to improve and clarify the paper.

M.\,B. acknowledges support from the International Fulbright Science and Technology Award, and from NASA Headquarters under the NASA Earth and Space Science Fellowship Program, grant NNX14AQ07H. This research was supported in part by the Department of Energy Contract DE-AC02-76SF00515 to the SLAC National Accelerator Center. G.\,M. and A.\,F. acknowledge the support via NASA grant NNX13AO97G. D.\,B. acknowledges support from the French Space Agency (CNES) for financial support.

This work was supported under NASA Contract No.~NNG08FD60C, and made use of data from the {\em NuSTAR}\ mission, a project led by the California Institute of Technology, managed by the Jet Propulsion Laboratory, and funded by the National Aeronautics and Space  Administration. We thank the {\em NuSTAR}\ Operations, Software and  Calibration teams for support with the execution and analysis of these observations. This research has made use of the {\em NuSTAR}\ Data Analysis Software (NuSTARDAS) jointly developed by the ASI Science Data Center (ASDC, Italy) and the California Institute of Technology (USA). 

\veritas is supported by grants from the U.S. Department of Energy Office of Science, the U.S. National Science Foundation and the Smithsonian Institution, by NSERC in Canada, and by STFC in the U.K. We acknowledge the excellent work of the technical support staff at the Fred Lawrence Whipple Observatory and at the collaborating institutions in the construction and operation of the instrument. The \veritas Collaboration is grateful to Trevor Weekes for his seminal contributions and leadership in the field of VHE gamma-ray astrophysics, which made this study possible.

The \magic collaboration would like to thank the Instituto de Astrof\'{\i}sica de Canarias for the excellent working conditions at the Observatorio del Roque de los Muchachos in La Palma. The financial support of the German BMBF and MPG, the Italian INFN and INAF, the Swiss National Fund SNF, the ERDF under the Spanish MINECO, and the Japanese JSPS and MEXT is gratefully acknowledged. This work was also supported by the Centro de Excelencia Severo Ochoa SEV-2012-0234, CPAN CSD2007-00042, and MultiDark CSD2009-00064 projects of the Spanish Consolider-Ingenio 2010 programme, by grant 268740 of the Academy of Finland, by the Croatian Science Foundation (HrZZ) Project 09/176 and the University of Rijeka Project 13.12.1.3.02, by the DFG Collaborative Research Centers SFB823/C4 and SFB876/C3, and by the Polish MNiSzW grant 745/N-HESS-MAGIC/2010/0.

The {\em Fermi}\,-LAT\ Collaboration acknowledges generous ongoing support
from a number of agencies and institutes that have supported both the
development and the operation of the LAT as well as scientific data analysis.
These include the National Aeronautics and Space Administration and the
Department of Energy in the United States, the Commissariat \`a l'Energie Atomique
and the Centre National de la Recherche Scientifique / Institut National de Physique
Nucl\'eaire et de Physique des Particules in France, the Agenzia Spaziale Italiana
and the Istituto Nazionale di Fisica Nucleare in Italy, the Ministry of Education,
Culture, Sports, Science and Technology (MEXT), High Energy Accelerator Research
Organization (KEK) and Japan Aerospace Exploration Agency (JAXA) in Japan, and
the K.~A.~Wallenberg Foundation, the Swedish Research Council and the
Swedish National Space Board in Sweden. Additional support for science analysis
during the operations phase is gratefully acknowledged from the Istituto
Nazionale di Astrofisica in Italy and the Centre National d'\'Etudes Spatiales in France.

This research has made use of the XRT Data Analysis Software (XRTDAS) developed under the responsibility of the ASI Science Data Center (ASDC), Italy.

St.\,Petersburg University team acknowledges support from Russian RFBR
grant 15-02-00949 and St.\,Petersburg University research
grant 6.38.335.2015.

The IAC team acknowledges the support from the group of support astronomers and telescope operators of the Observatorio del Teide.

G.\,D. and O.\,V. gratefully acknowledge
the observing grant support from the Institute of Astronomy
and Rozhen National Astronomical Observatory,
Bulgaria Academy of Sciences. This work is a part of the
Projects No 176011 (Dynamics and kinematics of celestial
bodies and systems), No 176004 (Stellar physics) and No
176021 (Visible and invisible matter in nearby galaxies: theory
and observations) supported by the Ministry of Education,
Science and Technological Development of the Republic
of Serbia.

This research was partially supported by Scientific Research Fund of the
Bulgarian Ministry of Education and Sciences under grant DO 02-137 (BIn-13/09).

The Abastumani team acknowledges financial support of the project
FR/638/6-320/12 by the Shota Rustaveli National Science Foundation under contract 31/77.

T.\,G. acknowledges support from Istanbul University (Project numbers 49429 and 48285), Bilim Akademisi (BAGEP program) and TUBITAK (project numbers 13AT100-431, 13AT100-466, and 13AT60-430).

The Boston University effort was supported in part by NASA grants NNX12AO90G and NNX14AQ58G.

Data from the Steward Observatory spectropolarimetric monitoring project were used in this paper. This program is supported by  {\em Fermi} Guest Investigator grants NNX08AW56G, NNX09AU10G, NNX12AO93G, and NNX15AU81G.

The OVRO 40-m monitoring program is supported in part by NASA grants
NNX08AW31G and NNX11A043G, and NSF grants AST-0808050 and AST-1109911.

The Mets\"ahovi team acknowledges the support from the Academy of Finland
to our observing projects (numbers 212656, 210338, 121148, and others).

This research has made use of NASA's Astrophysics Data System, and of Astropy, a community-developed core Python package for astronomy \citep{astropy-2013}.

{\em Facilities:} {\em NuSTAR}, \magic$\!\!$, \veritas$\!\!$, {\em Fermi}, {\em Swift}

{}


\begin{thebibliography}{}

\bibitem[Abdo \etal(2011)]{abdo+2011} Abdo, A.\,A., \etal\ 2011, ApJ, 736, 131 
\bibitem[Acciari \etal(2009)]{acciari+2009} Acciari, V.\,A., \etal\ 2009, ApJ, 703, 169 
\bibitem[Acciari \etal(2011)]{acciari+2011} Acciari, V.\,A., \etal\ 2011, ApJ, 738, 25 
\bibitem[Acciari \etal(2014)]{acciari+2014} Acciari, V.\,A., \etal\ 2014, APh, 54, 1 
\bibitem[Ackermann \etal(2012)]{ackermann+2012} Ackermann, M., \etal\ 2012, ApJS, 203, 4 
\bibitem[Aleksi{\' c} \etal(2012)]{aleksic+2012} Aleksi{\' c}, J., \etal\ 2012, A\&A, 542, 100 
\bibitem[Aleksi{\'c} \etal(2014)]{aleksic+2014} Aleksi{\'c}, J., \etal\ 2014, A\&A, 572, A121 
\bibitem[Aleksi{\' c} \etal(2015a)]{aleksic+2015a} Aleksi{\' c}, J., \etal\ 2015a, A\&A, 573, 50 
\bibitem[Aleksi{\' c} \etal(2015b)]{aleksic+2015b} Aleksi{\' c}, J., \etal\ 2015b, A\&A, 576, 126 
\bibitem[Aleksi{\' c} \etal(2015c)]{aleksic+2015c} Aleksi{\' c}, J., \etal\ 2015c, A\&A, 578, 22 
\bibitem[Aleksi{\' c} \etal(2016a)]{aleksic+2016a} Aleksi{\' c}, J.,  \etal\ 2016a, APh, 72, 61 
\bibitem[Aleksi{\' c} \etal(2016b)]{aleksic+2016b} Aleksi{\' c}, J., \etal\ 2016b, APh, 72, 75 
\bibitem[Aliu \etal(2013)]{aliu+2013} Aliu E., \etal\ 2013, ApJ, 779, 92 
\bibitem[Angelakis \etal(2008)]{angelakis+2008} Angelakis, E., Fuhrmann, L., Marchili, N., \etal\ 2008, MmSAI, 79, 1042 
\bibitem[Angelakis \etal(2010)]{angelakis+2010} Angelakis, E., Fuhrmann, L., Nestoras, I., \etal\ in Proceedings of the Workshop {\it Fermi meets Jansky - AGN in Radio and Gamma-Rays} (eds. Savolainen, T., Ros, E., Porcas, R.\,W. and Zensus, J.\,A.), Bonn, 2010 ({\tt arXiv:1006.5610}) 
\bibitem[Angelakis \etal(2015)]{angelakis+2015} Angelakis, E., Fuhrmann, L., Marchili, N., \etal\ 2015, A\&A, 575, 55 
\bibitem[Astropy Collaboration(2013)]{astropy-2013} Astropy Collaboration 2013, A\&A, 558, 33 
\bibitem[Atwood \etal(2009)]{atwood+2009} Atwood, W. B., \etal\ 2009, ApJ, 697, 1071
\bibitem[Baars \etal(1977)]{baars+1977} Baars, J.\,W.\,M., Genzel, R., Paulini-Toth, I.\,I.\,K., \& Witzel, A. 1977, A\&A, 61, 99 
\bibitem[Balokovi{\'c} \etal(2013a)]{april-flare-atel1} Balokovi\'{c}, M., Furniss, A., Madejski, G., \& Harrison, F.\,A. 2013, ATel~\#4974 
\bibitem[Balokovi{\'c} \etal(2013b)]{balokovic+2013-granada} Balokovi{\'c}, M., \etal\ 2013, EPJ Web of Conf., Vol. 61, 04013 ({\tt arXiv:1309.4494}) 
\bibitem[Bartoli \etal(2011)]{bartoli+2011} Bartoli, B., \etal\ 2011, ApJ, 734, 110
\bibitem[Bartoli \etal(2012)]{argo-ybj-flare-2012} Bartoli, B., \etal\ 2012, ATel~\#4272
\bibitem[Blandford \& Eichler(1987)]{blandford+eichler-1987} Blandford, R.\,D., \& Eichler, D. 1987, Phys. Rep., 154, 1 
\bibitem[Blasi \etal(2013)]{blasi+2013} Blasi, M.\,G., Lico, R., Giroletti, M., \etal\ 2013, A\&A, 559, 75 
\bibitem[B{\l}a{\. z}ejowski \etal(2005)]{blazejowski+2005} B{\l}a{\. z}ejowski, M., Blaylock, G., Bond, I.\,H., \etal\ 2005, ApJ, 630, 130 
\bibitem[Blumenthal \& Gould(1970)]{blumenthal+gould-1970} Blumenthal, G.\,R., \& Gould, R.\,J. 1970, Rev. Mod. Phys., 42, 237 
\bibitem[B\"{o}ttcher \& Chiang(2002)]{bottcher+chiang-2002} B\"{o}ttcher, M., \& Chiang, J. 2002, ApJ, 581, 127 
\bibitem[Breeveld \etal(2011)]{breeveld+2011} Breeveld, A.\,A., Landsman, W., Holland, S.\,T., \etal\ 2011, AIPC, 1358, 373 
\bibitem[Burrows \etal(2005)]{burrows+2005} Burrows, D.\,N., Hill, J.\,E., Nousek, J.\,A. \etal\ 2005, SSRv, 120, 165 
\bibitem[Cao \& Wang(2013)]{cao+wang-2013} Cao, G., \& Wang, J. 2013, PASJ, 65, 109
\bibitem[Celotti \etal(1998)]{celotti+1998} Celotti, A., Fabian, A.\,C., Rees, M.\,J. 1998, MNRAS, 293, 239 
\bibitem[Cogan(2008)]{cogan-2008} Cogan, P., 2008, in Proceedings of the 30th ICRC, Vol.~3, p.~1385
\bibitem[Cortina \etal(2013)]{april-flare-atel2} Cortina, J., \etal\ 2013, ATel~\#4976
\bibitem[D'Ammando \etal(2012)]{fermi-flare-2012} D'Ammando, F., \etal\ 2012, ATel~\#4261
\bibitem[Daniel(2008)]{daniel-2008} Daniel, M.\ 2008, in Proceedings of the 30th ICRC, Vol.~3, p.~1325
\bibitem[Edelson \& Krolik(1989)]{edelson+krolik-1988} Edelson, R.\,A., \& Krolik, J.\,H.\ 1988, ApJ, 333, 646 
\bibitem[Fitzpatrick(1999)]{fitzpatrick-1999} Fitzpatrick, E.\,L. 1999, PASP, 111, 63
\bibitem[Fomin \etal(1994)]{fomin+1994} Fomin, V.\,P., Stepanian, A.\,A., Lamb, R. \etal\ 1994, APh, 2, 137 
\bibitem[Fossati \etal(2008)]{fossati+2008} Fossati, G., Buckley, J.\,H., Bond, I.\,H.\ \etal\ 2008, ApJ, 677, 906 
\bibitem[Franceschini \etal(2008)]{franceschini+2008} Franceschini, A., Rodighiero, G., Vaccari, M. 2008, A\&A, 487, 837 
\bibitem[Fuhrmann \etal(2007)]{fuhrmann+2007} Fuhrmann, L., Zensus, J.\,A., Krichbaum, T.\,P., Angelakis, E., Readhead, A.\,C.\,S. 2007, AIP Conf. Ser., Vol. 921 (eds. S. Ritz, P. Michelson, \& C.\,A. Meegan), 249--251 
\bibitem[Fuhrmann \etal(2014)]{fuhrmann+2014} Fuhrmann, L., Larsson, S., Chiang, J., \etal\ MNRAS, 441, 1899 
\bibitem[Furniss \etal(2015)]{furniss+2015} Furniss, A., Koda, N., Madejski, G., \etal\ 2015, ApJ, 812, 65 
\bibitem[Gaidos \etal(1996)]{gaidos+1996} Gaidos, J.\,A., Akerlof, C.\,W., Biller, S., \etal\ 1996, Nature, 383, 319 
\bibitem[Galante(2011)]{galante+2011} Galante, N.\ 2011, Proceedings of the 32nd ICRC, Vol.~8, p.~63 
\bibitem[Georganopoulos \& Kazanas(2003)]{2003ApJ...594L..27G} Georganopoulos, M., \& Kazanas, D.\ 2003, ApJL, 594, L27 
\bibitem[Ghisellini \etal(1985)]{ghisellini+1985} Ghisellini, G., Maraschi, L., \& Treves, A. 1985, A\&A, 146, 204 
\bibitem[Ghisellini \etal(2005)]{2005A&A...432..401G} Ghisellini, G., Tavecchio, F., \& Chiaberge, M.\ 2005, A\&A, 432, 401
\bibitem[Giebels \etal(2007)]{giebels+2007} {Giebels}, B. and {Dubus}, G. and {Kh{\'e}lifi}, B. 2007, A\&A, 462, 29
\bibitem[Guainazzi \etal(1999)]{guainazzi+1999} Guainazzi, M., Vacanti, G., Malizia, A., \etal 1999, A\&A, 342, 124 
\bibitem[Harrison \etal(2013)]{harrison+2013} Harrison, F.\,A., \etal\ 2013, ApJ, 770, 103 
\bibitem[Henri \& Saug{\'e}(2006)]{2006ApJ...640..185H} Henri, G., \& Saug{\'e}, L.\ 2006, ApJ, 640, 185 
\bibitem[Hillas(1985)]{hillas-1985} Hillas, A.\,M.\ 1985, Proceedings of the 19th ICRC, Vol.~3, p.~445 
\bibitem[Holder \etal(2006)]{holder+2006} Holder, J., \etal\ 2006, APh, 25, 391
\bibitem[Hovatta \etal(2012)]{ovro-flare-2012} Hovatta, T., \etal\ 2012, ATel~\#4451
\bibitem[Hovatta \etal(2015)]{hovatta+2015} Hovatta, T., Petropoulou, M., Richards, J.\,L., \etal\ 2015, MNRAS ,448, 3121 
\bibitem[Janiak \etal(2012)]{janiak+2012} Janiak, M., Sikora, M., Nalewajko, K., \etal\ 2012, ApJ 760, 129 
\bibitem[Jones \etal(1974)]{jones+1974} Jones, F.\,C., O'Dell, S.\,L., Stein, W.\,A. 1974, ApJ, 188, 353 
\bibitem[Jones \& Ellison(1991)]{jones+ellison-1991} Jones, F.\,C., Ellison, D.\,C. 1991, Space Sci. Rev. 58, 259 
\bibitem[Jorstad \etal(2010)]{2010ApJ...715..362J} Jorstad, S.\,G., Marscher, A.\,P., Larionov, V.\,M., \etal\ 2010, ApJ, 715, 362 
\bibitem[Kalberla \etal(2005)]{kalberla+2005} Kalberla, P.\,M.\,W., Burton, W.\,B., Hartmann, D., \etal 2005, A\&A, 440, 775 
\bibitem[Katarzy\'{n}ski \etal(2005)]{katarzynski+2005} Katarzy\'{n}ski, K., Ghisellini, G., Tavecchio, F., \etal\ 2005, A\&A, 433, 479 
\bibitem[Landau \etal(1986)]{landau+1986} Landau, R., Golisch, B., \& Jones, T.\,J.\ \etal\ 1986, ApJ, 308, 78 
\bibitem[Larionov \etal(2008)]{2008A&A...492..389L} Larionov, V.\,M., Jorstad, S.\,G., Marscher, A.\,P., \etal\ 2008, A\&A, 492, 389 
\bibitem[Li \& Ma(1983)]{li+ma-1983} Li, T. \& Ma, Y.\ 1983, ApJ, 272, 317 
\bibitem[Li \etal(2006)]{li+2006} Li, W., Jha, S., Filippenko, A.\,V., \etal\ 2006, PASP, 118, 37 
\bibitem[Lichti \etal(2008)]{lichti+2008} Lichti, G.\,G., Bottacini, E., Ajello, M., \etal\ 2008, A\&A, 486, 721 
\bibitem[Lister \etal(2013)]{lister+2013} Lister, M.\,L., Aller, M.\,F., Aller, H.\,D., \etal\ 2013, AJ, 146, 120 
\bibitem[Madhavan(2013)]{madhavan-2013} Madhavan, A. 2013, PhD Thesis, Iowa State University 
\bibitem[Madsen \etal(2015)]{madsen+2015} Madsen, K.\,K., Harrison, F.\,A., Markwardt, C.\,B., \etal\ 2015, ApJS, 220, 8 
\bibitem[Makino \etal(1987)]{makino+1987} Makino, F., Tanaka, Y., Matsuoka, M., \etal\ 1987, ApJ, 313, 662 
\bibitem[Malizia \etal(2000)]{malizia+2000} Malizia, A., Capalbi, M., Fiore, F., \etal\ 2000, MNRAS, 312, 123 
\bibitem[Mankuzhiyil \etal(2011)]{mankuzhiyil+2011} Mankuzhiyil, N., Ansoldi, S., Persic, \& M., Tavecchio, F.\ 2011, ApJ, 733, 14 
\bibitem[Marscher \& Gear(1985)]{marscher+gear-1985} Marscher, A.\,P., \& Gear, W.\,K.\ 1985, ApJ, 198, 114 
\bibitem[Marscher \etal(2008)]{marscher+2008} Marscher, A.\,P., Jorstad, S.\,G., D'Arcangelo, F.\,D., \etal\ 2008, Nature, 452, 966 
\bibitem[Marscher(2014)]{2014ApJ...780...87M} Marscher, A.\,P.\ 2014, ApJ, 780, 87  
\bibitem[Massaro \etal(2004)]{massaro+2004} Massaro, E., Perri, M., Giommi, P., \& Nesci, R.\ 2004, A\&A, 413, 489 
\bibitem[Max-Moerbeck \etal(2014)]{maxMoerbeck+2014} Max-Moerbeck, W., Hovatta, T., Richards, J.\,L., \etal\ 2014, MNRAS, 445, 428 
\bibitem[Moralejo \etal(2009)]{moralejo+2009} Moralejo, A., Gaug, M., Carmona, E., \etal\ 2009, Proceedings of the 31st ICRC ({\tt arXiv:0907.0943}) 
\bibitem[Nalewajko \etal(2011)]{nalewajko+2011} Nalewajko, K., Giannios, D., Begelman, M.\,C., \etal\ 2011, MNRAS, 413, 333 
\bibitem[Nalewajko \etal(2015)]{nalewajko+2015} Nalewajko, K., Uzdensky, D.\,A., Cerutti, B., \etal\ 2015, ApJ, submitted ({\tt arXiv:1508.02392}) 
\bibitem[Nilsson \etal(2007)]{nilsson+2007} Nilsson, K., Pasanen, M., Takalo, L.~O., \etal\ 2007, A\&A, 475, 199
\bibitem[Nolan \etal(2012)]{nolan+2012} Nolan, P.~L., \etal\ 2012, ApJS, 199, 31 
\bibitem[Paneque \etal(2013)]{april-flare-atel3} Paneque, D., D'Ammando, F., Orienti, M., \etal\ 2013, ATel~\#4977 
\bibitem[Pian \etal(2014)]{pian+2014} Pian, E., T\"{u}rler, M., Fiocchi, M., \etal\ 2014, A\&A, 570, 77 
\bibitem[Piner \etal(2010)]{Piner10} Piner, B.~G., Pant, N., \& Edwards, P.~G.\ 2010, ApJ, 723, 1150 
\bibitem[Ade \etal(2014)]{planck-cosmology-2014} Ade, P.\,A.\,R., \etal\ 2014, A\&A, 571, 16 
\bibitem[Poutanen \etal(2008)]{poutanen+2008} Poutanen, J., Zdziarski, A.\,A., \& Ibragimov, A.\ 2008, MNRAS, 389, 1427 
\bibitem[Punch \etal(1992)]{punch+1992} Punch, M., \etal, 1992, Nature, 358, 477 
\bibitem[Ravasio \etal(2004)]{ravasio+2004} Ravasio, M., Tagliaferri, G., Ghisellini, G., \& Tavecchio, F.\ 2004, A\&A, 424, 841 
\bibitem[Richards \etal(2011)]{richards+2011} Richards, J.~L., Max-Moerbeck, W. Pavlidou, V., \etal\ 2011, ApJS, 194, 29 
\bibitem[Richards \etal(2013)]{richards+2013-granada} Richards, J.~L., Hovatta, T., Lister, M.\,I., \etal\ 2013, EPJ Web of Conf., Vol. 61, 04010 
\bibitem[Roming \etal(2005)]{roming+2005} Roming, P.\,W.\,A., Kennedy, T.\,E., Mason, K.\,O., \etal\ 2005, SSRv, 120, 95 
\bibitem[Schlafly \etal(2011)]{schlafly+2011} Schlafly, E.\,F., \& Finkbeiner, D.\,P.\ 2011, ApJ, 737, 103S 
\bibitem[Schlegel \etal(1998)]{schlegel+1998} Schlegel, D.\,J., Finkbeiner, D.\,P., \& Davis, M. 1998, ApJ, 500, 525 
\bibitem[Sikora \etal(2009)]{sikora+2009} Sikora, M., Stawarz, {\L}., Moderski, R., \etal\ 2009, ApJ, 704, 38 
\bibitem[Sironi \etal(2015)]{sironi+2015} Sironi, L., Petropoulou, M., Giannios, D. 2015, MNRAS, 450, 183 
\bibitem[Smith \etal(2009)]{smith+2009} Smith, P.\,S., Montiel, E., Rightley, S., \etal\ 2009, Fermi Symposium eConf Proceedings C091122 ({\tt arXiv:0912.3621}) 
\bibitem[Stroh \& Falcone(2013)]{stroh+falcone-2013}  Stroh, M.\,C. \& Falcone,A.\,D.\ 2013, ApJS, 207, 28 
\bibitem[Takahashi \etal(1996)]{takahashi+1996} Takahashi, T., Tashiro, M., Madejski, G.\,M., \etal 1996, ApJ, 470, 89 
\bibitem[Tanihata \etal(2004)]{tanihata+2004} Tanihata, C., Kataoka, J., Takahashi, T., Madejski, G.\,M. 2004, ApJ, 601, 759 
\bibitem[Tanihata \etal(2003)]{tanihata+2003} Tanihata, C., Takahashi, T., Kataoka, J., Madejski, G.\,M. 2003, ApJ, 584, 153 
\bibitem[Tavecchio \etal(2010)]{tavecchio+2010} Tavecchio, F., Ghisellini, G., Ghirlanda, G., \etal\ 2010, MNRAS, 401, 1570 
\bibitem[Ter{\"a}sranta \etal(1998)]{terasranta+1998} Ter{\"a}sranta, H., Tornikoski, M., Mujunen, A., \etal\ 1998, A\&AS, 132, 305  
\bibitem[Tramacere \etal(2007a)]{tramacere+2007a} Tramacere, A., Giommi, P., Massaro, E., \etal\ 2007, A\&A, 467, 501 
\bibitem[Tramacere \etal(2007b)]{tramacere+2007b} Tramacere, A., Massaro, F., \& Cavaliere, A.\ 2007, A\&A, 466, 521 
\bibitem[Tramacere \etal(2009)]{tramacere+2009} Tramacere, A., Giommi, P., Perri, M., \etal\ 2009, A\&A, 501, 879 
\bibitem[Tramacere \etal(2011)]{tramacere+2011} Tramacere, A., Massaro, E., \& Taylor, A.\,M.\ 2011, ApJ, 739, 66 
\bibitem[Ulrich \etal(1975)]{ulrich+1975} Ulrich, M.~H., Kinman, T.\,D., Lynds, C.\,R., \etal\ 1975, ApJ, 198, 261 
\bibitem[Ulrich \etal(1997)]{ulrich+1997} Ulrich, M.-H., Maraschi, L., \& Urry, C.\,M.\ 1997, ARA\&A, 35, 445
\bibitem[Urry \& Padovani(1995)]{urry+padovani-1995} Urry, C.~M., \& Padovani, P.\ 1995, PASP, 107, 803 
\bibitem[Ushio \etal(2009)]{ushio+2009} Ushio, M., Tanaka, T., Madejski, G., \etal\ 2009, ApJ, 699, 1964 
\bibitem[Ushio \etal(2010)]{ushio+2010} Ushio, M., Stawarz, {\L}, Takahashi, T., \etal\ 2010, ApJ, 724, 1509 
\bibitem[Vaughan \etal(2003)]{vaughan+2003} Vaughan, S., Edelson, R., Warwick, R.\,S., \& Uttley, P.\ 2003, MNRAS, 345, 1271 
\bibitem[Villata \etal(1998)]{Villata1998} Villata, M., Raiteri, C.\,M., Lanteri, L., Sobrito, G., \& Cavallone, M.\ 1998, A\&AS, 130, 305
\bibitem[Villata \etal(2008)]{Villata2008} Villata, M., Raiteri, C.\,M., Larionov, V.\,M., \etal\ 2008, A\&A, 481, L79
\bibitem[Villata \etal(2009)]{Villata2009} Villata, M., Raiteri, C.\,M., Gurwell, M.\,A., \etal\ 2009, A\&A, 504, L9
\bibitem[Weekes \etal(2002)]{weekes+2002} Weekes, T.~C., \etal, 2002, APh, 17, 221
\end{thebibliography}
\end{document}